\documentclass[a4paper,11pt]{article}
\usepackage{jcappub}

\pdfoutput=1

\usepackage{tabularx,booktabs}
\newcolumntype{Y}{>{\centering\arraybackslash}X}
\usepackage{fontawesome}
\usepackage[utf8]{inputenc}
\usepackage[font=small,labelfont=bf,tableposition=top]{caption}
\usepackage{float}
\usepackage{adjustbox}
\usepackage{arydshln}
\usepackage{empheq}

\usepackage{amsmath,amssymb,amsbsy,amstext, amsthm, simplewick, amsfonts, bm}
\usepackage{hyperref}
\usepackage[normalem]{ulem}
\usepackage{graphicx}
\usepackage{siunitx}
\usepackage{upgreek}
\usepackage{framed}
\usepackage{wrapfig}
\usepackage{multirow}
\usepackage{subfig}
\usepackage{bbm}
\usepackage[nameinlink, capitalise]{cleveref}
\usepackage[svgnames,dvipsnames,x11names]{xcolor}
\usepackage{wrapfig}
\usepackage{soul}
\usepackage[htt]{hyphenat}
\usepackage{orcidlink}

\newcommand{\CLASS}{\texttt{CLASS}}

\title{Spectral distortions from acoustic dissipation with non-Gaussian (or not) perturbations}

\author[1]{Devanshu Sharma\,\orcidlink{0009-0002-3302-2153},}
\author[1]{Julien Lesgourgues\,\orcidlink{0000-0001-7627-353X},}
\author[2]{Christian T.~Byrnes\,\orcidlink{0000-0003-2583-6536}}

\affiliation[1]{Institute for Theoretical Particle Physics and Cosmology (TTK), RWTH Aachen University, \\ D-52056 Aachen, Germany}
\affiliation[2]{ Department of Physics and Astronomy, University of Sussex, Brighton BN1 9QH, UK\\}

\emailAdd{drsharma@physik.rwth-aachen.de}
\emailAdd{lesgourg@physik.rwth-aachen.de}
\emailAdd{C.Byrnes@sussex.ac.uk}

\date{}

\abstract{A well-known route to form primordial black holes in the early universe relies on the existence of unusually large primordial curvature fluctuations, confined to a narrow range of wavelengths that would be too small to be constrained by Cosmic Microwave Background (CMB) anisotropies. This scenario would however boost the generation of $\mu$-type spectral distortions in the CMB due to an enhanced dissipation of acoustic waves. Previous studies of $\mu$-distortion bounds on the primordial spectrum were based on the assumptions of Gaussian primordial fluctuations. In this work, we push the calculation of $\mu$-distortions to one higher order in photon anisotropies. We discuss how to derive bounds on primordial spectrum peaks obeying non-Gaussian statistics under the assumption of local (perturbative or not) non-Gaussianity. We find that, depending on the value of the peak scale, the bounds may either remain stable or get tighter by several orders of magnitude, but only when the departure from Gaussian statistics is very strong. Our results are translated in terms of bounds on primordial supermassive black hole mass in a companion paper.} 

\begin{document}

%\hfill{\small TTK-23-30}
\begin{flushleft}
TTK-24-15
\end{flushleft}

\maketitle
\flushbottom
\section{Introduction}

Formulated in the late 70s, the theory of cosmic inflation beautifully predicts the origin of the fluctuations that provided the seeds for all cosmological structures, on top of answering the limitations of the standard Big Bang model \cite{Starobinsky:1980te,Guth:1980zm,Albrecht:1982wi, Guth:1980zm, Linde:1981mu}. In its elementary form, inflation predicts adiabatic and Gaussian primordial perturbations with a nearly scale-invariant scalar power spectrum. These predictions have been confirmed to high accuracy by the space missions COBE \cite{COBE:1992syq}, WMAP \cite{WMAP:2003elm}, and PLANCK \cite{Planck:2018vyg, Planck:2018jri, Planck:2018nkj}. Within the range of scales probed by CMB maps, which spans about three orders of magnitude in wavenumber space, primordial curvature perturbations are thus established to be nearly Gaussian, with a small power spectrum amplitude of the order of ${\cal O}(10^{-9})$. However, the amplitude of fluctuations on smaller scales is yet poorly constrained from observations and will remain so even with next-generation anisotropy measurements from e.g. CMB-S4 \cite{CMB-S4:2016ple, Abazajian:2019eic} or LiteBIRD \cite{Hazumi:2019lys, Matsumura:2013aja}. 
 
It is well known that special features in the inflationary evolution could enhance the small scales in the power spectrum, thereby potentially resulting in the formation of Primordial Black Holes (PBHs) upon the gravitational collapse of these scales after horizon re-entry \cite{1967SvA....10..602Z, Hawking:1971ei, 1974MNRAS.168..399C, Chapline:1975ojl}. PBHs have been studied extensively since the groundbreaking discovery of gravitational waves by LIGO-VIRGO collaborations as one of the viable dark matter candidates \cite{LIGOScientific:2016aoc, LIGOScientific:2016dsl, Bird:2016dcv}. Any observation allowing constraints on the primordial spectrum on scales smaller than those probed by CMB anisotropy maps would bring invaluable information on these scenarios.

% parameter $\mathcal{R}$, defined in the Newtonian gauge as

% \begin{align}
%     \mathcal{R} &= \delta \psi - \sum_{i} \frac{\delta \rho_i}{ 3(\rho_{b_i} + p_{b_i})}
% \end{align}

% where, $\delta \psi, \delta \rho$ and $\rho_b$ denote the metric perturbation, density fluctuation and background energy density, summed over all the possible energy components. 

A promising direction to get such information is related to another side of the CMB: its Spectral Distortions (SDs), first discussed in the pioneering works by Zeldovich and Sunyaev \cite{Zeldovich:1969ff, Chluba:2011hw, 1975SvA....18..413I, Sunyaev:1970er}. CMB photons are expected to behave as a nearly perfect blackbody with a Planckian spectral energy distribution. This was verified more than three decades ago by the FIRAS instrument on the COBE satellite, which measured the frequency spectrum of the CMB \cite{Fixsen:1996nj, Mather:1993ij, Mather:1990tfx}. On the other hand, tiny deviations from a perfect blackbody can be produced in the CMB by either introducing some extra energy in the photon bath or altering the number density of the CMB photons. Some of the processes that could serve this purpose are the dissipation of acoustic waves, the adiabatic cooling of electrons, some possible energy injection from PBHs via Hawking evaporation and accretion in the early universe, and the Sunyaev–Zeldovich effect coming from clusters of galaxies in the late universe \cite{Chluba:2016bvg, DeZotti:2015awh, Sunyaev:1972eq, Birkinshaw:1998qp}. The efficiency of different processes contributing to the thermalization of photons and baryons before and after recombination depends on time. Therefore, the impact of energy injection should have been redshift-dependent. This results in an epoch-based classification of spectral distortions, with  $y$-type distortions generated mainly at $z \lesssim 5 \times 10^{4}$ and $\mu$-type at $5 \times 10^{4} \lesssim z \lesssim 2 \times 10^{6}$.

Owing to the quantum origin of the density fluctuations, one would expect them to be Gaussian. This would, of course, not be true for very large density perturbations at small scales, causing the modes to interact and behave away from Gaussianity. At the scales probed by the CMB, there are stringent bounds on non-Gaussianity from PLANCK \cite{Planck:2015sxf, Planck:2019kim}, however, there are no restrictions as such for the PBH scales. 

An enhancement of the primordial power spectrum on small scales -- of the type required to generate PBHs -- would enhance the CMB spectral distortions created by the dissipation of acoustic waves. This mechanism has been studied and used to derive FIRAS bounds on the scale and amplitude of a peak in the primordial spectrum, and thus, on the density of PBHs of a given mass. 

Recently, a lot of studies have focused on PBH formation including non-Gaussian effects, implemented either as a perturbative expansion parameterised by $\tilde{f}_{\rm NL}$, $\tilde{g}_{\rm NL}$, etc~\cite{Lyth:2005fi,Byrnes:2012yx,Shandera:2012ke, Young:2013oia,Nakama:2017xvq,Atal:2019cdz,Yoo:2019pma,Ferrante:2022mui,Gow:2022jfb}.However, to the best of our knowledge, the impact on sky-averaged spectral distortions on the type of non-Gaussianity typically assumed in these works has never been quantified yet.

Since non-Gaussianity can take virtually any form, it is important to be precise about which assumptions are made before comparing different works. For instance, a form of non-Gaussianity commonly assumed in the literature is local non-Gaussianity, parameterised in real space as ${\cal R}={\cal R}_{\rm G} + f_{\rm NL} ({\cal R}_{\rm G}^2 - \langle {\cal R}_{\rm G}^2 \rangle)$, where ${\cal R}$ is the primordial curvature field and ${\cal R}_{\rm G}$ an underlying Gaussian field. This restricts primordial fluctuation models up to the choice of a primordial curvature power spectrum $P_{\cal R}(k)$ and a value of $f_{\rm NL}$. However, when writing the relation between ${\cal R}$ and ${\cal R}_{\rm G}$ in Fourier space, model builders often allow $f_{\rm NL}$ to be a function of wavenumbers, in order to cover a wider class of models with scale-dependent non-Gaussianity. The mechanisms creating primordial curvature perturbations may lead to widely different assumptions for $P_{\cal R}(k)$ and $f_{\rm NL}(k)$ and vastly different observable consequences.

Many non-Gaussian models imply some level of mode-mode coupling between the large scales that contribute the most to the CMB anisotropies observed e.g. by the {\it Planck} satellite, which range typically from ${\cal O}(1)\,$Mpc to ${\cal O}(10^3)\,$Mpc, and the small scales that contribute the most to the production of $\mu$-distortions through the dissipation of acoustic waves, which span from ${\cal O}(10^{-4})\,$Mpc to ${\cal O}(1)\,$Mpc. Therefore, it is possible to derive interesting bounds on $f_{\rm NL}(k)$ from the combination of CMB  temperature, polarisation and $\mu$-anisotropy maps. Theoretical predictions for the cross-correlation spectra are usually derived assuming the power-law curvature spectrum of the standard cosmological model, $P_{\cal R}(k)=A_s^2 (k/k_0)^{n_s-4}$, with parameters ($A_s$, $n_s$) and pivot scale $k_0$ suggested by {\it Planck} observations, in combination with a power-law non-Gaussianity function $f_{\rm NL}(k)=f_{\rm NL} (k/k_0)^{n_{f_{\rm NL}}}$ \cite{Pajer:2012vz,Ganc:2012ae,Biagetti:2013sr,Ota:2014iva,Emami:2015xqa,Khatri:2015tla,Chluba:2016aln,Ravenni:2017lgw,Cabass:2018jgj,CMB-S4:2023zem}.\footnote{Some authors also studied the correlation between temperature/polarisation and $\mu$-distortion in models with a power spectrum and bispectrum motivated by inflation with an ultra-slow roll phase \cite{Ozsoy:2021qrg,Zegeye:2021yml,Ozsoy:2021pws}.} For this model, the most recent analysis of {\it Planck} data provides an upper bound $|f_{\rm NL}(k_\mu)|<6800$ on the scales contributing the most to $\mu$-type distortions \cite{Rotti:2022lvy} (see also \cite{Bianchini:2022dqh}).

It was however pointed out in \cite{Tada:2015noa,Young:2015kda,vanLaak:2023ppj} that any significant mode-mode coupling between the large scales observed in CMB maps and the small scales potentially responsible for PBH formation is excluded by {\it Planck} unless the fraction of PBHs in dark matter is very small. Indeed, such a coupling would be effectively interpreted as very large CDM isocurvature perturbations on CMB scales, and would ruin the agreement with the observed CMB temperature and polarisation spectrum. 

Thus, if our Universe features PBHs resulting from a peak in the small-scale power spectrum that obeys non-Gaussian statistics, it is appropriate to assume that non-Gaussianity is confined to small scales only. The underlying model should lead to mode-mode coupling between small scales, but not between those scales and CMB anisotropy scales. Since these two ranges of scales are distinct, this assumption is not implausible. In such case, there are no correlations between large and small scales, or in other words, no highly squeezed configurations in the bispectrum. Thus, bounds from correlations between temperature/polarisation and $\mu$-distortion do not apply.

A simple way to fulfil this requirement is to view primordial curvature perturbations as the sum of two uncorrelated components: small Gaussian perturbations with a spectrum covering a wide range of scale (with $P_{\cal R}(k)=A_s^2 (k/k_0)^{n_s-4}$ at least over CMB anisotropy scales), 
plus very large and non-Gaussian perturbations covering a narrow range of scales around a peak value $k_*$. Both components should produce $\mu$ distortions but, for large peaks in the range of interest, the distortions will be dominated by the latter component. In order to derive FIRAS bounds on that component, it is sufficient to consider it alone, setting the former component to zero for simplicity.   

The effect on sky-averaged spectral distortions of such non-Gaussianity confined to small scales has never been studied quantitatively (apart from a crude estimate in the appendix of \cite{Juan:2022mir}). This calculation is the main purpose of the present work. In a companion paper \cite{paper2}, we extend this discussion to show the impact of such non-Gaussianity on bounds on the PBH fraction $f_{\rm PBH}$, and we generalise existing constraints on the mass and abundance of PBHs coming from $\mu$-distortions.

In \cref{distortions_theory}, we recall how to express the spectral distortions generated by the dissipation of acoustic waves as a function of photon temperature (and to a lesser extent, polarisation) anisotropies. For the first time, we perform this calculation at next-to-leading order in photon anisotropies, that is, at the level of the three-point correlation function. In \ref{Leading_order_mu}, we explain how to perform this calculation accurately at leading order in the Gaussian case. We perform advanced calculations with the Einstein-Boltzmann solve \CLASS{} to re-check the accuracy of several approximation schemes previously discussed in the literature. In \cref{sec:BGF}, we present a preliminary discussion of the possible impact of primordial local non-Gaussianity on this calculation. We then perform a detailed estimate of spectral distortions in the limit of perturbative non-Gaussianity in \cref{sec:PNG_SD} and non-perturbative non-Gaussianity in \cref{sec:NPNG}. We summarize our findings in \cref{sec:conclusions}. In \cref{app:averaging}, we provide details on the averaging of the contribution to spectral distortions, and \cref{app:gauge} we establish the gauge invariance of the heating rate sourcing spectral distortions. In \cref{app:two_point}, we recall how to expand the photon two-point correlation function in multipoles. In \cref{app:pert_NG_calc}, we present a calculation of the bispectrum of photon temperature multipoles which, to the extent of our knowledge, is new. Finally, \cref{app:PNG_Integrand_precision_check} provides details on the numerical calculation of the contribution to the heating rate arising from the photon three-point correlation function.

\section{Distortions induced by dissipation of acoustic waves}\label{distortions_theory}

The theoretical basis for spectral distortions induced by the dissipation of acoustic waves was introduced by Sunyaev and Zeldovich \cite{Sunyaev:1970er} and pushed to high precision level in references \cite{Chluba:2012gq,Khatri:2012rt,Pajer:2012qep,Chluba:2013dna}.
%If this paper was only adressing leading-order spectral distortions, we would simply refer to these papers for an in-depth description of the underlying physics. However, our main goal is to 
Here, we will push the formalism to one more order in the temperature anisotropy field. In order to root this calculation on solid ground, it is necessary to provide a summary of the derivation of spectral distortions induced by acoustic wave dissipation and to insist on a few non-trivial points relevant for higher-order calculations.

\subsection{Definition of the heating rate  \label{sec:theo_base}}

We are interested in spectral distortions that form at the approach of recombination, even in standard cosmology. Since the photon fluid features temperature anisotropies, when the Thomson mean free path increases at the approach of recombination, the photon distribution in each point can be described as a sum of blackbodies with slightly different temperatures, which is no longer a locally isotropic blackbody distribution (as illustrated e.g. in Figure 19 of \cite{Chluba:2015bqa}). Any deviation from a local isotropic blackbody distribution is processed by photon interactions to create actual spectral distortions. The most relevant interactions are Compton Scattering (CS), Double Compton scattering (DC) and Bremsstrahlung (BS). Depending on the efficiency of each of these reactions at a given time, the reprocessing of the energy contained in such deviations will end up creating either $\mu$-distortions (when thermalisation by CS is still efficient, but number-changing processes like DC and BS are not) or $y$-distortions (when CS is still active, but not at the point of maintaining perfect thermal equilibrium).

From the fluid rather than particle point of view, the fact that the Thomson mean free path increases corresponds to the dissipation of acoustic waves (that is, the gradual damping of photon fluctuations on small scales). This dissipation is caused, firstly, by heat conduction, that is, the fact that photons and baryons have a growing velocity difference $(\vec{v}_\gamma-\vec{v}_{\rm b})$, and secondly, viscosity, that is, a growing anisotropic pressure, encoded into the quadrupole and higher-order multipoles of the local temperature anisotropy field $\Theta=\delta T/\bar{T}$.

Thus, the cause of the spectral distortions we are interested in is twofold: on the one hand, heat conduction and viscosity are responsible for the superposition of different blackbodies in each point, and on the other hand, CS is responsible for the reprocessing of the local excess of energy (beyond that of a locally isotropic blackbody) into spectral distortions.

Let us denote as $s$ the dimensionless parameter describing each type of spectral distortion, spatially averaged over the whole universe at a given time $t$. More specifically, $s$ may stand for either $\mu$ or $y$ if we neglect higher-order distortions, which are even smaller. Sunyaev and Zeldovich showed that each $s$ is sourced in each point by the fraction of the photon energy that is stored in local deviations from an isotropic blackbody, $\rho_{\rm ex}|_{\rm b}(\eta,\vec{x})/\rho_\gamma|_{\rm b}(\eta,\vec{x})$, where $\eta$ is conformal time, $\rho_\gamma$ is the total photon energy, $\rho_{\rm ex}$ the energy density in excess of a locally isotropic blackbody density (with `ex' for `excess'), and the subscript b indicates that a quantity is measured in the baryon rest frame. The growth rate of the average $s$ is thus proportional to the time derivative of the spatial average of this ratio,
\begin{equation}
\dot s \propto \frac{d}{d\eta} \left\langle \frac{ \rho_{\rm ex} |_{\rm b} }{\rho_\gamma |_{\rm b}}\right\rangle_{\rho(\vec{x})}~,
\label{eq:x_dot_energy}
\end{equation}
where a dot stands for a derivative with respect to conformal time and  $\langle ... \rangle_{\rho(\vec{x})}$ denotes spatial averaging over the volume $V$ of the observable Universe weighted by the local photon density $\rho_\gamma |_{\rm b}$. Indeed, a distortion created in a given point contributes more to the average one in places where photons are denser (see \cref{app:averaging}). Thus,
\begin{equation}
\dot s \propto \frac{d}{d\eta} \left[
\frac{
\int_V d^3\vec{x} 
\,\,\, \rho_{\gamma} |_{\rm b} \,\,\, 
\frac{ \rho_{\rm ex} |_{\rm b} }{\rho_\gamma |_{\rm b}}
}{
\int_V d^3\vec{x} 
\,\,\, \rho_{\gamma} |_{\rm b}
}
\right]
=
\frac{d}{d\eta} \frac{\langle \rho_{\rm ex} |_{\rm b} \rangle_{\vec{x}}}{\bar{\rho}_\gamma}
~,
\label{eq:x_dot_energy2}
\end{equation}
with $\langle ... \rangle_{\vec{x}} = \frac{1}{V} \int_V d^3x \,\, (...)$ and $\bar{\rho}_\gamma \equiv \langle \rho_{\gamma} |_{\rm b} \rangle_{\vec{x}}$.
The proportionality factor depends on time and accounts for the fact that, depending on the efficiency of the CS, DC or BS interactions, the excess energy tends to be reprocessed dominantly into $\mu$- or $y$-distortions. Thus, it plays the role of a time-dependent branching ratio. The authors of \cite{Chluba:2013pya} performed a highly accurate calculation of such branching ratios for $\mu$ and $y$-distortions (and even higher-order distortions that we neglect in this work), denoted as ${\cal I}_s(\eta)$, such that
\begin{equation}
\dot s = {\cal I}_s(\eta) \frac{d}{d\eta}  \frac{\langle \rho_{\rm ex}|_{\rm b} \rangle_{\vec{x}}}{\bar{\rho}_\gamma}
%~~~{\color{teal} \dot s = {\cal I}_s(\eta) \frac{d}{d\eta} \left\langle \frac{ \rho_{\rm ex} |_{\rm b} }{\rho_\gamma |_{\rm b}}\right\rangle_{\vec{x}} }
%~~~{\color{DarkOrange} \dot s = {\cal I}_s(\eta) \frac{d}{d\eta} \left\langle \frac{ \rho_{\rm ex} |_{\rm b} }{\rho_\gamma |_{\rm b}}\right\rangle_{\vec{x}} }
~.
\label{eq:x_dot_branching}
\end{equation}
%A even more rough but commonly used approximation consists in approximating ${\cal I}_\mu(z)$ as a constant in Eq.~(\ref{eq:mu_Q_I}). Over the redshift range relevant for $\mu$-distortion generation, the average  value of ${\cal I}_\mu(z)$ can be shown to be 1.4, giving the simple approximation
%\begin{equation}
%    \mu \simeq - 1.4 \int dz \dot{Q} / \rho_\gamma~,
%\end{equation}
%which is equivalent to the Sunyaev \& Zeldovich (70) result.
It is important to evaluate $\rho_{\rm ex}$
%{\color{teal} and $\rho_\gamma$} 
in the baryon rest frame. As a matter of fact, the property of ``having an isotropic blackbody distribution'' is not by construction frame-invariant: if photons had a perfectly isotropic blackbody distribution in a given point in a reference frame, an observer with a peculiar motion with respect to this frame would no longer see an isotropic blackbody due to the Doppler effect -- and, after averaging the photon distribution over all directions, would actually interpret their measurement as a $y$-distortion. This is however irrelevant for the reprocessing of the photon distribution into actual spectral distortions due to CS. Such a reprocessing happens in the frame in which CS occurs, that is, the baryon rest frame. The velocity of photons is relevant for this process only when it contributes to heat conduction, that is, when there is a bulk velocity of photons relative to baryons, $(\vec{v}_\gamma-\vec{v}_{\rm b})$. Note that seminal papers like \cite{Sunyaev:1970er,Chluba:2012gq,Khatri:2012rt} do not mention explicitly that \cref{eq:x_dot_branching} must be evaluated in the baryon rest frame, but this assumption always appears implicitly at a later stage in their derivation. The fact that the ratio in \cref{eq:x_dot_branching} is to be evaluated in this frame makes the definition of $\dot{s}$ explicitly frame-invariant (or, in the language of perturbation theory, gauge-invariant), as justified in details in \cref{app:gauge}.

To express equation \cref{eq:x_dot_branching} in a different way, one may also define the heating rate 
\begin{equation}
    \dot{Q}\equiv \bar{\rho}_\gamma \frac{d}{d\eta} 
    \frac{\langle \rho_{\rm ex}|_{\rm b} \rangle_{\vec{x}}}{\bar{\rho}_\gamma}
    %}~~{\color{teal} \dot{Q}\equiv \bar{\rho}_\gamma \frac{d}{d\eta} \left\langle \frac{ \rho_{\rm ex} |_{\rm b} }{\rho_\gamma |_{\rm b}}\right\rangle_{\vec{x}}}
    %~~{\color{DarkOrange} \dot{Q}\equiv \bar{\rho}_\gamma \frac{d}{d\eta} \left\langle \frac{ \rho_{\rm ex} |_{\rm b} }{\rho_\gamma |_{\rm b}}\right\rangle_{\vec{x}}}
    ~,
    \label{eq:heating_rate_def}
\end{equation}
such that
\begin{equation}
\dot{s} = {\cal I}_s(z) \frac{\dot{Q}}{\bar{\rho}_\gamma}~.
\end{equation}
%Since $\bar{\rho}_\gamma a^4$ is constant,\footnote{This is only approximation, because only one third of the energy released by the dissipation of acoustic wave produces spectral distortions. The rest is used to reheat photons. However, we will see later that the deviation from a constant $\bar{T} \, a$ and a constant $\bar{\rho}_\gamma \, a$ can be neglected even in the next-to-leading order calculation of $\dot{s}$.} we can write a simpler expression for the heating rate,
%\begin{equation}
%\dot{Q} \simeq a^{-4} \frac{d}{dt} \left\{ \langle \rho_{\rm ex}|_{\rm b} \rangle_{\vec{x}} \,\, a^4\right\}.
%\end{equation}
Once this rate is known as a function of time or redshift, each type of distortion can be evaluated as an integral over time or redshift. 
The well-known formula by Sunyaev and Zeldovich, $\mu \simeq - 1.4 \int d\eta \dot{Q} / \rho_\gamma$, comes from approximating the branching ratio ${\cal I}_\mu(z)$ as a constant factor ${\cal I}_\mu=1.4$ within the redshift range of interest. Otherwise, the total $\mu$-distortion reads
\begin{equation}
    \mu = \int dz \,\, {\cal I}_\mu(z) \frac{dQ/dz}{\bar{\rho}_\gamma} 
    = - \int dz\,\, {\cal I}_\mu(z) \frac{\dot{Q}}{a_0 H \bar{\rho}_\gamma}
    = - \int dz\,\, {\cal I}_\mu(z) \frac{dQ/dt}{(1+z) H \bar{\rho}_\gamma}~,
    \label{eq:mu_integral_z}
\end{equation}
where $t$ stands for proper (or cosmological) time.

\subsection{Properties of the heating rate \label{sec:theo_discuss}}

Before proceeding with more detailed calculations, it is useful to discuss some general properties of \cref{eq:x_dot_branching}, or equivalently of the heating rate definition \cref{eq:heating_rate_def}. We describe the blackbody component of photons at each time $\eta$, point $\vec{x}$ and direction $\hat{n}$ with a temperature field $T(\eta, \vec{x},\hat{n})$. We expand it in mean temperature $\bar{T}(\eta)=\langle T \rangle_{\hat{n},\vec{x}}$
and anisotropies $\Theta(\eta,\vec{x},\hat{n})$, such that $T=\bar{T}(1+\Theta)$ with $|\Theta| \ll 1$. In this work, $|\Theta|$ will not always be as small as ${\cal O}(10^{-5})$, since we are interested in peaks in the primordial power spectrum creating significant inhomogeneities on scales smaller than CMB scales. Nevertheless, we will always assume  $|\Theta|\ll 1$ in order to remain compatible with constraints on the amplitude of spectral distortions, as well as on the density of primordial black holes.

Spectral distortions should vanish asymptotically in two limits. Firstly, they should remain negligible at early times when all photon interactions are very efficient and enforce a perfect isotropic blackbody distribution in every point. Let us check that this is consistent with \cref{eq:x_dot_branching}. At high redshift, in the tight-coupling limit, the Boltzmann equation forces $\Theta(\eta,\vec{x},\hat{n})$ to reduce to a monopole component $\Theta_{(0)}\equiv \langle \Theta \rangle_{\hat{n}}$ (accounting for local over/underdensities) and a dipole component $\hat{n}\cdot\vec{v}_\gamma$ (accounting for the local bulk velocity of photons relative to the observer). The dipole could in principle contribute to an apparent mixture of blackbodies, but according to \cref{eq:x_dot_branching}, it should be evaluated in the baryon rest frame. However, the tight-coupling limit enforces $\vec{v}_\gamma=\vec{v}_{\rm b}$, that is, $\vec{v}_\gamma=\vec{o}$ in the frame comoving with baryons. One is only left with a pure monopole in each point, and thus, with an isotropic blackbody distribution, such that $\rho_{\rm ex}|_{\rm b}$ vanishes. Thus, as expected, spectral distortions cannot be sourced in this limit. Instead, at the approach of decoupling, the growth of the dipole component in the baryon rest frame, $\hat{n}\cdot(\vec{v}_\gamma-\vec{v}_{\rm b})$, and of higher multipoles -- that is, heat conduction and viscosity -- leads to an increasing value of $\rho_{\rm ex}|_{\rm b}$ that sources spectral distortions.

The second limit in which we expect that spectral distortions cannot be sourced is when CS is totally inefficient, such that photons free stream. Along each geodesic, free-streaming photons have a conserved phase-space distribution (up to a trivial rescaling of each momenta due to global expansion and metric fluctuations, which do not change the shape of the distribution). A blackbody remains a blackbody, and so does any other shape. In the idealised limit of instantaneous photon decoupling, an observer living today would measure in every point a superposition of blackbodies with a different temperature in each direction. In principle, this observer could average the information received from all directions and interpret the result as a spectrum with y-distortions, but this is not what we are after. What the observer would actually claim is that they measure a perfect blackbody in each direction. A relevant distortion means that the spectrum measured in each direction is no longer a blackbody. This can only happen if, over the past photon history, some interactions reprocessed the energy density $\rho_{\rm ex}$ into actual distortions for the ensemble of photons travelling along each geodesic. This property is actually encoded in \cref{eq:x_dot_branching}, as long as we take into account the fact that, when the time derivative is moved inside the averaging operator, it stands for a total derivative along each photon geodesic. Then, we can use the Boltzmann equation to compute the evolution of the available energy in each point and along each direction. When Thomson scattering becomes totally inefficient and photons free stream, the collisionless Boltzmann equation just accounts for photon number conservation in phase space. We will see explicitly in the next sections that, in turn, this implies that the average
$\langle \rho_{\rm ex}|_{\rm b} / \bar{\rho}_\gamma\rangle_{\vec{x}}$
remains constant in time. Then, each spectral distortion $s$ freezes out at the value acquired during the epoch at which CS was still active.

\subsection{Perturbative expansion}

To proceed, one needs to compute how much of the energy stored in photons in each point $\vec{x}$ matches that of an isotropic blackbody distribution, and how much contributes to spectral distortions. At each point $\vec{x}$, the photons have an energy and number density\footnote{The expressions given in this section neglect corrections to the photon occupation number coming from polarisation, which are described in \cite{Chluba:2014qia}. These corrections are very small and only affect $y$-distortions at the approach of recombination, while in this work we are mostly interested in $\mu$-distortions sourced a few e-folds before recombination.}
\begin{align}
    \rho_\gamma(\eta,\vec{x}) 
    &= a_{\rm R} \langle T(\eta,\vec{x},\hat{n})^4 \rangle_{\hat{n}} 
    = a_{\rm R} \bar{T}^4 \langle (1+\Theta)^4 \rangle_{\hat{n}} ~,
    \\
    n_\gamma(\eta,\vec{x}) 
    &= b_{\rm R} \langle T(\eta,\vec{x},\hat{n})^3 \rangle_{\hat{n}} 
    = b_{\rm R}  \bar{T}^3 \langle (1+\Theta)^3 \rangle_{\hat{n}}~,
\end{align}
with $a_{\rm R}=\frac{\pi^2}{15}$, $b_{\rm R}=\frac{2 \xi(3)}{\pi^2}$, and the angular average is defined as
\begin{equation}
\langle f(\hat{n}) \rangle_{\hat{n}} = \frac{1}{4\pi} \int d\hat{n} f(\hat{n}).
\end{equation}
%\sout{For simplicity, we work first in a coordinate frame where the photon temperature is uniform on equal-time slices (with vanishing monopole $\Theta_0=\langle \Theta \rangle_{\hat{n}}=a_0^0 Y_0^0= 0$) and fixed coordinate observers are at rest with respect to photons (with vanishing dipole $\vec{v}_\gamma \cdot \hat{n} = \sum a_1^m Y_1^m=0$). If $\Theta \ll 1$, we can use perturbation theory and reformulate the last statement as: we work in the gauge comoving with photons and of uniform photon density (gauge transformation have just enough degrees of freedom to cancel simultaneously the fields $\delta \rho_\gamma=\rho_\gamma-\bar{\rho}_\gamma$ and $\vec{v}_\gamma$).}
The local blackbody photon temperature is inferred from the number density, $T_{\rm bb}=(n_\gamma/b_{\rm R})^{1/3}$, and the corresponding blackbody energy reads
$\rho_{\rm bb}=a_{\rm R} T_{\rm bb}^{4}$. Thus, the energy density stored in photons and contributing respectively to blackbody radiation and to spectral distortions  reads
\begin{align}
 \rho_{\rm bb}(\eta,\vec{x}) &= a_{\rm R}\left(\frac{n_\gamma(\eta, \vec{x})}{b_R}\right)^{4/3}~,\\
 \rho_{\rm ex}(\eta, \vec{x}) &= \rho_\gamma(\eta, \vec{x}) - a_{\rm R}\left(\frac{n_\gamma(\eta, \vec{x})}{b_R}\right)^{4/3}~.
\end{align}
In terms of anisotropies, the latter density reads (see e.g. \cite{Chluba:2012gq})
\begin{align}
    \rho_{\rm ex}(\eta, \vec{x})
    &= a_{\rm R} \bar{T}^4 \left[ \langle (1+\Theta)^4 \rangle_{\hat{n}} -  \langle (1+\Theta)^3 \rangle_{\hat{n}}^{4/3}\right]~.
    \label{eq:rho_ex_1}
\end{align}
We can then use the fact that $|\Theta| \ll 1$ and perform a Taylor expansion. At order three in $\Theta$, one gets:
\begin{align}
\rho_{\rm ex}(\eta, \vec{x}) = a_R \bar{T}^4 \left[ 2\langle \Theta^2 \rangle_{\hat{n}}
    -2 \Theta_{(0)}^2
    +\frac{8}{3} \langle\Theta^3\rangle_{\hat{n}}
    -4 \Theta_{(0)} \langle \Theta^2 \rangle_{\hat{n}}
    + \frac{4}{3} \Theta_{(0)}^3
    + {\cal O}(\Theta^4)
    \right]~.
    \label{eq:rho_ex_2}
\end{align}
The order-two terms account for the leading order, and the order-three terms for the next-to-leading order. In this work, the latter terms will be potentially relevant when evaluating corrections arising from non-Gaussian primordial perturbations.  

\Cref{{eq:rho_ex_2}} confirms that for a temperature distribution that is isotropic in each point, $\Theta=\Theta_{(0)}$, the energy contributing to spectral distortions vanishes as it should (since there is no superposition of blackbodies with different temperatures). This is true order-by-order in the previous expression, as can be checked by replacing each $\langle\Theta^p\rangle_{\hat{n}}$ by $\Theta_{(0)}^{p}$. However, as soon as anisotropies are present, $\Theta \neq \Theta_{(0)}$, the monopole does contribute to $\rho_{\rm ex}$. It does not contribute to the leading-order term, since the orthogonality of spherical harmonics allows us to rewrite $(\langle \Theta^2 \rangle_{\hat{n}}- \Theta_{(0)}^2)$ as $\langle (\Theta-\Theta_{(0)})^2 \rangle_{\hat{n}}$, but it remains present in the next-to-leading terms. The physical explanation of the role of the monopole is given in \cref{app:gauge_monopole}.

%\sout{As expected, this expression is frame-invariant (or in perturbation theory, gauge-invariant) and vanishes for anisotropies only consisting in a monopole and a dipole component, as in the tight-coupling limit.}
%We note that for an isotropic temperature distribution in each point, $\Theta=\Theta_{(0)}$, the energy in spectral distortions vanishes, as it should (since we then have a perfect blackbody). This is true order-by-order in the previous expression (it can be checked by replacing each $\langle\Theta^p\rangle_{\hat{n}}$ by $\Theta_{(0)}^p$). \julien{However, physically, we would also expect the result to vanish for a pure dipole, $\Theta=\Theta_1$, and this is not the case. This is because I must change something in the initial definition of $\rho_{\rm ex}$, to make it frame-invariant (or gauge-invariant) from the beginning.}
%Evaluating the density of \cref{eq:rho_ex_2} in the baryon rest frame just means that the monopole and dipole are computed in a frame where $\vec{v}_{\rm b}=\vec{o}$. 
Next, we expand locally the ratio sourcing spectral distortions,
\begin{align}
    \frac{\rho_{\rm ex}(\eta, \vec{x})|_{\rm b}}{\bar{\rho}_\gamma(\eta)}
    &
    = 
    \frac{a_{\rm R} \bar{T}^4 \left[ \langle (1+\Theta)^4 \rangle_{\hat{n}} -  \langle (1+\Theta)^3 \rangle_{\hat{n}}^{4/3}\right]_{\rm b}}
    {a_{\rm R} \bar{T}^4 }
    \nonumber \\
    &
    = \left[ \left\langle 2 \Theta^2
    -2 \Theta_{(0)}^2
    +\frac{8}{3} \Theta^3
    -4 \Theta_{(0)} \Theta^2
    + \frac{4}{3} \Theta_{(0)}^3 \right\rangle_{\hat{n}} 
    + {\cal O}(\Theta^4)
    \right]_{\rm b}
   ~.
\end{align}
 Finally, the heating rate that enters into the calculation of spectral distortions is given by
\begin{align}
\dot{Q}
    =& \bar{\rho}_\gamma \frac{d}{dt} \left[
    \left\langle
    2 \Theta^2
    -2 \Theta_{(0)}^2
    +\frac{8}{3} \Theta^3
    - 4
    %~{\color{teal} 12}
    %~{\color{DarkOrange} 6}~
    \Theta_{(0)} \Theta^2 
    + \frac{4}{3}
    %~{\color{teal} \frac{28}{3}}~{\color{DarkOrange} \frac{10}{3}}~ 
    \Theta_{(0)}^3
    \right\rangle_{\hat{n},\vec{x}}
    + {\cal O}(\Theta^4) \right]_{\rm b} \nonumber \\
    =& 4 \bar{\rho}_\gamma \left[ \left\langle
    \Theta \dot\Theta
    - \Theta_{(0)} \dot\Theta_{(0)}
    +2 \Theta^2 \dot\Theta
    - 
    %{\color{purple}1}~{\color{teal}3}~{\color{DarkOrange}\frac{3}{2}}~
    \dot\Theta_{(0)} \Theta^2 
    -2
    %{\color{purple}2}~{\color{teal}6}~{\color{DarkOrange}3}~ 
    \Theta_{(0)} \Theta \dot\Theta 
    %{\color{teal}-4 \langle \Theta_{(0)} \langle \Theta \dot\Theta \rangle_{\hat{n}} \rangle_{\vec{x}}} 
    %\right. \right. 
    %\nonumber \\
    %&\left. \left.
    + 
    %{\color{purple}1}~{\color{teal}7}~{\color{DarkOrange}\frac{5}{2}}~
    \Theta_{(0)}^2 \dot\Theta_{(0)} 
    \right\rangle_{\hat{n},\vec{x}} 
    + {\cal O}(\Theta^4) \right]_{\rm b}~,
    \label{eq:q_time_der}
\end{align}
where $\dot \Theta$ is assumed to be a total derivative along photon geodesics, and some averages are now both on position and direction:
\begin{equation}
\langle f(\vec{x},\hat{n}) \rangle_{\hat{n},\vec{x}} = \frac{1}{V} \int_V d^3\vec{x} \,\, \frac{1}{4\pi} \int d\hat{n} \,\, f(\vec{x},\hat{n}).
\end{equation}
We now use the second-order Boltzmann equation, which reads in the baryon rest frame \cite{Bartolo_2006,Bartolo_2007,Pitrou_2009,Pitrou_2009b,Pitrou_2010,Beneke_2010,Naruko:2013aaa}:
\begin{equation}
\dot{\Theta}|_{\rm b} = \dot{\tau} \left[ \Theta^{{\rm C}1} + \Theta^{{\rm C}2}\right]_{\rm b}~,
\label{eq:boltzmann}
\end{equation}
where $\dot{\tau}$ is the Thomson scattering rate (with respect to conformal time), while the first-order and second-order collision terms read in the baryon rest frame
\begin{align}
\Theta^{{\rm C}1}|_{\rm b} &= - \Theta + \Theta_{(0)}-\Pi^{1}_{(2)}~,
\nonumber \\
\Theta^{{\rm C}2}|_{\rm b} &= (\Theta-\Theta_{(0)})^2- \Pi^{2}_{(2)}+ \delta_{\rm b} \Theta^{{\rm C}1}|_{\rm b}~,
\end{align}
where $\delta_{\rm b}$ stands for the baryon density fluctuations measured in this frame.\footnote{The exact second-order Boltzmann equation involves the electron number density fluctuations $\delta n_{\rm e}/n_e$ instead of $\delta_{\rm b}$, but we assume that at the redshift of interest Coulomb scattering is sufficient to enforce $\frac{\delta n_{\rm e}}{n_{\rm e}} = \frac{\delta n_{\rm b}}{n_{\rm b}} = \frac{\delta \rho_{\rm b}}{\rho_{\rm b}}=\delta_{\rm b}$.}
Upper indices refer to orders in the perturbative expansion, while lower indices between parenthesis refer to the angular dependence (monopole, dipole, quadrupole, etc).
The first-order and second-order polarisation corrections have a pure quadrupole geometry, whose expression will be specified later when relevant.
Thus, importantly, in the baryon rest frame (and actually any other frame), $\Theta^{\rm C1}$ has no monopole component, that is, $\langle \Theta^{{\rm C}1} \rangle_{\hat{n}}=0$.
%while $\langle \Theta^{{\rm C}2} \rangle_{\hat{n}}=\langle (\Theta-\Theta_{(0)})^2 \rangle_{\hat{n}}$. 
Thus, the angular average of \cref{eq:boltzmann} gives $\dot\Theta_{(0)}= \dot{\tau} \langle \Theta^{{\rm C}2} \rangle_{\hat{n}}$. 
Metric fluctuations do not appear explicitly in \cref{eq:boltzmann} but are present implicitly in the total derivative on the left-hand side. Using this equation, we can replace $\dot \Theta$ in eq.~\cref{eq:q_time_der}:
\begin{align}
\dot{Q}
    &= 4 \bar{\rho}_\gamma \dot{\tau} \left[ \left\langle
    \Theta \Theta^{{\rm C}1}
    + \Theta \Theta^{{\rm C}2}
    - \Theta_{(0)} \Theta^{{\rm C}2} 
    +2 \Theta^2 \Theta^{\rm C1}
    %\right. \right.
    %\nonumber \\
    %& \left. \left.
    - 2 
    %{\color{teal}6}~{\color{DarkOrange}3}~ 
    \Theta_{(0)} \Theta \Theta^{{\rm C}1}
    %{\color{teal} -4 \langle \Theta_{(0)} \langle \Theta \Theta^{\rm C} \rangle_{\hat{n}} \rangle_{\vec{x}}}
    \right\rangle_{\hat{n}, \vec{x}}
    + {\cal O}(\Theta^4) \right]_{\rm b}~.
\end{align}
After rearranging the terms and using the orthogonality of multipoles, $\langle \Theta_{(0)} \Theta^{{\rm C}1}\rangle_{\hat{n}}=0$, we get an equivalent expression that shows explicitly that the monopole is absent from all terms but one (which is consistent with the discussion in \cref{app:gauge_monopole}):
\begin{align}
\dot{Q}
    =& 4 \bar{\rho}_\gamma \dot{\tau} \left[ \left\langle
    (\Theta-\Theta_{(0)})\Theta^{{\rm C}1}
    + (\Theta-\Theta_{(0)}) \Theta^{{\rm C}2}
    +2 (\Theta-\Theta_{(0)})^2 \Theta^{{\rm C}1}
    %\right. \right.
    %\nonumber \\
    %& \left. \left.
    %+ 2
    %{\color{purple}(+2)}~{\color{teal}(-2)}~{\color{DarkOrange}(+1)}~
    \Theta_{(0)} (\Theta-\Theta_{(0)}) \Theta^{{\rm C}1}
    %{\color{teal}-4 \langle \Theta_{(0)} (\Theta-\Theta_{(0)}) \Theta^{\rm C} \rangle_{\hat{n},\vec{x}}}
    \right\rangle_{\hat{n},\vec{x}}
    \right. 
    \nonumber \\
    & \left.~~~~~~
    + {\cal O}(\Theta^4) \right]_{\rm b}~.
\end{align}
We can now relax the condition that this expression must be evaluated in the baryon rest frame by noticing that a transformation to another frame shifts the photon monopole by $\frac{\dot{a}}{a} \, v_b$, the photon dipole by $- \hat{n}\cdot\vec{v}_{\rm b}$, and leaves higher multipoles unaffected (see \cref{app:gauge}). Therefore, in an arbitrary frame,
\begin{empheq}[box=\fbox]{align}
\dot{Q}
    = 4 \bar{\rho}_\gamma \dot{\tau} & \left[  \left\langle
    \Theta^{\rm B} \Theta^{{\rm C}1}
    + \Theta^{\rm B} \Theta^{{\rm C}2} 
    +2 (\Theta^{\rm B})^2 \Theta^{{\rm C}1}
    %\right. \right.
    %\nonumber \\
    %& \left. \left.
    +2
    %{\color{purple}(+2)}~{\color{teal}(-2)}~{\color{DarkOrange}(+1)}
    (\Theta_{(0)} + \frac{\dot{a}}{a} v_{\rm b}) \Theta^{\rm B} \Theta^{{\rm C}1}
    \right\rangle_{\hat{n},\vec{x}}
    + {\cal O}(\Theta^4) \right]~,
    \label{eq:Q_dot_full}
\end{empheq}
where we defined 
\begin{equation}
\Theta^{\rm B} \equiv \Theta-\Theta_{(0)}-\hat{n}\cdot\vec{v}_{\rm b} = \hat{n}\cdot(\vec{v}_\gamma - \vec{v}_{\rm b})
+ \Theta_{(\geq2)}~,
\end{equation}
in which $\Theta_{(\geq2)}$ stands for the photon anisotropies beyond the monopole and dipole contributions, that is, all multipoles with $\ell \geq2$ in a spherical harmonic expansion. This definition is valid in any frame, but the letter `B' for baryons hints that $\Theta^{\rm B}$ represents the photon temperature anisotropies as seen by an observer comoving with baryons.
Still, in an arbitrary frame, the first- and second-order collision terms read
\begin{align}
\Theta^{{\rm C}1} &= - \Theta^{\rm B} - \Pi^{1}_{(2)}~,
\nonumber \\
\Theta^{{\rm C}2} &= (\Theta^{\rm B})^2 - \Pi^{2}_{(2)}~.
\end{align}
A few checks are in order. First, this heating rate is gauge-invariant, since it only relies on the two gauge-invariant terms $\hat{n}\cdot(\vec{v}_{\gamma}-\vec{v}_{\rm b})$ and $(\Theta_{(0)} + \frac{\dot{a}}{a} v_{\rm b})$, and on multipoles of order two or higher, which are automatically gauge-invariant. Second, as expected, it vanishes in the tight-coupling limit ($\dot{\tau}\rightarrow \infty$) where $\Theta^{\rm B}$, $\Pi^{1}_{(2)}$ and $\Pi^{2}_{(2)}$ all tend towards zero. Third, it vanishes in the free-streaming limit ($\dot{\tau}\rightarrow 0$). Thus, we recover all the properties expected from the physical discussion of \cref{sec:theo_base,sec:theo_discuss}. 

We raise however a word of caution. The derivation of acoustic dissipation effects from the average of the ratio $\rho_{\rm ex}/\bar{\rho}_\gamma$ is presented in the specialised literature on spectral distortions as a heuristic approach. A fully rigorous derivation of the leading-order (quadratic) terms requires a careful study of the second-order Boltzmann equation \cite{Chluba:2012gq}. For next-to-leading-order (cubic) terms, one should use the third-order Boltzmann equation, which is beyond the scope of this paper. It is worth stressing that, for leading-order terms, our simplified approach is able to fully reproduce the results derived using the second-order Boltzmann equation -- mainly because we followed a path where we computed $\rho_{\rm ex}/\bar{\rho}_\gamma$ first in the baryon frame, and then performed a gauge transformation to an arbitrary frame. The excellent match between the two derivations can be checked at the end of the next section, where the final expression derived from the expansion of $\langle \Theta^{\rm B} \Theta^{\rm C1}\rangle$, \cref{eq:S_sd}, matches exactly the full result presented e.g. in \cite{Chluba:2012gq,Khatri:2012rt,Chluba:2013dna}. However, there is no guarantee that this would also be the case at higher order: the third-order Boltzmann equation could in principle reveal additional corrections to \cref{eq:Q_dot_full}. Since \cref{sec:PNGC} shows that the contribution from next-to-leading-order terms to spectral distortions is always negligible in the context of this work, we don't believe that our conclusions would be altered in case we missed small corrections to it.

\section{Leading-order contribution to spectral distortions}\label{Leading_order_mu}

\subsection{Acoustic dissipation source function\label{sec:Q_order_two}}

At leading order, $\mu$-distortions are given by \cref{eq:mu_integral_z,eq:Q_dot_full} where $\dot{Q}$ features only quadratic terms in $\Theta$, that is,
\begin{equation}
\dot{Q} = 4 \dot{\tau} \bar{\rho}_\gamma 
\langle \Theta^{\rm B} \Theta^{{\rm C}1}\rangle_{\hat{n},\vec{x}}~.
\label{eq:QBC1}
\end{equation}
A standard calculation, detailed in \cref{app:two_point}, allows to express the two-point correlation function $\langle \Theta^{\rm B} \Theta^{{\rm C}1}\rangle_{\hat{n},\vec{x}}$ as an integral over the transfer function of the Legendre multipoles $\Theta^{\rm B}_\ell(k)$, $\Theta^{{\rm C}1}_{\ell}(k)$ and over the primordial curvature power spectrum $P_{\cal R} (k)$, which is defined as
\begin{equation}
\langle {\cal R}(\vec{k}) {\cal R}^*(\vec{k}') \rangle = 
P_{\cal R} (k) \, \delta^{(3)}(\vec{k}-\vec{k'})~.
\label{eq:def_PR}
\end{equation}
In what follows, we will also frequently use the dimensionless curvature power spectrum ${\cal P}_{\cal R} (k)\equiv\frac{k^3}{2\pi^2} P_{\cal R} (k)$.
Then, the two-point correlation function reads \begin{align}
    \langle \Theta^{\rm B} \Theta^{{\rm C}1}  \rangle_{\hat{n},\vec{x}} &=
     \int \frac{dk \, k^2}{(2\pi)^{3}}
\sum_{\ell} 4\pi  (2\ell+1) 
\Theta^{\rm B}_\ell(k) \Theta^{{\rm C}1}_\ell(k)
  P_{\cal R} (k)~.
\label{eq:theta2}
\end{align}
The Legendre multipoles of $\Theta^{\rm B}$ and $\Theta^{{\rm C}1}$ relate to those of temperature anisotropies $\Theta$ as $ \Theta^{\rm B}_\ell = - \Theta^{{\rm C}1}_\ell = \Theta_\ell$, excepted for the monopole ($\Theta^{\rm B}_0 = \Theta^{{\rm C}1}_0=0$), for the dipole,
\begin{equation}
    \Theta^{\rm B}_1 = - \Theta^{{\rm C}1}_1 = \Theta_1 - \frac{k}{3} v_{\rm b} = \frac{k}{3} (v_\gamma - v_{\rm b})~,
\end{equation} 
and for the quadrupole,
\begin{equation}
   \Theta^{\rm B}_2 = \Theta_2~, \qquad \Theta^{{\rm C}1}_2 = -\Theta_2 +\frac{1}{10} \left( \Theta_2 + \Theta^{\rm P}_0 + \Theta^{\rm P}_2 \right)~,
\end{equation}
where $\Theta^{\rm P}_\ell$ are the Legendre momenta of polarisation anisotropies. Finally, this gives
\begin{equation}
\dot{Q} = - 4 \bar{\rho}_\gamma \dot{\tau} \int \frac{dk\,k^2}{2 \pi^2} P_{\cal R}(k)
S_{\rm ac}(k,\eta)
= - 4 \rho_\gamma \dot{\tau} \int \frac{dk}{k} {\cal P}_{\cal R}(k)
S_{\rm ac}(k,\eta)~,
\end{equation}
where we introduced the acoustic dissipation source function
\begin{align}
S_{\rm ac}(k,\eta) 
& \equiv - \sum_\ell (2 \ell + 1) \Theta^{\rm B}_\ell(k) \Theta^{{\rm C}1}_\ell(k) \nonumber \\
& = \frac{k^2({v}_{\gamma}-{v}_{\rm b})^2}{3}
+ \sum_{l\geq2} (2\ell + 1) \Theta_\ell^2
- \frac{1}{2} \Theta_2(\Theta_2+\Theta_0^{\rm P}+\Theta_2^{\rm P})
\nonumber \\
& = \frac{k^2(v_\gamma-v_b)^2}{3}
+ \frac{9}{2} \Theta_2^2 
- \frac{1}{2} \Theta_2(\Theta_0^{\rm P}+\Theta_2^{\rm P})
+ \sum_{l\geq3} (2\ell + 1) \Theta_\ell^2~.
\label{eq:S_sd}
\end{align}
This result was obtained for the first time in \cite{Chluba:2012gq}.

\subsection{Properties of the source function}

Different levels of approximation can be used to evaluate $S_{\rm ac}(k,\eta)$ with a Boltzmann code or a simpler analytical method. 

Since the dissipation of acoustic waves takes place before recombination, we can use results from the tight-coupling limit in which, for $\ell>2$, each higher multipole is suppressed by one more power of $\frac{k}{2 \dot \tau}$ (see e.g. \cite{Blas:2011rf}), which is larger than one excepted for very small wavelengths that have already been dissipated. Cutting the sum over multipoles at $\ell=3$ gives a very accurate approximation %initially derived by \cite{Chluba:2012gq,Khatri:2012rt} 
that we call (A1),
\begin{equation}
S_{\rm ac}^{\rm (A1)}(k,\eta) = 
\underbrace{\frac{(\theta_\gamma-\theta_b)^2}{3 k^2}}_{\rm dipole}
+ \underbrace{\frac{9}{2} \left(\frac{\sigma_\gamma}{2}\right)^2
- \frac{1}{8} 
\left(\frac{\sigma_\gamma}{2}\right) (G_{\gamma 0}+G_{\gamma 2})}_{\rm quadrupole}
+ \underbrace{7 \left(\frac{F_{\gamma 3}}{4}\right)^2}_{\rm octopole}~.
\label{eq:Sac_A1}
\end{equation}
Here we switched to the notations of Ma \& Bertschinger \cite{Ma:1995ey} for perturbations. These are also the notations used internally by \CLASS{} \cite{Lesgourgues:2011re,Blas:2011rf}. The temperature multipoles are denoted as $F_{\gamma\ell}=4\Theta_\ell$, the polarisation mutipoles as $G_{\gamma\ell}=4\Theta_\ell^{\rm P}$, the photon shear as $\sigma_\gamma=\frac{1}{2}F_{\gamma 2} = 2 \Theta_2$, and the velocity divergences as $\theta_\gamma=\frac{3k}{4} F_{\gamma 1} = k^2 v_\gamma = 3k \Theta_1$ for photons and $\theta_{\rm b}= k^2 v_{\rm b}$ for baryons. 

\begin{figure}[hbt!]
  %\vspace{-4cm}
        \centering
        %\hspace{-4cm} 
        \includegraphics[width=14.5cm,angle=0]{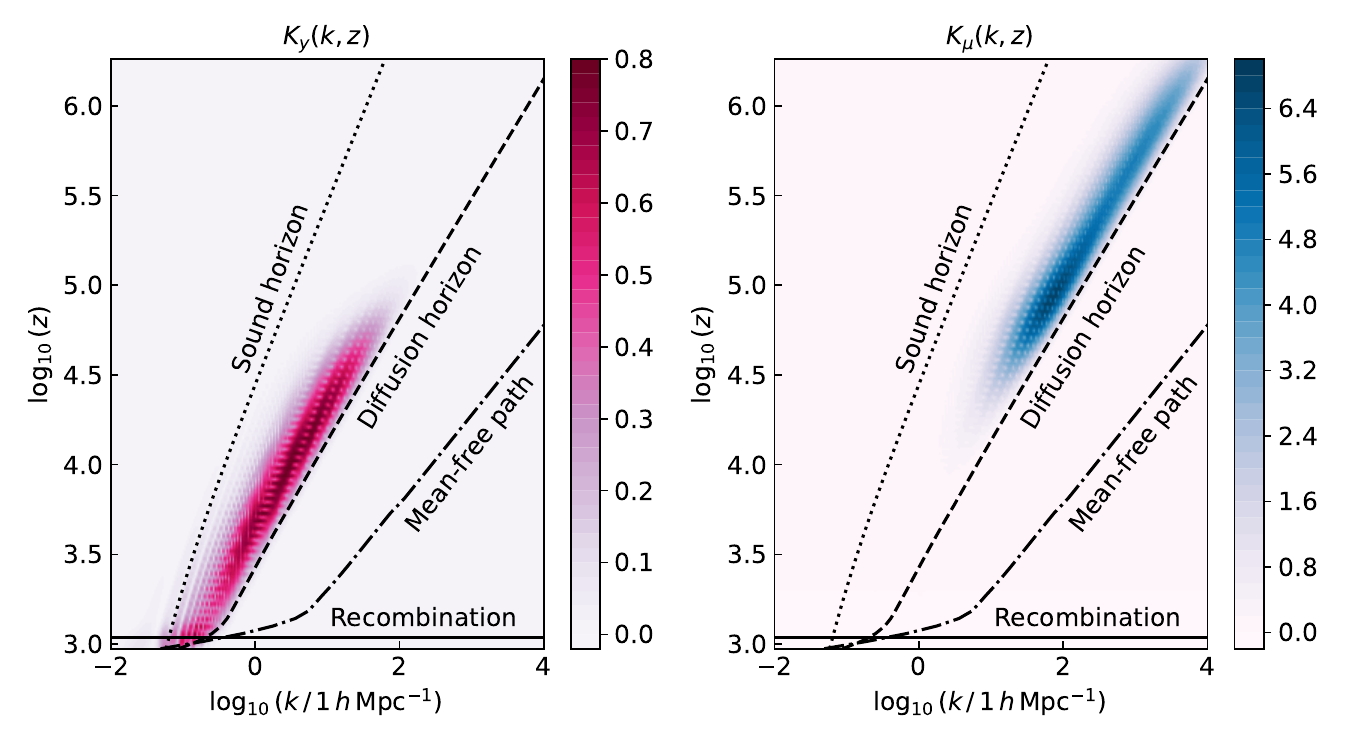}
        \caption{Dimensionless integration kernel for $y$-distortions (left) and $\mu$-distortions (right) in $(\ln k, \ln z)$ space. For a scale-invariant power spectrum, each pixel in this space would contribute proportionally to its color level. The various lines correspond to sound horizon crossing ($k r_{\rm s} = 2 \pi$), diffusion horizon crossing ($k r_{\rm D} = 2 \pi$), the Thomson mean-free-path scale ($k = 2 \pi \, \dot \tau$), and recombination time. All quantities were extracted from \CLASS{} (assuming a $\Lambda$CDM cosmology with {\it Planck} best-fit parameters \cite{Planck:2018vyg}).
        %\julien{I might try to extend to $\log_{10}k=4.5$ and $\log_{10}z=6.5$ if I have time.}
        \label{fig:K_contour}}
\end{figure}
To understand in which range of scales and times (or redshifts) it is important to get an accurate approximation for $S_{\rm ac}(k,\eta)$, we cast the calculation of $s=\mu,y$ in the form
\begin{equation}
s = \int \! d \ln \! z \, \! \int \!\! d \ln \! k \,\, K_s(k,z) \,\, {\cal P}_{\cal R}(k)~,
\end{equation}
with
\begin{equation}
    K_s(k,z) \equiv - {\cal I}_s(z) \, \frac{4 \dot \tau\,z}{H} \, S_{\rm ac}(k,z)~,
\end{equation}
and we plot the integration kernels $K_\mu(z,k)$ and $K_y(z,k)$ as a function of the logarithm of $k$ and $z$ in \cref{fig:K_contour}. 

The dark (magenta or blue) areas show the region in $(k,z)$ space that would contribute the most to the final distortion in the case of a nearly scale-invariant curvature spectrum. If instead the curvature spectrum is very peaked near one scale $k_*$, these dark regions show at which redshift $z(k_*)$ most of the distortions are generated. In particular, we see that, in order to compute the distortions generated by peaks at wavenumbers up to $k_*=10^4\,h\,{\rm Mpc}^{-1}$, we need to compute the source function up to $z=10^{6.5}$.

The baryon-to-photon ratio $R$, sound speed $c_{\rm s}$ and comoving sound horizon $r_{\rm s}$ are defined as 
\begin{equation}
R=\frac{3\bar{\rho}_{\rm b}}{4 \bar{\rho}_\gamma}~, \qquad
    c_{\rm s}^2 = \frac{1}{3(1+R)}~,
    \qquad
    r_s=\int_0^\eta c_{\rm s}(\tilde{\eta})d\tilde{\eta}~.
\end{equation}
As discussed by Hu and Sugiyama \cite{Hu:1995en},  the comoving diffusion horizon can be approximated as
\begin{equation}
r_{\rm D}
= \frac{2\pi}{k_{\rm D}} 
= \left[ \int dz \frac{c_s^2}{2 \dot{\tau} H} \left(\frac{R^2}{1+R} + \frac{16}{15} \right)
\right]^{1/2}~.
\label{eq:k_D}
\end{equation}
\Cref{fig:K_contour} shows the sound horizon crossing line, $k r_{\rm s}=2\pi$, the diffusion horizon crossing line, $k r_{\rm D}=2\pi$, the Thomson mean-free-path scale, $k/(2 \pi) = 1/\dot{\tau}$, and the time of recombination. The regions contributing to the distortions are located between the sound and diffusion horizon crossing lines, that is, when the modes are well inside the Hubble radius but have not yet experienced full dissipation.
As expected, the $\mu$-distortions are seeded roughly at redshifts in the range $10^{4.5} < z < 10^{6.2}$, when DC and BS are no longer efficient but CS maintains thermal equilibrium, while $y$-distortions get contributions from $10^{2.9} < z < 10^{4.8}$, when CS also becomes gradually inefficient. Since the modes contributing to the distortions fulfil $k / \dot \tau \gg 2\pi$, that is, have a wavelength much larger than the Thomson mean-free-path, it is legitimate to assume $\frac{k}{2 \dot \tau}\ll 1$ and neglect high multipoles in the expression of $S_{\rm ac}^{\rm (A1)}$.

\begin{figure}[hbt!]
  %\vspace{-4cm}
        \centering
        %\hspace{-4cm} 
        \includegraphics[width=14cm,angle=0]{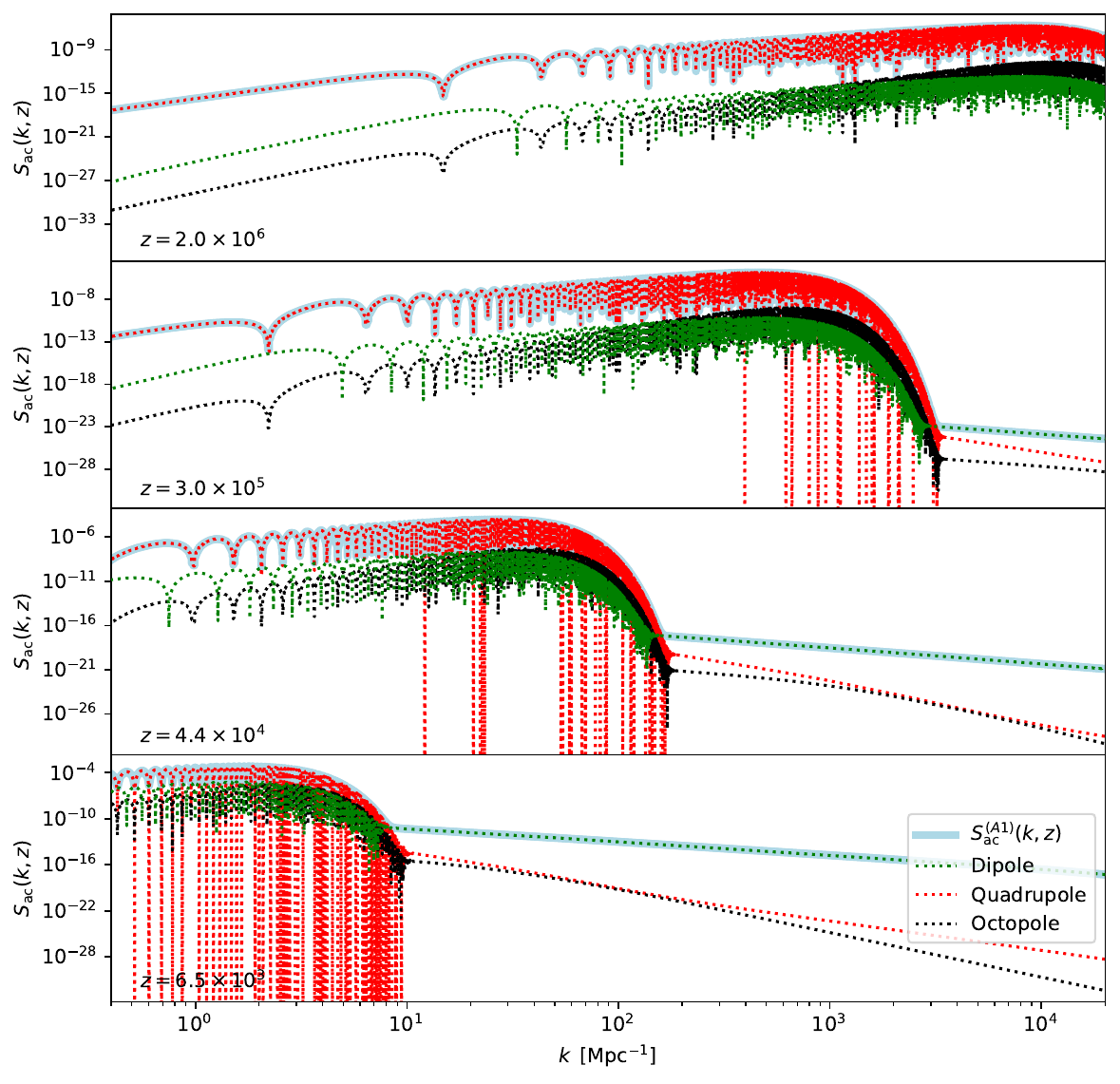}
        \caption{Individual contribution of the dipole (dotted green), quadrupole (dotted red) and octopole (dotted blue) terms to the source function $S_{\rm ac}^{(A1)}(k,z)$ of \cref{eq:Sac_A1} (solid cyan) at four different redshifts. Obtained with \CLASS.}\label{fig:S_sd_contributions}
\end{figure}

We plot $S_{\rm ac}^{\rm (A1)}(k,\eta)$ as returned by \CLASS{} as a function of $k$  at four different times (or redshifts) in \cref{fig:S_sd_contributions} (thick solid cyan lines). The figure also shows the contribution
to the source function coming from dipole, quadrupole and octopole terms. Most of the distortions are seeded by quadrupole terms;  the octopole term is always subdominant, and the dipole term only dominates the large-$k$ tail of the source function. 

These results can be understood analytically using the tight-coupling approximation (TCA), which corresponds to the limit where $\dot \tau \gg aH$ and $\dot \tau \gg k$ \cite{Ma:1995ey,Blas:2011rf}. In the TCA limit, one expects the photon and baryon fluid to track each other, while the photon shear is suppressed by efficient Thomson scattering. Both $(\theta_\gamma-\theta_b)$ and $\sigma_\gamma$ are then very small, of order one in the Thomson interaction time $1/\dot{\tau}$. However, the solution of the TCA equations gives $(\theta_\gamma-\theta_b)/k \sim \frac{R}{\sqrt{1+R}} \sigma_\gamma$, such that the dipole contribution to the source function is subdominant compared to the quadrupole one before decoupling \cite{Chluba:2012gq}. The octopole is also sub-dominant because, in the TCA limit, $F_{\gamma 3} \simeq \frac{6}{7} \frac{k}{\dot \tau} \sigma_\gamma$. Finally, the polarization terms bring a significant contribution to the total quadrupole contribution, since $G_{\gamma_0} = \frac{1}{5} G_{\gamma 2} = \frac{\sigma_\gamma}{2}$, which implies
\begin{equation}
S_{\rm ac}(k,\eta) \simeq 
\frac{9}{2} \left(\frac{\sigma_\gamma}{2}\right)^2
- \frac{1}{8} 
\left(\frac{\sigma_\gamma}{2}\right) (G_{\gamma 0}+G_{\gamma 2})
\simeq 
\frac{15}{4} \left(\frac{\sigma_\gamma}{2}\right)^2~.
\label{eq:Sac_quadrupole}
\end{equation}
However, this approximation does not account for the large-$k$ asymptotic behaviour of the source function, not captured by the TCA. An analytic study of diffusion damping shows that once the effective photon temperature perturbation $(\Theta+\psi)$ has been driven to zero, residual photon perturbations of the order of $R\psi$ (where $\psi$ is the Newtonian potential) persist due to baryonic infall \cite{Hu:1995en,Hu:1995em}. In this limit, the dipole term dominates over the quadrupole, with
\begin{equation}
S_{\rm ac}(k,\eta) \simeq 
\frac{(\theta_\gamma-\theta_b)^2}{3 k^2}
\simeq 
\frac{(k^2 R \psi / \dot \tau)^2}{3 k^2}~.
\label{eq:damping_tail}
\end{equation}

\subsection{Approximation schemes\label{sec:Sac_approx}}

\begin{figure}[hbt!]
  %\vspace{-4cm}
        \centering
        %\hspace{-4cm} 
        \includegraphics[width=15cm,angle=0]{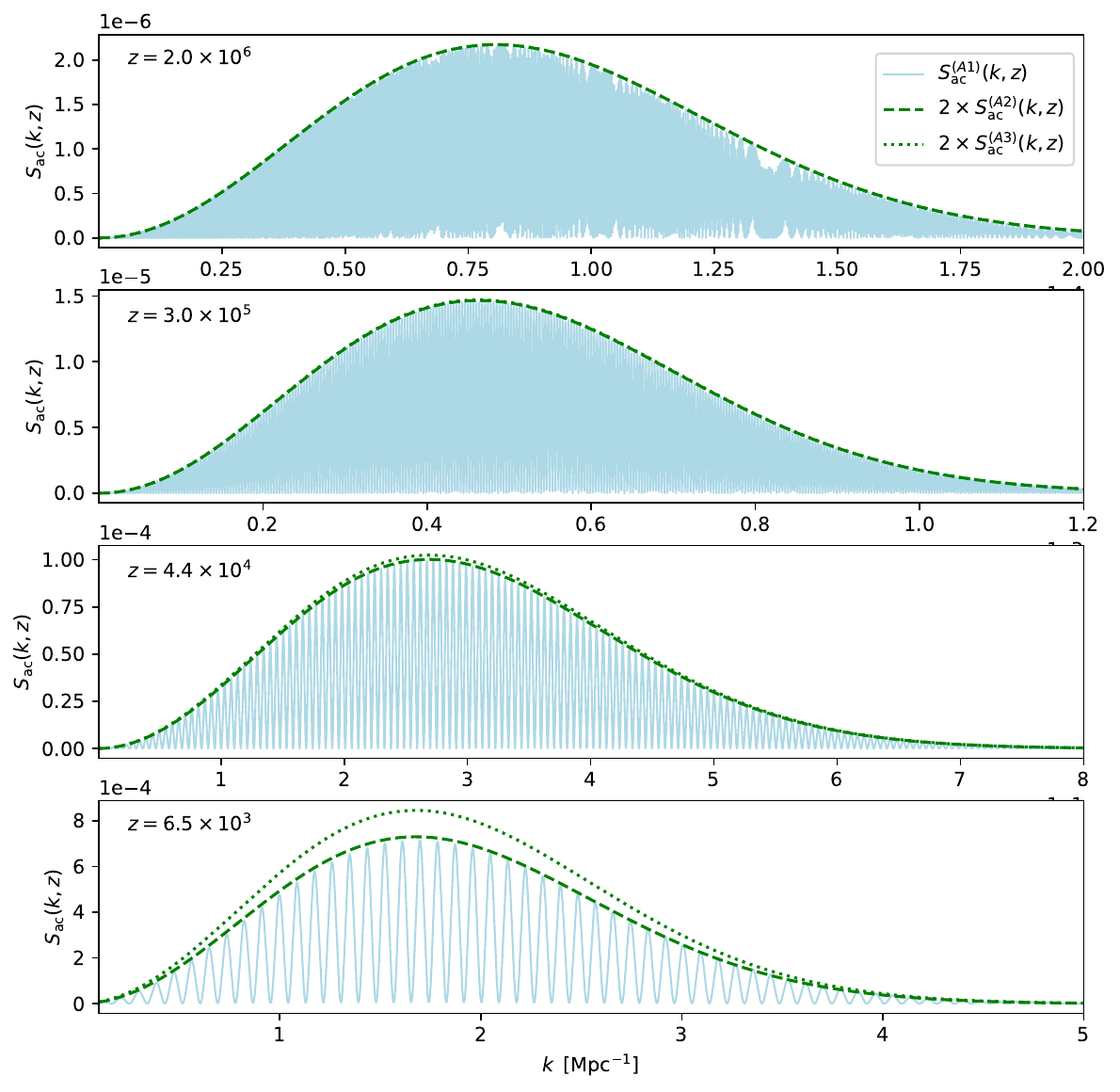}
        \caption{Comparison of the oscillatory source function $S_{\rm ac}^{(A1)}(k,z)$ (solid cyan) with two oscillation-averaged approximations, $S_{\rm ac}^{(A2)}(k,z)$ (dashed green) and $S_{\rm ac}^{(A3)}(k,z)$ (dotted green), multiplied by a factor two. At all redhsift, $2\times S_{\rm ac}^{(A2)}(k,z)$ is a very good approximation for the enveloppe of $S_{\rm ac}^{(A1)}(k,z)$, which proves that $S_{\rm ac}^{(A2)}(k,z)$ accurately describes the average of $S_{\rm ac}^{(A1)}(k,z)$ over one period of oscillation. Obtained with \CLASS.\label{fig:S_sd_enveloppe}}
\end{figure}

In the TCA limit, the photon velocity divergence in the Newtonian gauge, $\theta_\gamma^N$, and the photon shear in any gauge, $\sigma_\gamma$, are related through $\sigma_\gamma \simeq \frac{16}{45 \dot{\tau}} \theta_\gamma^N$. Chluba et al. \cite{Chluba:2012gq} propose to express the source function in the range where the quadrupole term dominates and \cref{{eq:Sac_quadrupole}} applies as
\begin{equation}
S_{\rm ac}(k,\eta) \simeq 
\frac{16}{135 \dot{\tau}^2} (\theta_\gamma^N)^2~.
\label{eq:S_sd_theta_gamma}
\end{equation}
In order to get a convenient expression not requiring a Boltzmann solver, one can replace $\theta_\gamma$ by analytical approximations. For wavelengths above the sound horizon, the transfer function of the photon velocity divergence grows and approaches the oscillatory regime according to the leading-order WKB solution \cite{Hu:1995em}
\begin{equation}
\theta_\gamma^N = A_{\rm WKB} \frac{k c_s}{(1+R)^{1/4}} \sin(k r_s)~, \qquad
k r_{\rm s} \ll 2 \pi~.
\label{eq:theta_gamma_super}
\end{equation}
The normalisation factor $A$ depends on initial conditions. Given that we normalise all transfer functions to the growing mode of adiabatic perturbations with unit curvature perturbation ${\cal R}=1$, this factor is given by \cite{Ma:1995ey}
\begin{equation}
A_{\rm WKB} \simeq \left( 1+\frac{4}{15} f_\nu \right)^{-1}~,
\label{eq:A}
\end{equation}
with $f_\nu = \bar{\rho}_\nu/(\bar{\rho}_\gamma+\bar{\rho}_\nu)$. Around horizon crossing, the oscillation amplitude gets boosted by a factor of three due to the decay of metric fluctuations.\footnote{The ``gravity boost'' effect experienced by the velocity divergence is very well approximated (up to the one percent level) by a factor of three in the oscillation amplitude. It also produces a small phase shift that we neglect here. We get a direct confirmation of the accuracy of this analytic approximation in \cref{fig:S_sd_enveloppe}.}  On sub-horizon scales, the leading-order WKB solution becomes \cite{Hu:1995em}
\begin{equation}
\theta_\gamma^N = A_{\rm WKB} \frac{3 k c_s}{(1+R)^{1/4}} \sin(k r_s) e^{-(k/k_{\rm D})^2}\qquad
k r_{\rm s} \gg 2 \pi~,
\label{eq:theta_gamma}
\end{equation}
where the last factor accounts for diffusion damping. With such an expression for $\theta_\gamma^N$,
the $k$-integral over the source function could be very cumbersome to evaluate numerically due to the quickly oscillating term $\sin^2(k r_s)$: at $z \sim 10^6$, the sound horizon is of the order of $r_s \sim 10 \, {\rm Mpc}^{-1}$, such that integrating up to $k \sim 5 \times 10^4 \,{\rm Mpc}^{-1}$  would require to resolve ${\cal O}(10^3)$ oscillations. Fortunately, as noticed e.g. in references \cite{Chluba:2012gq,Chluba:2013dna}, these oscillations are not playing any role in concrete problems, because any realistic power spectrum should be much smoother than this oscillating term. Thus, the oscillations always get averaged out, and one can safely replace $\sin^2(k r_s)$ by its average over one period, $\langle \sin^2(kr_s) \rangle = \frac{1}{2}$.\footnote{Even in the purely academic case of a primordial spectrum consisting in a Dirac delta-function, $\dot{Q}(z)$ is an oscillatory function of $z$, but these oscillations are averaged over when integrating over $z$ to get the final spectral distortion amplitude, giving the same result as if the substitution $\langle \sin^2(kr_s) \rangle = \frac{1}{2}$ had been performed from the beginning.}
One can then substitute \cref{eq:theta_gamma} inside \cref{eq:S_sd_theta_gamma} and average the result, getting an approximation that we call (A2):
\begin{equation}
    S_{\rm ac}^{(A2)}(k,\eta) \simeq 
    \frac{8 A_{\rm WKB}^2 k^2}{45 \dot{\tau}^2 (1+R)^{3/2}} e^{-2(k/k_{\rm D})^2}~, \qquad
k r_{\rm s}(\eta) \gg 2 \pi~.
\label{eq:Sac_A2}
\end{equation}
As a final approximation, Chluba et al. \cite{Chluba:2013dna} take the limit $R \ll 1$ and $c_s^2 \simeq \frac{1}{3}$, valid deep inside radiation domination. They notice that, in this limit, the time derivative $\partial_\eta r_{{\rm D}, R=0}^{2}$ is equal to $\frac{8}{45 \dot{\tau}}$. With all these simplifications, the source function reduces to a compact expression that we call (A3),
\begin{equation}
S_{\rm ac}^{(A3)}(k,\eta) =
\frac{A_{\rm WKB}^2 k^2}{\dot{\tau}}
    (\partial_\eta r_{{\rm D}, R=0}^{2})  \exp^{- 2 (k/k_D)^2}~,
\end{equation}
with $A_{\rm WKB}$ given by \cref{eq:A} and $r_{{\rm D}, R=0}$ by the $R \rightarrow 0$ limit of \cref{eq:k_D}.\footnote{The public version of \CLASS{} (since version {\tt v2.10.6}) assumes (A3), but with $r_D$ evaluated using the actual function $R(z)$. This has no actual justification and we will replace it soon by (A2). This makes anyway no difference for $\mu$-distortions and only a tiny difference for $y$-distortions, given that $R\ll1$ deep inside the radiation-dominated regime.}

In \cref{fig:S_sd_contributions}, $S_{\rm ac}^{(A1)}(k,\eta)$ was computed with a fine $k$-grid that is prohibitively expensive computationally. To get an efficient computation, one needs to use a coarser grid, but then one cannot resolve each oscillation of $\sin^2(k r_s)$ at large $k$. However, there is a wide range of scales on which the envelope of the oscillations is accurately given by the analytic approximation of $S_{\rm ac}^{(A2)}(k,\eta)$ up to a factor two, as shown in \cref{fig:S_sd_enveloppe}. 
In this range, we get a confirmation that the use of (A2) is accurate up to the level of a few percent for $z>10^3$. \Cref{fig:S_sd_enveloppe} also shows that $S_{\rm ac}^{(A3)}(k,\eta)$ coincides with $S_{\rm ac}^{(A2)}(k,\eta)$ at  $z>10^5$, but does not provide a good approximation at  $z<10^5$, when $R$ starts to be significant. 

\begin{figure}[hbt!]
  %\vspace{-4cm}
        \centering
        %\hspace{-4cm} 
        \includegraphics[width=14cm,angle=0]{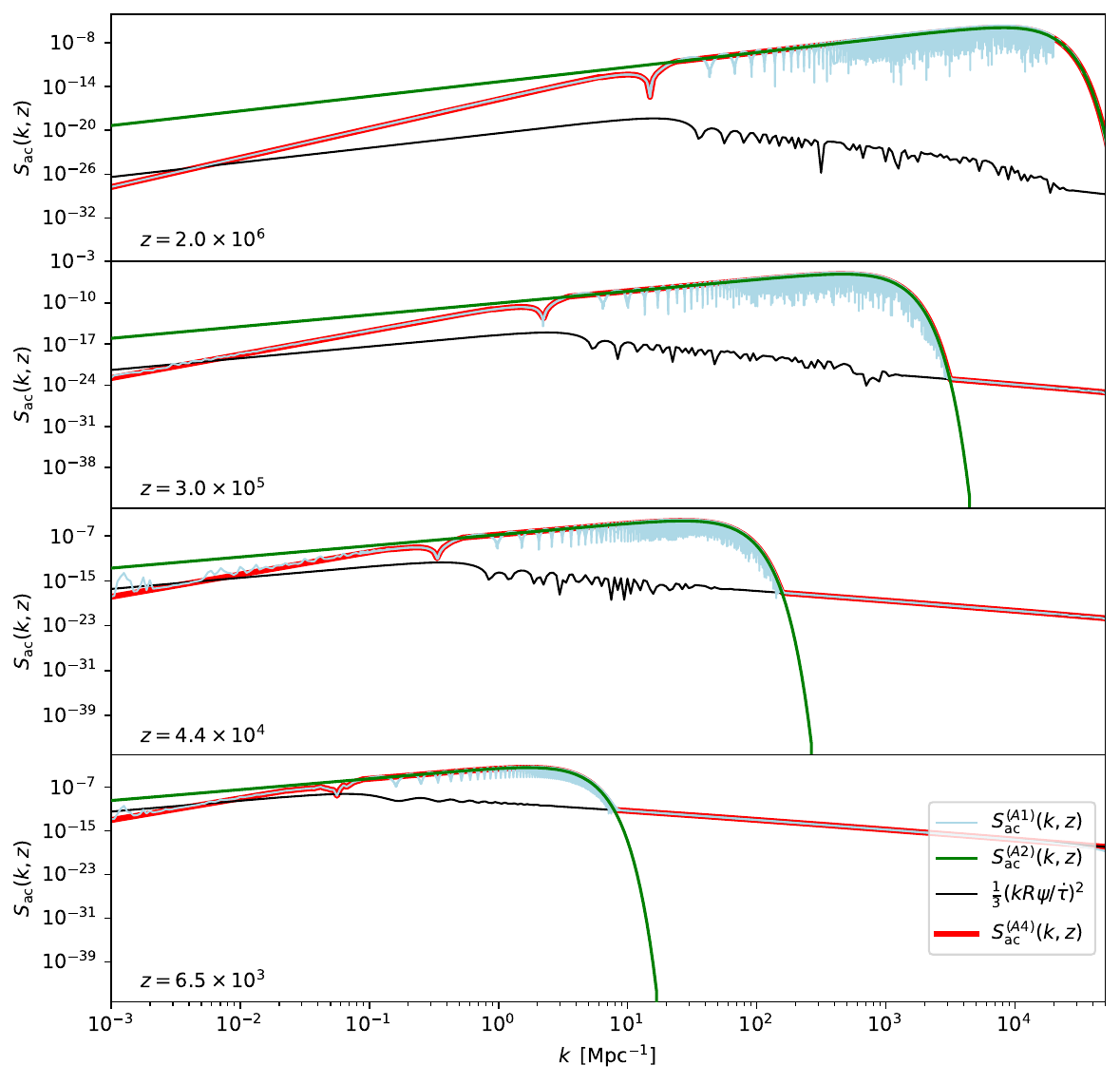}
        \caption{Comparison of the oscillatory source function $S_{\rm ac}^{(A1)}(k,z)$ (cyan) at four redshifts with: (i) the oscillation-averaged approximation $S_{\rm ac}^{(A2)}(k,z)$ (green), which works very well in the central part, (ii) the function $\frac{1}{3}(kR\psi/\dot{\tau})^2$ (black), which accounts for the large-$k$ asymptote of $S_{\rm ac}^{(A1)}(k,z)$, 
        and (iii) the oscillation-averaged approximation $S_{\rm ac}^{(A4)}(k,z)$ (red), which correctly accounts for the small-$k$ and large-$k$ asymptotes of $S_{\rm ac}^{(A1)}(k,z)$. Obtained with \CLASS.\label{fig:S_sd_comparision}}
\end{figure}

We compare various approximation schemes for the source function in \cref{fig:S_sd_comparision}. The curve corresponding to (A1) can be considered as the nearly exact result. The approximation (A2) provides a very good oscillation-averaged version of (A1), as already proved in \cref{fig:S_sd_enveloppe}, excepted in the asymptotic high-$k$ and low-$k$ limits.
At large $k$, (A2) neglects the residual dipole contribution mentioned at the end of the previous subsection. The comparison of the black and red curves in \cref{fig:S_sd_comparision} shows that \cref{eq:damping_tail} provides a very good approximation to the high-$k$ asymptote.
At low $k$, the asymptote reflects the behaviour $\sin(kr_{\rm s}) \simeq k r_{\rm s}$, while (A2) would replace the sine function with $1/\sqrt{2}$. Finally, between these two asymptotic regimes, the numerical calculation based on the oscillation-averaged version of (A2) should actually be more accurate and robust than the full (A1) calculation, especially when the $k$-grid over which the source function is tabulated is too coarse.

These considerations lead us to define the most accurate and efficient approximation that can be used within \CLASS{} for the source function, which we call (A4):
\begin{equation}
S_{\rm ac}^{(A4)}(k,\eta) \equiv \left\{
\begin{tabular}{ll}
$\frac{2}{9}\,\sin^2(k r_{\rm s})\,S_{\rm ac}^{(A2)}(k,\eta)$ & for $k r_{\rm s}<\frac{\pi}{4}$~, \\
$\min\left[S_{\rm ac}^{(A1)},S_{\rm ac}^{(A2)}\right]$ & for $\frac{\pi}{4} \leq k r_{\rm s}<\pi$~, \\
$\max\left[S_{\rm ac}^{(A2)},\frac{1}{3}(k R \psi / \dot \tau)^2\right]$ & for $k r_{\rm s}\geq\pi$~,
\end{tabular}
\right.
\label{eq:s4}
\end{equation}
where the first line is justified by \cref{eq:theta_gamma_super} and the third one by \cref{eq:damping_tail}.
The result is shown as the red line in \cref{fig:S_sd_comparision}. It provides an excellent approximation to the oscillation-averaged source function in all regimes. The approximation (A4) is more stable numerically than the full expression (A1), as can be checked in \cref{fig:S_sd_comparision} (where, for instance, one can see glitches in (A1) at small $k$). It is also much more efficient since it only requires a coarse grid of $k$ values.  In the next sections, we will check whether this more accurate expression leads to differences with respect to the most common approximation usually performed in the literature when computing $\mu$-distortions.

\subsection{Redshift integral\label{sec:W_approx}}

We can perform the integral over $z$ of \cref{eq:mu_integral_z} to express the final $\mu$-distortion induced by the dissipation of acoustic waves as
\begin{equation}
 \mu=\int
 %_{k_{\rm min}}^{\infty} 
 dk \frac{k^2}{2\pi^2} P_{\cal R}(k) W_\mu(k) = \int
 %_{k_{\rm min}}^{\infty} 
 \frac{dk}{k} {\mathcal P}_{\cal R}(k) W_\mu(k)~,
 \label{eq:mu_P_W}
 \end{equation}
where we defined the window function
\begin{equation}
    W_\mu(k) \equiv 
    \int d \! \ln \! z \,\, K_\mu(k,z)
    =
    - \int dz \, {\cal I}_\mu(z) \, \frac{4 \dot \tau(z)\, S_{\rm ac}(k,z)}{H(z)}~.
    \label{eq:z_integral}
\end{equation}
This window function is slightly cosmology-dependent, but if we just need an estimate of $\mu$-distortions assuming a standard cosmology, we can use a reference function $W_\mu(k)$ computed for the $\Lambda$CDM model best fitting current data. Reference~\cite{Chluba:2013dna} performed a numerical fit -- that we will call (F1) -- to the result of this integral and reports\footnote{although \cite{Cyr:2023pgw} have a factor of 360 instead of 340 in the exponent, which makes a negligible difference} 
\begin{equation}
W_\mu^{\rm (F1)}(k)=2.8 A_{\rm WKB}^2 \left[\exp\left\{-\frac{(\hat k/1360)^2}{1+(\hat k/260)^{0.3}+\hat k/340}\right\}-\exp\left\{-(\hat k/32)^2\right\}\right], \label{eq:Wmu-long}
\end{equation}
where $\hat k=k/[1 \mathrm{Mpc}^{-1}]$, still with $A_{\rm WKB}$ given by \cref{eq:A}.
An even simpler approximation,
\begin{equation}
W_\mu^{\rm (F2)}(k)=2.8 A_{\rm WKB}^2 \left[\exp\left\{-\frac{\hat k}{5400}\right\}-\exp\left\{-(\hat k/31.6)^2\right\}\right]~, \label{eq:Wmu-short}
\end{equation}
was used in e.g.~\cite{Chluba:2012we,Juan:2022mir}, which results in a slightly weaker $\mu$-constraint than \cref{eq:Wmu-long} (with the difference typically of order 10\%). Finally, we produced our own fit $W_\mu^{\rm(F3)}(k)$, using \cref{eq:s4} and \cref{eq:z_integral} while assuming the {\it Planck} best-fit values of the cosmological parameters\footnote{We use as input to \CLASS{} cosmological parameters
$\omega_{\rm b}=0.02229$,
$\omega_{\rm cdm}=0.1189$,
$h=0.6781$,
close to the mean values reported by \cite{Tristram:2023haj} for Planck Release 4 in combination with Baryon Acoustic Oscillation data,
$N_{\rm eff}=3.044$, as suggested by standard neutrino decoupling \cite{Drewes:2024wbw},
$Y_{\rm He}=0.24672$, as suggested by Nucleosynthesis combined with Planck \cite{Planck:2018vyg},
and 
$T_{\rm cmb}=2.7255\,{\rm K}$, as suggested by FIRAS \cite{Fixsen:2009ug}. Other parameters are irrelevant.
} and a PCA decomposition of spectral distortions at order two based on the FIRAS frequency range:\footnote{For the calculation of spectral distortions, we use the input \CLASS{} parameters: 
{\tt 'sd\_branching\_approx':'exact'},
{\tt 'sd\_detector\_name':'FIRAS'},
{\tt 'sd\_detector\_file':'FIRAS\_nu\_delta\_I.dat'},
{\tt 'sd\_PCA\_size':2}, 
all described in \cite{Lucca:2019rxf}.}
\begin{equation}
W_\mu^{\rm (F3)}(k)= \exp\left[\sum_{n=0}^6 w_n \ln\left(\frac{k}{1\, {\rm Mpc}^{-1}}\right)\right]~,
\label{eq:W_F3}
\end{equation}
with coefficients $z_n$ provided in \cref{tab:W_F3}. 
%$\{z_n\}=\{-2.78,1.58,-1.33\,10^{-1},1.477\,10^{-2}, -1.1864\,10^{-2},2.0739\,10^{-3},-1.0633\,10^{-4} \}$. 
These approximations will be compared in the next section.

\begin{table}[hbt!]
\begin{center}
\begin{tabular}{cr|cr}
$n$&$z_n~~~~~~~$&$n$&$z_n~~~~~~~$\\
\hline
0 & $-2.78~~~~~~$ & 4 & $-1.1864\times10^{-2}$\\
1 & $1.58~~~~~~$ & 5 & $2.0739\times10^{-3}$\\
2 & $-1.33\times10^{-1}$ & 6 & $-1.0633\times10^{-4}$ \\
3 & $1.477\times10^{-2}$ &&\\
\end{tabular}   
\end{center}
\caption{Coefficients of the numerical fit to $W_\mu(k)$ of \cref{eq:W_F3}.
\label{tab:W_F3}}
\end{table}

%Figure~\ref{fig:mu_ratio} comapres this approximation to the ones of the previous section. [...]

\subsection{Distortions induced by a peaked primordial spectrum}

\begin{figure}[hbt!]
  %\vspace{-4cm}
        \centering
        %\hspace{-4cm} 
        \includegraphics[width=7.5cm,angle=0]{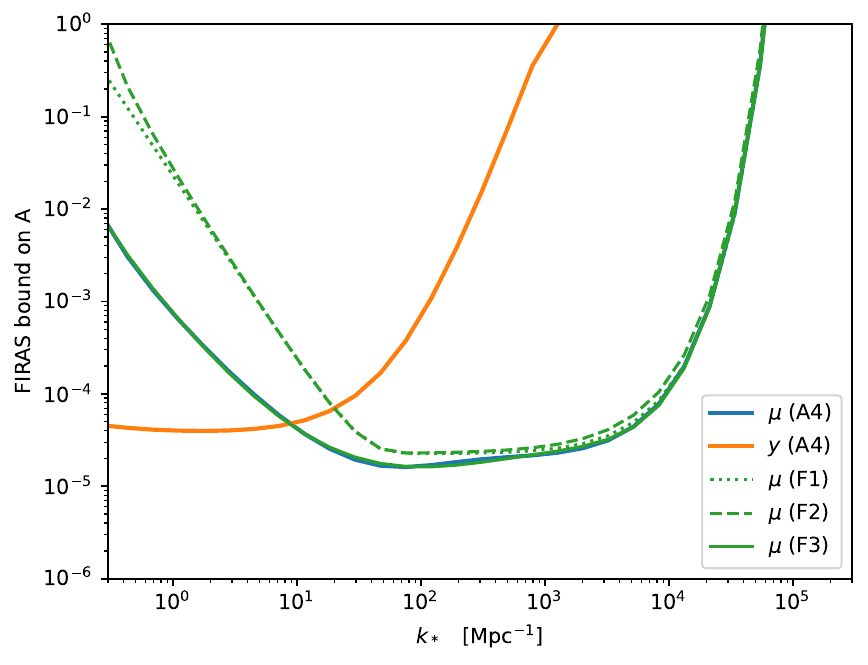}
        \includegraphics[width=7.5cm,angle=0]{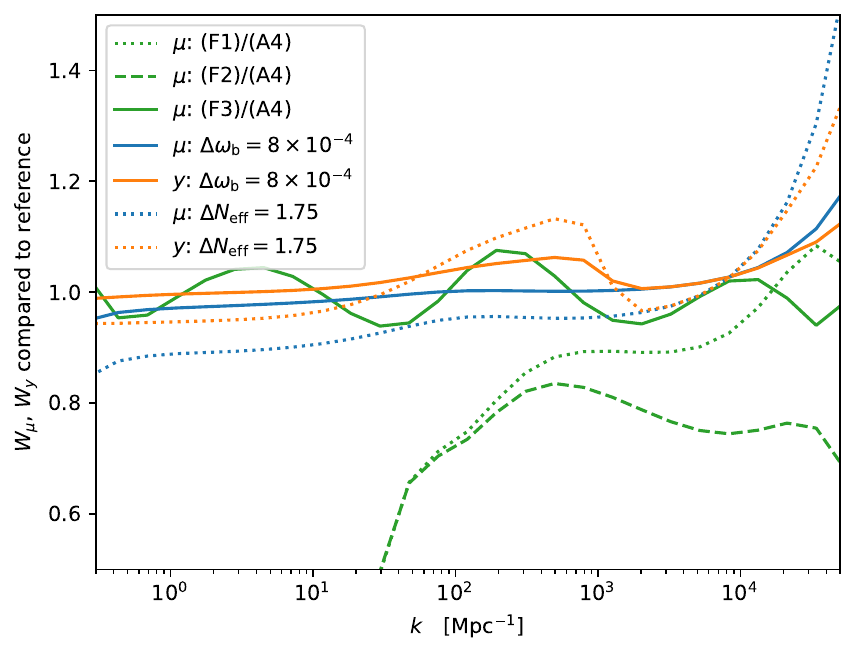}
        \caption{(Left) FIRAS bounds on the amplitude $A$ of a Dirac delta-function curvature spectrum (or, equivalently, on the variance of ${\cal R}$) as a function of the peak scale $k_*$, when the window functions $W_y$, $W_\mu$ are given by an integral over the accurate source function (A4) (solid orange and blue), or when $W_\mu$ is approximated by the fitted functions (F1), (F2) and (F3) (green curves). The (F3) fit (solid green) is nearly indistinguishable by eye from the accurate prediction (solid blue). We use $\mu_{\rm FIRAS}=4.7\times 10^{-5}$ \cite{Bianchini:2022dqh} and  $y_{\rm FIRAS}=1.5\times 10^{-5}$ \cite{Fixsen:1996nj}. The accurate window functions $W_\mu$, $W_y$ are computed with the {\it Planck} reference cosmology.  (Right) Ratio
        of the fitted window functions (F1), (F2), (F3) over the accurate one (green curves), and relative variation of the accurate window functions $W_\mu(k)$, $W_y(k)$ when varying $\omega_{\rm b}$ (solid lines) or $N_{\rm eff}$ (dotted lines) by about 5$\sigma$ of the current observational bounds.
        \label{fig:mu_ratio}}
\end{figure}

We conclude this section by showing how the different approximations discussed in \cref{sec:Sac_approx,sec:W_approx} compare at the level of $\mu$-distortions. For simplicity, we assume that distortions are mainly generated by a sharp peak in the primordial spectrum, which we approximate as a Dirac delta-function
\begin{equation}
    {\cal P}_{\cal R}(k)= \frac{k^3}{2\pi^2} P_{\cal R}(k) = A k_* \delta(k-k_*)~.
    \label{eq:pr_dirac}
\end{equation}
The factor $A$ represents at the same time the amplitude of the curvature power spectrum in Fourier space and the variance of curvature perturbations in real space, since
\begin{equation}
    \left\langle {\cal R}^2(\vec{x}) \right\rangle = \int_0^\infty \frac{dk}{k} {\cal P}_{\cal R}(k)=A~.
    \label{eq:pr_norm}
\end{equation}
In this case, the $\mu$-distortion is simply given by $\mu(k_*) = A \, W_\mu(k_*)$. We show in \cref{fig:mu_ratio} (left panel) the bound on $A$ implied by FIRAS for each value of $k_*$, which is given by $A \leq \mu_{\rm FIRAS}/W_\mu(k_*)$. We use the updated calculation of the FIRAS 95\%CL upper bound on $\mu$ from reference \cite{Bianchini:2022dqh}, $\mu_{\rm FIRAS} = 4.7 \times 10^{-5}$, which is stronger than the original FIRAS bound by a factor two \cite{Fixsen:1996nj}. 
Similarly, the $y$-distortion is given by $y(k_*) = A \, W_y(k_*)$.
We also show in the left panel of \cref{fig:mu_ratio} the bound $A \leq y_{\rm FIRAS}/W_y(k_*)$ inferred from the original FIRAS 95\%CL upper bound $y_{\rm FIRAS} = 1.5 \times 10^{-5}$ \cite{Fixsen:1996nj}.

The left panel of \cref{fig:mu_ratio} compares the 95\%CL FIRAS bounds on $A$ as a function of $k_*$ obtained in the accurate case (A4), that is when integrating numerically the source function $S_{\rm ac}^{(A4)}(k,z)$ within \CLASS{} following \cref{eq:s4} and \cref{eq:z_integral} (solid blue curve), with those derived from the fitted window functions $W_\mu^{\rm (F1)}(k)$, $W_\mu^{\rm (F2)}(k)$ and  $W_\mu^{\rm (F3)}(k)$ (respectively dotted, dashed and solid green curves). The right panel also shows the ratio of these fits to the accurate window function (A4). The fits (F1) and (F2) are very inaccurate in the small $k$ limit because, as explained in \cite{Cyr:2023pgw}, they do not take accurately into account the branching ratios between $\mu$, $y$ and other types of distortions based on the PCA approach. However, for $k>100\,{\rm Mpc}^{-1}$, they are 30\% accurate. Instead, our fit $W_\mu^{\rm (F3)}(k)$ reaches 10\% accuracy in the whole displayed range.

We don't show the result based on $S_{\rm ac}^{(A2)}(k,z)$, which is indistinguishable from the one based on $S_{\rm ac}^{(A4)}(k,z)$ in the range probed by the figure. This shows that, given the sensitivity of FIRAS, our improved modelling of the source functions in the small-$k$ and large-$k$ limits is irrelevant for $\mu$-distortions. The difference between $S_{\rm ac}^{(A2)}(k,z)$ and $S_{\rm ac}^{(A4)}(k,z)$ only appears in the small- and large-$k$ tails of the source function, and is thus only potentially relevant for very small or very large values of $k_*$ compared to the range $k \in [0.5,5\times10^4] \, {\rm Mpc}^{-1}$ where $\mu$-distortions matter. In other words, to perceive the difference between the asymptotes of $S_{\rm ac}^{(A2)}$ and $S_{\rm ac}^{(A4)}$, one would need to choose extreme values of $k_*$ such that $\mu$ would anyway be negligible with respect to the FIRAS observational bound. With much more sensitive data than FIRAS, it would in principle be possible to get bounds on $A$ even when $k_*$ is very large (e.g. $k_* \sim 10^6\, {\rm Mpc}^{-1}$), and our point regarding the modelling of the large-$k$ tail of the source function may gain more relevance. 

We also checked that the bounds displayed in the left panel of \cref{fig:mu_ratio} are in excellent (10\% level) agreement with those in Figures 1 to 4 of \cite{Cyr:2023pgw}.

Finally, we checked the dependence of our window functions on cosmological parameters. The acoustic dissipation source function only needs to be integrated until some time soon after photon decoupling. Therefore, in the $\Lambda$CDM model and simple extensions of it, the window functions should depend at most on the parameters $\{\omega_{\rm b}, \omega_{\rm m}, h, N_{\rm eff}, Y_{\rm He}, T_{\rm cmb}\}$ that play a role in the evolution of perturbations in the early universe. We checked the impact on $(W_\mu, W_y)$ of shifting each of these parameters five standard deviations away from the {\it Planck} best-fit cosmology, using confidence limits from {\it Planck}, BAO, primordial element abundances and FIRAS.\footnote{For this exercise, we took $\sigma(\omega_{\rm b})=0.00012$, $\sigma(\omega_{\rm cdm})=0.00090$, $\sigma(h)=0.0038$ from \cite{Tristram:2023haj}, $\sigma(N_{\rm eff})=0.35$, $\sigma(Y_{\rm He})=0.0006$ from \cite{Planck:2018vyg}, and $\sigma(T_{\rm cmb})=0.00057$ from \cite{Fixsen:2009ug}.} The largest variations are observed when shifting $\omega_{\rm b}$ and $N_{\rm eff}$, as shown in the right panel of \cref{fig:mu_ratio}. However, even with such extreme parameter shifts, the window functions only vary by 20\% in the range $k \in [0.5, 2\times10^4] \, {\rm Mpc}^{-1}$. Thus, for most current purposes, it is reasonable to neglect the dependence of the $\mu$- and $y$-bounds on cosmology.

\section{Beyond Gaussian fluctuations \label{sec:BGF}}

\subsection{Local non-Gaussianity: perturbative and non-perturbative limits \label{sec:LNG}}

Primordial non-Gaussianity of the local type, which arises when deviations from Gaussianity are local in real space, can be characterised by a Taylor expansion of the comoving curvature perturbation, 
\begin{equation}    
{\cal R}(\vec{x})=
    {\cal R}_{\rm G}(\vec{x})
    + \tilde{f}_{\rm NL} 
    \left( {\cal R}_{\rm G}(\vec{x})^2
    - \langle {\cal R}_{\rm G}(\vec{x})^2 \rangle 
    \right) + \tilde{g}_{\rm NL} 
     {\cal R}_{\rm G}(\vec{x})^3    
     + ...
    \label{eq:Rg}
\end{equation}
where $\tilde{f}_{\rm NL}=3f_{\rm NL}/5$ and $\tilde{g}_{\rm NL}=9g_{\rm NL}/25$ are the first two expansion coefficients and ${\cal R}_{\rm G}$ is a Gaussian random field with $|{\cal R}_{\rm G}(\vec{x})| \ll \mathcal{O}(1)$ and $\langle{\cal R}_{\rm G}(\vec{x})\rangle=0$. Thus, each new term in the expansion is expected to be suppressed with respect to the previous one, unless its coefficient is unusually large.
%The order of magnitudes for $\tilde{f}_{\rm NL}$ and  $g_{\rm NL}$ is highly dependent on the mechanism that gave rise to the curvature perturbations. Ref. \cite{Suyama:2010uj} categorises inflationary models into three primary groups: linear $g_{\rm NL}$, where both $g_{\rm NL}$ and $\tilde{f}_{\rm NL}$ are of the same order, enhanced $g_{\rm NL}$, where $g_{\rm NL}$ is proportionate to the $nth$ power of $\tilde{f}_{\rm NL}$ with $n$ being 2 at the most \cite{Takahashi:2014bxa}, and finally, suppressed $g_{\rm NL}$, where $g_{\rm NL}$ is suppressed with respect to $\tilde{f}_{\rm NL}$. 
In a majority of inflationary models, $\tilde{g}_{\rm NL}$ cannot be unduly large relative to $\tilde{f}_{\rm NL}$ \cite{Suyama:2010uj,Takahashi:2014bxa}, although exceptions could arise due to fine-tuning \cite{Nurmi:2013xv,Suyama:2013nva}.  Therefore, there is enough motivation to disregard the $\tilde{g}_{\rm NL}$ term in \cref{eq:Rg}, keeping only the leading order deviation from Gaussianity. In Fourier space,
\begin{equation}
    {\cal R}(\vec{k})=
    f_{\rm L}{\cal R}_{\rm G}(\vec{q})
    + \tilde{f}_{\rm NL}
    \int \frac{d^3 \vec{q}}{(2 \pi)^{3 / 2}} {\cal R}_{\rm G}(\vec{q}) {\cal R}_{\rm G}(\vec{k}-\vec{q})~,
    \label{eq:local_ng_fourier}
\end{equation}
where $f_{\rm L}=0$ or 1 is just a switch introduced for the clarity of the discussion.
The Gaussian case corresponds to $f_{\rm L}=1$ and $\tilde{f}_{\rm NL}=0$. By perturbative non-Gaussianity we mean $f_{\rm L}=1$ and $|\tilde{f}_{\rm NL}| \langle {\cal R}_{\rm G}^2 \rangle^{1/2} \ll 1$, such that the first term in  the r.h.s. of \cref{eq:Rg} dominates. By non-perturbative non-Gaussianity we mean the contrary: $f_{\rm L}=1$ and $|\tilde{f}_{\rm NL}| \langle {\cal R}_{\rm G}^2 \rangle^{1/2} \gg 1$, such that the second term dominates. Pushing this limit to its extreme, we get the case of a pure $\chi^2$ distribution. A simple way to express the fact that the first term is sub-dominant is to define this limit as $f_{\rm L}=0$. In this case, the parameter $\tilde{f}_{\rm NL}$ is not interesting by itself because it can be absorbed in the power spectrum of ${\cal R}_{\rm G}$, but we will keep it anyway for clarity.  

As explained in the introduction, we are interested in models in which non-Gaussianity applies only to the curvature fluctuations that generate PBHs. We assume that these fluctuations are strongly enhanced around a scale associated to the wavenumber $k_*$, with a sharp peak in the power spectrum and local non-Gaussian statistics. We further assume that they are not correlated with those explaining CMB anisotropies, which have a nearly scale-invariant power spectrum over a wide range of scales and remain very close to Gaussian. In this set-up, there is no coupling between the modes probed by CMB maps and those potentially giving rise to PBH formation. Thus, we can neglect the smooth Gaussian component and consider only the peaked non-Gaussian one. Since the curvature fluctuations of interest are significant only over a narrow range of wavenumbers, it is reasonable to assume that $f_{\rm NL}$ is constant over this range.

\subsection{Leading contributions to spectral distortions\label{sec:NG_SD}}

We assume that the root-mean-square of ${\cal R}_{\rm G}(\vec{x})$ is of the order of $A^{1/2}_{\rm G}$. Then,  neglecting numerical factors of order unity, we have 
\begin{equation}
\langle {\cal R}^2(\vec{x})\rangle 
\sim f_{\rm L} A_{\rm G} + \tilde{f}_{\rm NL}^2 A^2_{\rm G}~,
\quad
\langle {\cal R}^3(\vec{x}) \rangle
\sim f_{\rm L} \tilde{f}_{\rm NL} A^2_{\rm G} + \tilde{f}_{\rm NL}^3 A^3_{\rm G}~,
\label{eq:R_NG}
\end{equation}    
where we used $f_{\rm L}^2 = f_{\rm L}$. 

Spectral distortions are sourced by the heating rate $\dot{Q}$ given in real space by \cref{eq:Q_dot_full}. The leading and next-to-leading order contributions to the source function are given schematically by a sum over the two-point and three-point correlation function of photon temperature anisotropies $\Theta(\vec{x},\hat{n},\eta)$.

The Fourier modes of photon temperature multipoles $\Theta_\ell(\vec{k},\eta)$ and primordial curvature perturbations ${\cal R}(\vec{k})$ are related linearly at leading order, $\Theta_\ell(\vec{k},\eta) = T_{\Theta_\ell}(k,\eta) {\cal R}(\vec{k})$, with  $T_{\Theta_\ell}(k,\eta)$ being a linear transfer function. The temperature anisotropy $\Theta(\vec{x},\hat{n},\eta)$ can acquire non-Gaussian statistics for two reasons: (i) Even if initial conditions are Gaussian, the small degree of non-linearity in the evolution equation (Einstein and Boltzmann equations) will generate some non-Gaussianity. This effect can be captured by expanding the relation between $\Theta_\ell(\vec{k},\eta)$ and ${\cal R}(\vec{k})$ beyond linear order, but it is known to be very small (see e.g.~\cite{Pettinari:2013he}). (ii) The initial conditions created e.g.~by inflation could predict some initial non-Gaussian statistics for ${\cal R}(\vec{k})$.
In this work, we are interested exclusively in the latter type of non-Gaussianity. The idea is that some non-Gaussian statistics is generated very early at the level of curvature perturbations, and then propagates within radiation and matter perturbations, without experiencing further distortions. Thus, we will stick to a linear relation between the Fourier modes of $\Theta_\ell$ and of ${\cal R}$.

To keep this preliminary discussion simple, let us further assume that the relevant transfer functions are of order one,\footnote{We will come back to this assumption at the end of \cref{sec:PNGC}} such that $\langle \Theta^2\rangle \sim  \langle {\cal R}^2\rangle$ and $\langle \Theta^3\rangle \sim  \langle {\cal R}^3\rangle$. In this case, we have: 
\begin{equation}
\langle \Theta^2\rangle 
\sim f_{\rm L} A_{\rm G} + \tilde{f}_{\rm NL}^2 A^2_{\rm G}~,
\quad
\langle \Theta^3\rangle
\sim f_{\rm L} \tilde{f}_{\rm NL} A^2_{\rm G} + \tilde{f}_{\rm NL}^3 A^3_{\rm G}~.
\label{eq:T_NG}
\end{equation} 
%
%\julien{to be discussed: we shall recheck that we can always assume this linear relation. Otherwise we would need to discuss second-order solutions to the Bolztmann equation of the type $\Theta \sim \Theta^{(1)} {\cal R} + \Theta^{(2)} {\cal R}^2$. We can decide that we are interested only in models where the linear solution is enough because $\Theta$ is small enough. But we must be sure that it is consistent to consider simultaneously this assumption and the case where the $\Theta^3$ term matters in the perturbative NG case. Chris says: the non-linear evolution can be formulated as producing a small $\tilde{f}_{\rm NL} \sim 1$, while we are interested in a primordial fnl >> 1, so we should be fine.} 
Then, we can distinguish between the following sub-cases:
\begin{itemize}
    \item Gaussian limit: when $f_{\rm L}=1$ and $\tilde{f}_{NL}=0$, the statistics of temperature fluctuations 
    obeys $\langle\Theta^2\rangle \sim A_{\rm G}$,
    $\langle\Theta^3\rangle = 0$.
    The fact that curvature perturbations always remain small (to avoid an over-production of black holes) implies 
    $\langle {\cal R}^2\rangle \sim A_{\rm G} \ll 1$.
    \item Perturbative non-Gaussianity regime:
    when $f_{\rm L}=1$ and $|\tilde{f}_{\rm NL}| A^{1/2}_{\rm G}<1$, $\langle\Theta^2\rangle$ is dominated by the Gaussian contribution proportional to $A_{\rm G}$, and $A_{\rm G}\ll1$ must still be true to avoid exceedingly large curvature perturbations. The leading non-Gaussian correction to spectral distortions is given either by the contribution of $\tilde{f}_{\rm NL}^2 A^2_{\rm G}$ to $\langle\Theta^2\rangle$ or by the contribution of $\tilde{f}_{\rm NL} A^2_{\rm G}$ to $\langle\Theta^3\rangle$. The latter dominates the source function only if $|\tilde{f}_{\rm NL}|<1$, in which case the non-Gaussian correction is expected to be very small. For completeness, we will calculate the contribution of both terms to the $\mu$-distortion. 
    \item Non-perturbative non-Gaussian regime (with curvature obeying $\chi^2$ statistics): 
    in the limit $|\tilde{f}_{\rm NL}|A^{1/2}_{\rm G}\gg 1$, we have $\langle\Theta^2\rangle \sim \tilde{f}_{\rm NL}^2 A^2_{\rm G}$ and $\langle\Theta^3\rangle \sim \tilde{f}_{\rm NL}^3 A^3_{\rm G}$. The fact that curvature perturbations remain small requires $\langle {\cal R}(\vec{k})^2\rangle \sim \tilde{f}_{\rm NL}^2 A^2_{\rm G} \ll 1$, which in turn implies $\langle\Theta^2\rangle \gg \langle\Theta^3\rangle$. The conditions on $\tilde{f}_{\rm NL}$ are summarised as $A^{-1/2}_{\rm G} \ll |\tilde{f}_{\rm NL}| \ll A^{-1}_{\rm G}$ and are compatible with each other as long as $A_{\rm G} \ll 1$. For instance, one could have $A_{\rm G} \sim 10^{-3}$ and $\tilde{f}_{\rm NL}\sim 10^2$. 
    \item Note that a technically equivalent (but somewhat simpler) way to describe the non-perturbative non-Gaussian regime is to set $f_{\rm L}$ to zero and absorb $\tilde{f}_{\rm NL}^2$ in the definition of $A_{\rm G}$. We would then say that $\langle\Theta^2\rangle \sim A^2_{\rm G} \ll 1$ while $\langle\Theta^3\rangle \sim A^3_{\rm G} \ll \langle\Theta^2\rangle$.
\end{itemize}
 In summary, we expect that in the perturbative non-Gaussian case, the $\mu$ and $y$-distortions are mildly affected by corrections coming either from the two- or three-point correlation function, depending on the value of $|\tilde{f}_{\rm NL}|$; while in the $\chi^2$ limit they should arise dominantly from the two-point correlation function $\langle\Theta^2\rangle$. We will obtain the distortions in the limit of perturbative non-Gaussianity in \cref{sec:PNG_SD}, and in the case of $\chi^2$ (non-perturbative) non-Gaussianity in \cref{sec:NPNG}. 

\section{Distortions in the limit of local perturbative non-Gaussianity \label{sec:PNG_SD}}

We proceed with the calculations under the assumption of perturbative non-Gaussian of the local type, as defined in \cref{eq:Rg}, with $\tilde{g}_{\rm NL}=0$:
\begin{eqnarray}
    {\cal R}(\vec{x}) =  {\cal R}_{\rm G}(\vec{x}) + \tilde{f}_{\rm NL}\left[{\cal R}_{\rm G}^2(\vec{x}) - \langle {\cal R}_{\rm G}^2\rangle\right]~.
\end{eqnarray}
As already mentioned in \cref{sec:LNG}, we always need to assume $|\tilde{f}_{\rm NL} {\cal R}_{\rm G} | \sim |\tilde{f}_{\rm NL}| {\cal P}_{{\cal R}_{\rm G}}^{1/2}<1$ in order to remain in the perturbative non-Gaussianity regime.

\subsection{Next-to-leading contribution from two-point correlation function\label{sec:NG_SD_Q}}

As discussed in \cref{sec:NG_SD}, as long as we further assume $|\tilde{f}_{\rm NL}|>1$, the leading non-Gaussian correction to spectral distortions arises from extra contributions to the two-point correlation function $\langle \Theta^2 \rangle$, rather than from the three-point correlation function $\langle \Theta^3 \rangle$. Thus, to compute these corrections, we only need to focus on the primordial power spectrum. As discussed in \cite{Byrnes:2007tm}, corrections to this power spectrum come from a single 1-loop diagram which results in
\begin{equation}
    P_{\cal R}(k)=
    P_{{\cal R}_{\rm G}}(k) 
    + \tilde{f}_{\rm NL}^2
    P_{{\cal R}_{\rm G}^2}(k)=
    P_{{\cal R}_{\rm G}}(k) 
    + 2 \tilde{f}_{\rm NL}^2
    \int \frac{d^3q}{(2\pi)^3}P_{{\cal R}_{\rm G}}(q)P_{{\cal R}_{\rm G}}(|\vec k-\vec q|)~.
    \label{eq:P_PG_chi2}
\end{equation}
We assume that the Gaussian field ${\cal R}_{\rm G}$ has a delta-peaked power spectrum of the form of \cref{eq:pr_dirac} at scale $k_*$ and with amplitude $A_{\rm G}$. We write $\vec{q}$ in spherical coordinates as $(q,\theta,\phi)$, assuming that $\vec{k}$ points in the $z$ direction such that $\vec{k}\cdot\vec{q}=k\,q\cos\theta$. Then, we can analytically integrate \cref{eq:P_PG_chi2} to find
\begin{eqnarray}
    P_{{\cal R}^2_{\rm G}}(k)&=&  
    2\times 2\pi^2 A_{\rm G} k_*^{-2} \times(2\pi) \frac{1}{(2\pi)^3} \int_0^\pi d\theta \sin\theta \int_0^\infty dq q^2 \delta(q-k_*) P_{{\cal R}_{\rm G}}(|\vec k-\vec q|) \nonumber \\  
    %&=& -A k_* \int d(\cos\theta) \int \frac{dq}{q} \delta(q-k_*) P_{{\cal R}_{\rm G}}(|\vec k-\vec q|) \\ 
    &=& -A_{\rm G} \int_0^\pi d(\cos\theta) \,\, P_{{\cal R}_{\rm G}}(|\vec k- k_*\hat q|) \nonumber \\ 
    &=&  2\pi^2 A_{\rm G}^2 k_*^{-2} \int_{-1}^1 dy \,\,  \delta(|\vec k- k_*\hat q|-k_*)~,
    \end{eqnarray}
using the notation $\hat q = \vec q / q$ and the definition $y\equiv\cos\theta$, such that  
$\hat k\cdot\hat q=y$. Then
\begin{equation}
P_{{\cal R}^2_{\rm G}}(k)
= 2\pi^2 A_{\rm G}^2 k_*^{-2} \int_{-1}^1 dy \,\, \delta(f(y))
\qquad {\rm with} \qquad
f(y) \equiv \sqrt{k^2-2k k_*y+k_*^2}-k_*~.
\end{equation}
When $k>2k_*$, $f(y)$ is always non-zero and $P_{{\cal R}^2_{\rm G}}(k)$ vanishes. When $k\leq 2k_*$, $f(y)$ has one root $y_0=\frac{k}{2k_*}$ with $f'(y_0)=-k$, such that 
\begin{equation}
P_{{\cal R}^2_{\rm G}}(k)
= 2\pi^2 A_{\rm G}^2 k_*^{-2} \int_{-1}^1 dy \,\,
\frac{\delta(y-\frac{k}{2k_*})}{|k|}
= 2\pi^2 A_{\rm G}^2 k_*^{-2} k^{-1} H(2k_*-k)~,
\label{eq:P_Heaviside}
\end{equation}
where $H$ is the Heaviside function. Finally, the dimensionless power spectrum reads
\begin{equation} 
{\cal P}_{\cal R}(k)\equiv \frac{k^3}{2\pi^2}P_{\cal R}(k)
= {\cal P}_{{\cal R}_{\rm G}}(k) + \tilde{f}_{\rm NL}^2
{\cal P}_{{\cal R}^2_{\rm G}}(k)~,
\label{PS-chi-sq}
\end{equation}
with
\begin{equation} 
{\cal P}_{{\cal R}^2_{\rm G}}(k)
= A_{\rm G}^2 \frac{k^2}{k_*^2} H(2k_*-k).
\end{equation}
Due to the mode-mode coupling induced by non-Gaussianity, the spectrum of ${\cal R}^2_{\rm G}$ acquires a width and actually peaks at a scale $k=2k_*$ shifted with respect to the peak of ${\cal P}_{{\cal R}_{\rm G}}(k)$. The amplitude of the full curvature spectrum (which corresponds to the curvature variance in real space) is given by
\begin{equation}
A \equiv \langle {\cal R}^2(\vec{x}) \rangle = \int_0^\infty \frac{dk}{k} \,
{\cal P}_{\cal R}(k) = A_G + 2 \tilde{f}_{\rm NL}^2 A_G^2~. 
\label{eq:A_to_AG}
\end{equation}
The $s$-distortion can then be computed using \cref{eq:mu_P_W} with $P_{{\cal R}}(k)$ substituted with \cref{PS-chi-sq}. The result can be put in the form
\begin{equation}
    s
    %_{\rm CSNG}
    =
    A_{\rm G} W_s (k_*)
    +
    2 \tilde{f}_{\rm NL}^2 A^2_{\rm G} \, W^{\rm (NG2)}_s (k_*)~,
    %= 2 A_{\rm G}^2 F_2(k_*)
    \label{eq:def_W_CSNG}
\end{equation}
where NG2 refers to ``Non-Gaussian window function for $\langle\Theta^2\rangle$ contribution'' and
%the subscript CSNG stands for `chi-square non-Gaussian' and 
\begin{equation}
   W^{\rm (NG2)}_s(k_*) \equiv 
   \frac{1}{2} \int_0^{2 k_*} \frac{dk\,k}{k_*^2} W_s(k_*)
   =
   - \frac{1}{2} \int_0^{2 k_*} \frac{dk\,k}{k_*^2}  \, \int dz \, {\cal I}_s(z) \, \frac{4 \dot{\tau}(z) \, S_{\rm ac}(k_*, z)}{(1+z) \, H(z) }~.
   \label{eq:W_NG2}
\end{equation}   
Given the different $k$-dependence of $W_s$ and $W^{\rm (NG2)}_s$, we cannot write a linear relation between $s$ and the total curvature power spectrum amplitude $A$.

\begin{figure}[H]
    \centering
    \includegraphics[scale=0.8]{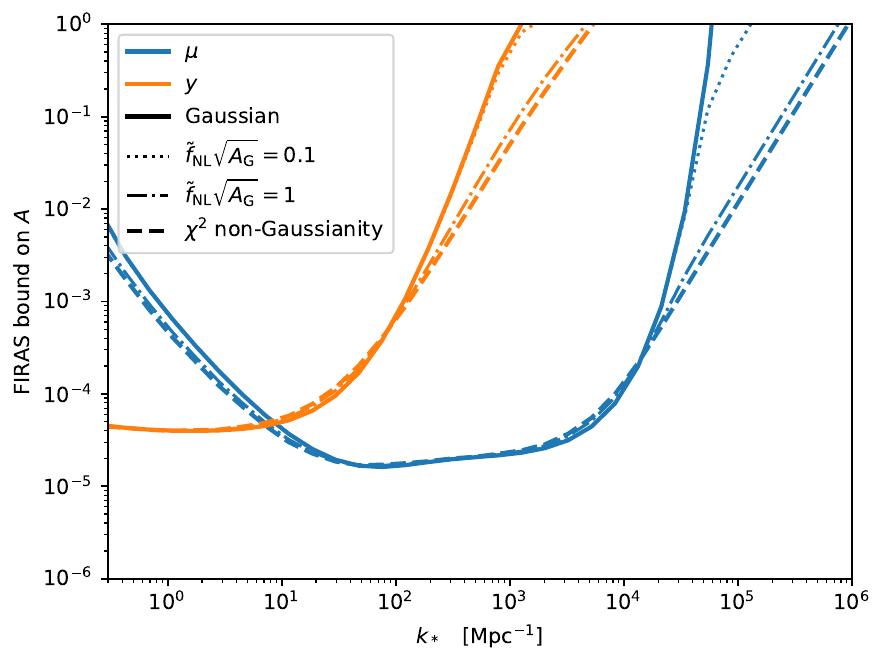}
    \caption{FIRAS bounds on the amplitude of the curvature power spectrum ${\cal P}_{\cal R}(k)$ (or, equivalently, on the variance of ${\cal R}$) in the Gaussian case (solid), in the case with $|\tilde{f}_{\rm NL}| \sqrt{A_{\rm G}} =0.1$ (small perturbative non-Gaussianity, dotted) or $|\tilde{f}_{\rm NL}| \sqrt{A_{\rm G}} =1$ (maximal perturbative non-Gaussianity, dashed-dotted), and finally in the non-perturbative non-Gaussian limit with $\chi^2$ statistics (dashed). Orange lines show the constraints from $y$-distortions and blue lines from $\mu$-distortions. Here, all window functions are given by an integral over the accurate source function (A4).
    }
    \label{fig:pert_mu_dist}
\end{figure}

In \cref{fig:pert_mu_dist}, we plot the FIRAS bound on the power spectrum amplitude $A$ obtained in two cases: when $|\tilde{f}_{\rm NL}| \sqrt{A_{\rm G}} = 1$ (maximum level of non-Gaussianity such that we remain in the perturbative non-Gaussianity regime), and when $|\tilde{f}_{\rm NL}| \sqrt{A_{\rm G}} = 0.1$ (arbitrarily chosen intermediate case). In these two cases, the $s$-distortions are given by:
\begin{equation}
s = \left\{
\begin{tabular}{ll}
$\frac{A}{1.02}  \left[ W_s(k_*) + 0.02 W^{\rm (NG2)}_s(k_*) \right]$
&
if $|\tilde{f}_{\rm NL}| \sqrt{A_{\rm G}} =0.1$~,
\\[3ex]
$\frac{A}{3}  \left[ W_s(k_*) + 2 W^{\rm (NG2)}_s(k_*) \right]$
&
if $|\tilde{f}_{\rm NL}| \sqrt{A_{\rm G}} = 1$~.
\end{tabular}
\right.
\end{equation}
The figure compares these constraints with the Gaussian case already displayed in \cref{fig:mu_ratio}. The difference between the two cases boils down to the shape of the curvature spectrum ${\cal P}_{\cal R}(k)$. This spectrum has a primary delta-function peak in both cases and an additional saw-tooth shaped peak in the non-Gaussian case, slightly offset with respect to the first peak. Thus, the constraints are unsurprisingly identical in the region where $W_{\mu,y}(k_*)$ are nearly independent of $k_*$. They become stronger at large $k$ due to the $k<k_*$ tail of the non-Gaussian spectrum ${\cal P}_{{\cal R}^2_{\rm G}}(k)$, and also a bit stronger at small $k$ due to the peak of ${\cal P}_{{\cal R}^2_{\rm G}}(k)$ being shifted from $k_*$ to $2k_*$. 

We only show in \cref{fig:mu_ratio} the constraints derived from the accurate (A4) approximation to the acoustic dissipation source function $S_{\rm ac}(k,z)$. Like in the Gaussian case, we checked that employing the slightly cruder approximation (A2), which does not capture the full asymptotic behaviour of the source function, makes no difference in practice in the range of scales relevant for FIRAS bounds. 

Finally, in \cref{tab:W_NG2}, we provide the coefficients of a fit to $W_\mu^{\rm (NG2)}(k)$ of the same form as that of \cref{eq:W_F3} (although of higher order). This fit is 10\% accurate in the range displayed in \cref{fig:pert_mu_dist}.

\begin{table}[hbt!]
\begin{center}
\begin{tabular}{cr|cr}
$n$&$z_n~~~~~~~$&$n$&$z_n~~~~~~~$\\
\hline
0 & $-2.31~~~~~~~$ & 5 & $-7.6855\times 10^{-3}$ \\
1 & $1.41~~~~~~~$ & 6 & $2.70326\times 10^{-3}$ \\
2 & $-1.25\times 10^{-1}$ & 7 & $-3.71457\times 10^{-4}$ \\
3 & $2.443\times 10^{-2}$ & 8 & $2.31035\times 10^{-5}$ \\
4 & $-1.405\times 10^{-3}$ & 9 & $-5.42652\times 10^{-7}$ \\
\end{tabular}   
\end{center}
\caption{Coefficients of a numerical fit to $W_\mu^{\rm (NG2)}(k)$, of the same form as that of \cref{eq:W_F3}.
\label{tab:W_NG2}}
\end{table}

% -5.42652e-07,2.31035e-05,-3.71457e-04,2.70326e-03,-7.6855e-03,-1.405e-03,2.443e-02,-1.25e-1,1.41,-2.31

For consistency, we checked that the values of $A$ saturating the FIRAS bounds in \cref{fig:pert_mu_dist} correspond to values of $|\tilde{f}_{\rm NL}|$ greater than one, as assumed throughout this section. In the next section, we discuss the case $|\tilde{f}_{\rm NL}|<1$.

\subsection{Next-to-leading contribution from three-point correlation function\label{sec:PNGC}}

\begin{figure}[hbt!]
  %\vspace{-4cm}
        \centering
        %\hspace{-4cm} 
        \includegraphics[width=14cm,angle=0]{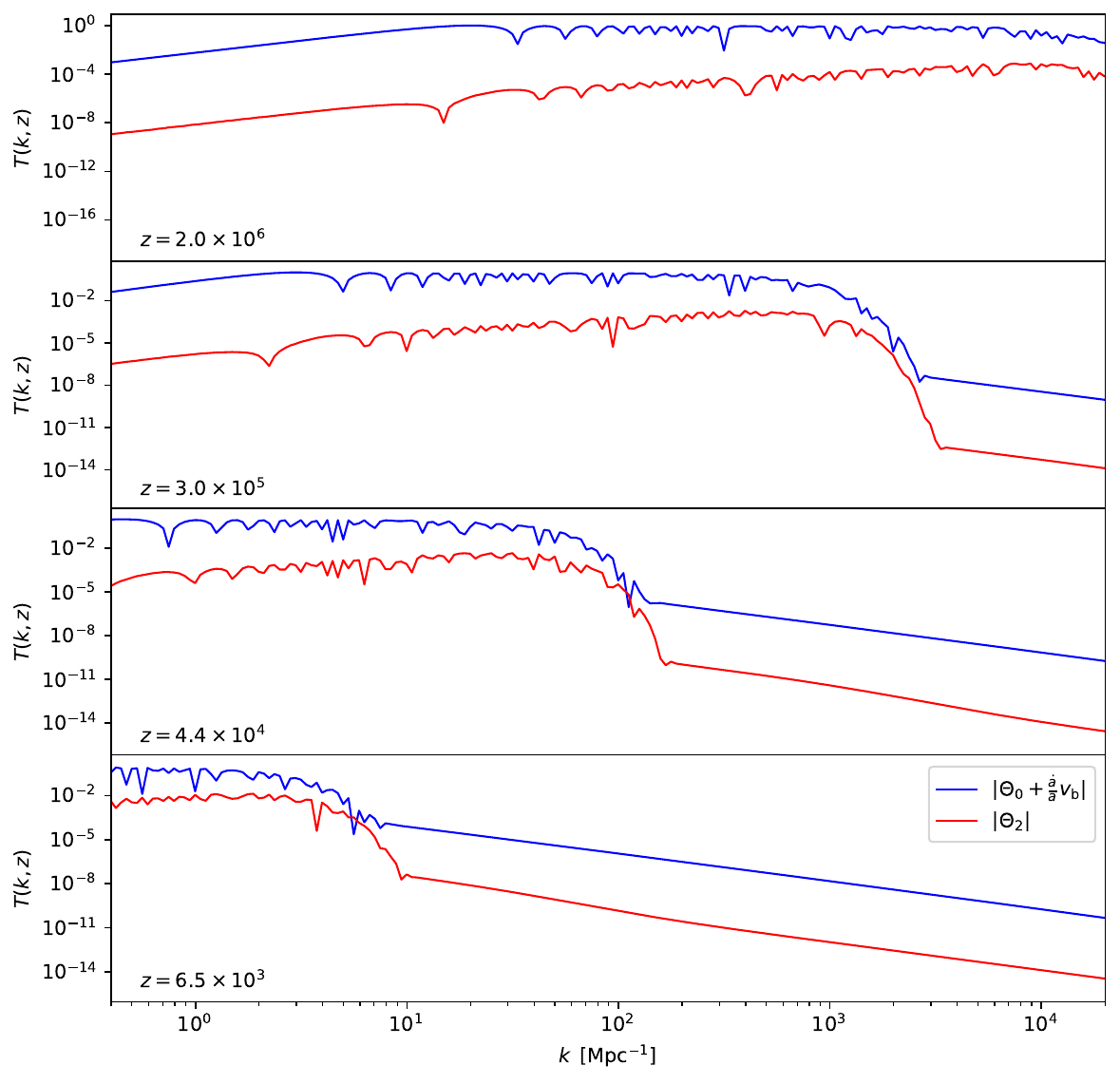}
        \caption{Absolute value of transfer functions (normalised to ${\cal R}(k)=1$) for the photon comoving monopole $\tilde{\Theta}_{0}\equiv(\Theta_{0}+\frac{\dot{a}}{a} v_{\rm b})$ and quadrupole $\Theta_{2}$ at four different redshifts, as a function of wavenumbers. At each redshift, in the regions where the functions are the largest, the monopole always dominates over the quadrupole by a few orders of magnitude. Note that these functions are both gauge invariant.}\label{fig:cubic_contributions}
\end{figure}

As explained in \cref{sec:NG_SD}, in the limit $|\tilde{f}_{\rm NL}|<1$, the leading non-Gaussian correction to spectral distortions are expected to arise from terms in the heating rate of \cref{eq:Q_dot_full} that are cubic in the temperature anisotropy $\Theta$, rather than from a correction to quadratic terms. The expression of the heating rate in \cref{eq:Q_dot_full} can be simplified by keeping only the dominant multipoles $\Theta_{(\ell)}$. In the range of times and scales relevant for the calculation of spectral distortions produced by the dissipation acoustic waves, the (gauge-invariant) comoving monopole $\tilde{\Theta}_{(0)}\equiv(\Theta_{(0)}+\frac{\dot{a}}{a} v_{\rm b})$ dominates over the quadrupole $\Theta_{(2)}$, see \cref{fig:cubic_contributions}.\footnote{This can also be seen analytically. In Fourier space, the WKB approximation for the quadrupole on sub-Hubble scale can be inferred from the one of the dipole, given by \cref{eq:theta_gamma}. That for the comoving monopole reads
$$
\tilde{\Theta}_0 = \frac{A_{\rm WKB}}{(1+R)^{1/4}}
\left[ 
\cos(k r_s) + \frac{aH}{k} c_s \sin(k r_s) 
\right] 
e^{-(k/k_{\rm D})^2}~.
$$ Comparing the two expressions, one gets $|\Theta_2| \sim \frac{8 k c_s}{45 \dot{\tau}} |\tilde{\Theta}_0|$. Thus $|\Theta_2| \ll  |\tilde{\Theta}_0|$ for relevant modes with $k \ll \dot{\tau}$.} We also know from previous sections that the quadrupole $\Theta_{(2)}$ is of the same order as the polarisation term $\Pi^1_{(2)}$ and dominates over the relative dipole $(\Theta_{(1)}-\hat{n}\cdot\vec{v}_{\rm b})$ and all higher multipoles $\Theta_{(\ell>2)}$. 
Thus, at next-to-leading order, the heating rate is given by
\begin{equation}
    \dot{Q} \simeq 4 \bar{\rho}_\gamma \dot{\tau} \left[ 
    \langle \Theta^{\rm B}_{(2)} \Theta^{\rm C1}_{(2)} \rangle
    +
    %{\color{purple}+}~{\color{teal}-}~ 
    2 \langle \tilde{\Theta}_{(0)} \Theta^{\rm B}_{(2)} \Theta^{\rm C1}_{(2)}
    \rangle
    \right]~.
    \label{eq:Q_cubic}
\end{equation}
In an expansion in Legendre multipoles, the term 
$\Theta^{\rm B}(\vec{x},\hat{n},\eta)$ gives a quadrupole moment $\Theta^{\rm B}_2 = -\Theta_2$, while the term $\Theta^{\rm C1}(\vec{x},\hat{n},\eta)$ gives in the tight-coupling approximation $\Theta^{\rm C1}_2 = \Theta_2-\frac{1}{10}(\Theta_2+\Theta_0^{\rm P}+\Theta_2^{\rm P}) \simeq \Theta_2-\frac{1}{10}(\Theta_2+\frac{6}{4} \Theta_2)=\frac{3}{4} \Theta_2$. Thus, the heating rate is approximately equal to
\begin{equation}
    \dot{Q} \simeq - 4 \bar{\rho}_\gamma \dot{\tau} \,\, \frac{3}{4} \left[ 
    \langle (\Theta_{(2)})^2 \rangle
    +2
    %{\color{purple}(+2)}~{\color{teal}(-2)}~{\color{DarkOrange}(+1)}
    \langle \tilde{\Theta}_{(0)} (\Theta_{(2)})^2
    \rangle
    \right]~.
    \label{eq:Q_cubic_two}
\end{equation}
The leading non-Gaussian contribution $\langle \tilde{\Theta}_{(0)} (\Theta_{(2)})^2 
    \rangle$ is sourced by the three-point correlation of primordial curvature,
\begin{equation} \label{eq:three-point-fn}
    \langle {\cal R}(\vec{k}_1) {\cal R}(\vec{k}_2) {\cal R}(\vec{k}_3) \rangle = \tilde{f}_{\rm NL} \langle {\cal R}_{\rm G}(\vec{k}_1) {\cal R}_{\rm G}(\vec{k}_2) {\cal R}_{\rm G}^2(\vec{k}_3)\rangle + {\rm permutations.}
\end{equation}
Applying Wick's theorem to the right-hand side of \cref{eq:three-point-fn} and using the definition of the power spectrum, one gets
\begin{eqnarray}
    \langle {\cal R}(\vec{k}_1) {\cal R}(\vec{k}_2) {\cal R}(\vec{k}_3) \rangle &=& \tilde{f}_{\rm NL} \left[P_{{\cal R}_{\rm G}}(k_1)P_{{\cal R}_{\rm G}}(k_2) + P_{{\cal R}_{\rm G}}(k_2)P_{{\cal R}_{\rm G}}(k_3) + P_{{\cal R}_{\rm G}}(k_3)P_{{\cal R}_{\rm G}}(k_1)\right] \nonumber \\
    && \delta^{(3)} (\vec{k}_1+\vec{k}_2+\vec{k}_3)~.
    \label{eq:three-point-fn-dirac}
\end{eqnarray}
We show in \cref{app:pert_NG_calc} that the relevant three-point correlation function can be expressed as
\begin{equation}
     \langle \tilde{\Theta}_{(0)} (\Theta_{(2)})^2 \rangle = \frac{15\sqrt{2} \tilde{f}_{\rm NL} }{\pi^{7/2}} \int^{\infty}_0 \!\!\!\!\!\! dr \, r^2 \left[ \int^{\infty}_0 \!\!\!\!\!\! dk_1 P_{{\cal R}_{\rm G}}(k_1) k^2_1 \Theta_{2}(k_1) j_{2}(rk_1)\right]^2  
     \int^{\infty}_0 \!\!\!\!\!\! dk_2  k^2_2 \tilde{\Theta}_{0}(k_2) j_{0}(rk_2)~, \label{eq:Integral-pert-NG}
\end{equation}
where $j_\ell(x)$ is the spherical Bessel function and the dependence of the multipoles $\Theta_{2}(k_1)$, $\tilde{\Theta}_{0}(k_2)$ on time (or redshift) is implicit. This term sources corrections to spectral distortion $s=y,\mu$ of the form
\begin{equation}
\Delta s = 
%{\color{purple}+}~{\color{teal}-}~ 
\int dz \, {\cal I}_s(z) \, \frac{4 \dot{\tau}(z)}{a_0 H(z)}\,\,\frac{3}{2} \langle \tilde{\Theta}_{(0)} (\Theta_{(2)})^2\rangle~,
\label{eq:source_correction}
\end{equation}
where $\langle \tilde{\Theta}_{(0)} (\Theta_{(2)})^2\rangle$ still depends on redshift $z$.
Assuming like in previous sections that the power spectrum of the Gaussian variable ${\cal R}_{\rm G}$ is a Dirac delta-function of amplitude $A_{\rm G}$, we get a simplified expression for the correlator,
\begin{equation}
     \langle \tilde{\Theta}_{(0)} (\Theta_{(2)})^2 \rangle = 60\sqrt{2\pi} \tilde{f}_{\rm NL} A_{\rm G}^2 \int^{\infty}_0 \!\!\!\!\!\! dr\, r^2 \left[ \Theta_{2}(k_*) j_{2}(rk_*)\right]^2  
     \int^{\infty}_0 \!\!\!\!\!\! dk \, k^2 \tilde{\Theta}_{0}(k) j_{0}(rk)~.
     \label{eq:theta3_Dirac}
\end{equation}
We can swap the integration order and integrate over $r$ analytically using \cref{eq:def_J,eq:res_J}. We obtain
\begin{equation}
     \langle \tilde{\Theta}_{(0)} (\Theta_{(2)})^2 \rangle = \frac{15\pi \sqrt{2\pi} \, 
\left[\Theta_{2}(k_*)\right]^2}{k_*^2}\,
\int_0^{\sqrt{2}k_*} \!\!\!\!\! dk \, k \,\tilde{\Theta}_{0}(k) \, P_2\left(1-\frac{k^2}{2 k_*^2} \right)~.
\label{eq:theta3_Dirac_analytic}
\end{equation}
We can then express the total distortion (including the main Gaussian contribution and the two next-to-leading-order corrections) as
\begin{equation}
s = A_{\rm G} W_s(k_*) + 2 \tilde{f}_{\rm NL}^2 A_{\rm G}^2 W^{\rm (NG2)}_s(k_*)
+ \tilde{f}_{\rm NL} A_{\rm G}^2 W^{\rm (NG3)}_s(k_*)~,
\label{eq:full_s}
\end{equation}
where NG3 refers to ``Non-Gaussian window function for $\langle\Theta^3\rangle$ contribution''. This new window function is defined as
\begin{equation}
W^{\rm (NG3)}_s(k_*) \equiv
%{\color{purple}+}~{\color{teal}-}~ 
\int dz \, {\cal I}_s(z) \,
\frac{4 \dot{\tau}(z)}{a_0 H(z)}\,
\frac{45}{2}\pi \sqrt{2\pi} \, 
\left[\Theta_{2}(k_*,z)\right]^2\,
\int_0^{\sqrt{2}k_*} \!\!\!\!\!\!\! d \ln k  \,\,\frac{k^2}{k_*^2} \tilde{\Theta}_{0}(k,z) \, P_2\left(1-\frac{k^2}{2 k_*^2} \right)~,
\end{equation}
where we restored the explicit dependence on $z$ in the arguments. Therefore, in order to compute $\Delta s = \tilde{f}_{\rm NL} A_{\rm G}^2 W^{\rm (NG3)}_s(k_*)$ with a Boltzmann code, one can proceed like in the standard Gaussian case, but with the source function $S_{\rm ac}(k,z)\simeq\frac{15}{4} [\Theta_{2}(k,z)]^2$ and primordial spectrum ${\cal P}_{\cal R}(k)=A_{\rm G} k_*\delta(k-k_*)$ replaced by
\begin{eqnarray}
    S_{\rm ac}(k,z)
    &\longrightarrow&
    %{\color{purple}(+1)}~{\color{teal}(-1)}~{\color{DarkOrange}(+\frac{1}{2})}\, 
     \frac{45}{2}\pi\sqrt{2\pi} \, 
\left[\Theta_{2}(k_*,z)\right]^2\, \tilde{\Theta}_{0}(k,z)~, \nonumber\\
{\cal P}_{\cal R}(k)
    &\longrightarrow&
    \tilde{f}_{\rm NL} A_{\rm G}^2
\, \frac{k^2}{k_*^2} \, P_2\left(1-\frac{k^2}{2 k_*^2} \right)
H(\sqrt{2}k_*-k)~,
\end{eqnarray}
where $H(x)$ is the Heaviside function.

\begin{figure}[H]
    \centering
    \includegraphics[scale=0.8]{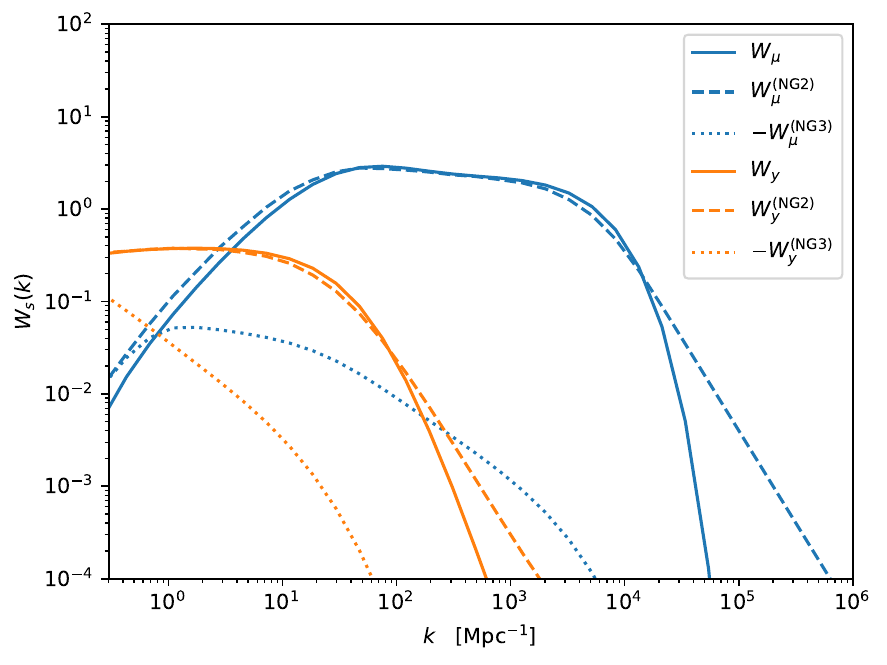}
    \caption{Comparison between the window function for $y$- and $\mu$-distorsions in the Gaussian case, $W_{y,\mu}(k)$, and in the case of a local non-Gaussian contribution sourced by a Dirac-delta spectrum ${\cal P}_{{\cal R}_{\rm G}}$ peaking at a scale $k$. In the latter case, $W^{\rm (NG2)}$ refers to the window function associated to the contribution of the two-point correlation function $\langle \Theta^2 \rangle$, and $W^{\rm (NG3)}$ to that from the three-point correlation function $\langle \Theta^3 \rangle$. We find that $W^{\rm (NG3)}$ is suppressed with respect to the other window functions by several orders of magnitude (excepted in the small-$k$ limit).
    %{\color{purple} [The legend should read $-W_s^{\rm (NG3)}$} {\color{teal} $W_s^{\rm (NG3)}]$} {\color{DarkOrange} $-2W_s^{\rm (NG3)}]$}
    }
    \label{fig:W123}
\end{figure}

The numerical calculation of $W^{\rm (NG3)}_s(k_*)$ with \CLASS{} is however more involved than that of the Gaussian window function $W_s(k)$ and of $W^{\rm (NG2)}_s(k_*)$. The latter functions only involve the square of oscillatory functions, such that one can substitute $\sin^2(x)$ by its average $1/2$. Instead, $W^{\rm (NG3)}_s(k_*)$ involves odd powers of oscillatory functions centred around zero, requiring a fine sampling of the integrands as a function of $k$. In the limit of a Dirac-delta spectrum ${\cal P}_{{\cal R}_{\rm G}}$, after performing the integrals over $k$ (and $r$) in \cref{eq:theta3_Dirac} or in the reduced form \cref{eq:theta3_Dirac_analytic}, one obtains a heating rate $\dot{Q}(z)$ which is again a quickly-oscillatory function of $z$ centred in zero, also requiring a fine sampling. We performed this calculation for a log-normal peak in ${\cal P}_{{\cal R}_{\rm G}}$ according to \cref{eq:Integral-pert-NG}, and for a Dirac-delta peak using either \cref{eq:theta3_Dirac} or \cref{eq:theta3_Dirac_analytic}. We checked the numerical convergence of our results, as well as the agreement between the two approaches for the Dirac-delta case (see \cref{app:PNG_Integrand_precision_check} for more details). Our result for $W^{\rm (NG3)}_s(k_*)$ in the Dirac-delta limit is reported and compared to $W_s(k_*)$ and of $W^{\rm (NG2)}_s(k_*)$ in \cref{fig:W123}. We note that $W^{\rm (NG3)}_y(k_*)$ and $W^{\rm (NG3)}_\mu(k_*)$ are both found to be 
negative
%{\color{purple} negative} {\color{teal} positive} {\color{DarkOrange} negative},
such that, with $\tilde{f}_{\rm NL}>0$, the cubic contribution would tend to slightly suppress
%{\color{purple} decrease} {\color{teal} enhance} {\color{DarkOrange} negative}
the $y$ and $\mu$ distortions, and vice versa. We will see however that the effect is negligible in practise.

%Using the approximate expression $S_{\rm ac}(k,z)\simeq \frac{15}{4} [\Theta_{2}(k,z)]^2$ of \cref{eq:Sac_quadrupole}, we further see that the cubic correction to spectral distortions is formally equivalent to an additional contribution to the curvature power spectrum of the form
%\begin{equation}
%\Delta {\cal P}_{\cal R}^{{\rm effective},k_*}(k,z)
%\equiv {\color{teal} (-)}  \, 6\pi\sqrt{2\pi} \, 
%\tilde{f}_{\rm NL} A_{\rm G}^2 \left[\frac{\Theta_{2}(k_*,z)}{\Theta_{2}(k,z)}\right]^2\, \tilde{\Theta}_{0}(k,z)
%\, \frac{k^2}{k_*^2} \, P_2\left(1-\frac{k^2}{2 k_*^2} \right)
%H(\sqrt{2}k_*-k)~,
%\end{equation}
%such that 
%\begin{equation}
%    W^{\rm (NG3)}_s(k_*) = \frac{1} {\tilde{f}_{\rm NL} A_{\rm G}^2}\int \frac{dk}{k} \,\,\, {\cal P}_{\cal R}^{{\rm eff},k_*}(k) \,\,\, W_s(k)~.
%\end{equation}

%

Our preliminary discussion in section \cref{sec:NG_SD}, which suggested that the contribution of the three-point correlation function $\langle \Theta^3 \rangle$ to the heating rate $\dot{Q}$ could dominate over that of the two-point correlation function $\langle \Theta^2 \rangle$ when $|\tilde{f}_{\rm NL}|<1$, was based on the simplifying assumption that all transfer functions are of order one. In reality, the order of magnitude of the transfer functions $\Theta_\ell(k,\eta)$ depends on the multipole $\ell$. It is true that, for the monopole, $\tilde{\Theta}_0(k,\eta)$ peaks at values of order one, as confirmed by \cref{fig:cubic_contributions}. Instead, for the quadrupole, $\tilde{\Theta}_2(k,\eta)$ peaks at much smaller values, as shown by \cref{fig:S_sd_contributions}. However, when comparing the the NG2 and NG3 contributions to the heating rate $\dot{Q}$, we find that the former is sourced by  $2\tilde{f}_{\rm NL}^2 A_{\rm G}^2[\Theta_2]^2$ and the later by $\tilde{f}_{\rm NL} A_{\rm G}^2 \tilde{\Theta}_0 [\Theta_2]^2$. Since $\tilde{\Theta}_0$ is typically of order one, we may really expect that the condition for NG3 to dominate is $|\tilde{f}_{\rm NL}|<1$. This argument is oversimplified. It neglects the fact that the transfer functions need to be integrated over $k$, like in \cref{eq:Integral-pert-NG,eq:theta3_Dirac,eq:theta3_Dirac_analytic}. The fact that $\tilde{\Theta}_0$ is a quickly oscillatory function of $k \eta$, with oscillations centred around zero, reduces the NG3 contribution by several orders of magnitude. Our explicit calculation of $W^{\rm (NG3)}(k)$ in \cref{fig:W123} shows this suppression explicitly, since $W^{\rm (NG3)}(k)\ll W^{\rm (NG2)}(k)$ (excepted for the smallest values $k$, which are anyway irrelevant for FIRAS bounds). Thus, the NG3 term would actually dominate over the NG2 term only for values $|\tilde{f}_{\rm NL}|\ll 1$ such that non-Gaussian corrections are negligible as a whole. This is immediately visible from \cref{eq:full_s}, which can be cast as
\begin{equation}
s = A_{\rm G} \left[W_s(k_*) + 2 \tilde{f}_{\rm NL}^2 A_{\rm G} W^{\rm (NG2)}_s(k_*)
+ \tilde{f}_{\rm NL} A_{\rm G} W^{\rm (NG3)}_s(k_*) \right]~.
\end{equation}
To avoid an over-production of PBHs, we must have $A\ll1$ and thus, according to \cref{eq:A_to_AG},  $|\tilde{f}_{\rm NL} A_{\rm G}|\ll 1$. Since additionally $|W^{\rm (NG3)}_s(k_*)|\ll W_s(k_*)$, we see that for any possible value of  $\tilde{f}_{\rm NL}$ and $A_{\rm G}$, the NG3 contribution is always suppressed by several orders of magnitude compared to the Gaussian one.

Our calculations in this section prove that in the case of local non-Gaussianity, corrections to spectral distortions arising from the three-point correlation function $\langle \Theta^3 \rangle$ are always negligible. The leading non-Gaussian corrections are always captured by the two-point correlation function described in \cref{sec:NG_SD_Q} in the limit $|\tilde{f}_{\rm NL}| A_{\rm G}\leq 1$ and \cref{sec:NPNG} in the limit $|\tilde{f}_{\rm NL}| A_{\rm G}\gg 1$.

\section{Distortions in the limit of local non-perturbative non-Gaussianity \label{sec:NPNG}}

\subsection{Chi-squared non-Gaussianity \label{sec:NPNG_CS}}

In this section we focus on a pure chi-squared non-Gaussianity, meaning that 
\begin{equation}
    {\cal R}(\vec x)={\cal R}_{\rm G}^2(\vec x)-\langle{\cal R}_{\rm G}^2\rangle.
\end{equation}
We have seen in \cref{sec:NG_SD} that in the $\chi^2$ limit, spectral distortions $s=\mu,y$ are dominated once more by the contribution of the two-point correlation function $\langle\Theta^2\rangle$ to the heating rate $\dot{Q}$. The primordial curvature spectrum $P_{\cal R}(k)=P_{{\cal R}_{\rm G}^2}(k)$ is given by \cref{PS-chi-sq}. We have already seen that, due to the mode-mode coupling induced by non-Gaussianity, the spectrum of ${\cal R}$ acquires a width and actually peaks at a scale $k=2k_*$ shifted with respect to the peak of ${\cal P}_{{\cal R}_{\cal G}}(k)$. Its amplitude (which corresponds to the curvature variance in real space) is given by
\begin{equation}
A \equiv \langle {\cal R}^2(\vec{x}) \rangle = \int_0^\infty \frac{dk}{k} \,
{\cal P}_{\cal R}(k) = 2 A_G^2~. 
\end{equation}
The $s$-distortion can then be computed using \cref{eq:mu_P_W} with $P_{{\cal R}}(k)$ substituted with \cref{eq:P_Heaviside}. The result can be put in the form
\begin{equation}
    s
    %_{\rm CSNG}
    =A \, W^{\rm (NG2)}_s (k_*)~,
    %= 2 A_{\rm G}^2 F_2(k_*)
\end{equation}
with $W^{\rm (NG2)}_s(k_*)$ given by \cref{eq:W_NG2}.

\begin{figure}[H]
    \centering
    \includegraphics[scale=0.8]{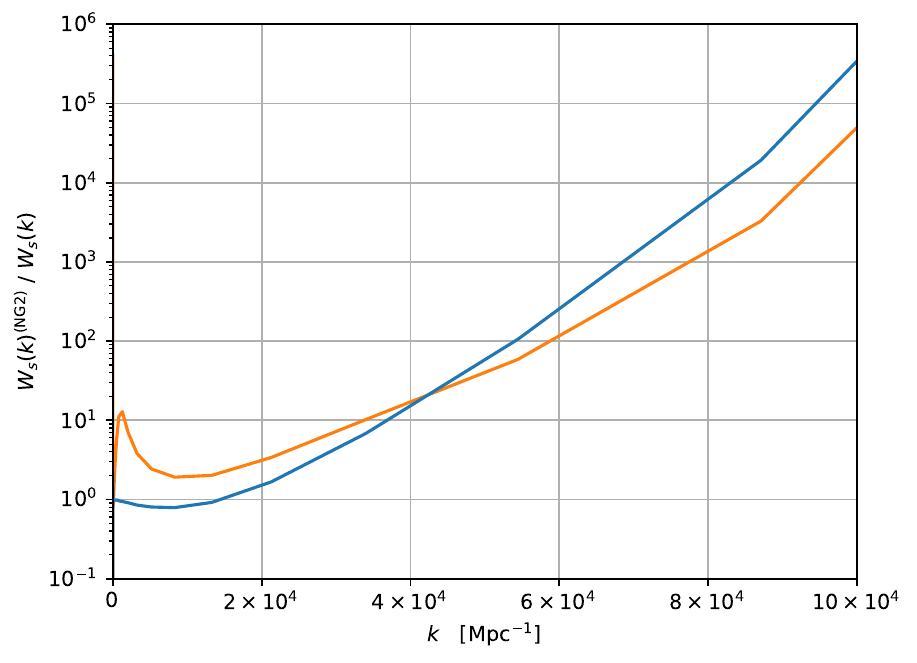}
    \caption{Ratio of window functions $W^{\rm (NG2)}_s(k_*)/W_s(k_*)$, for $y$-distortion (orange) and $\mu$-distortions (blue). These curves directly show how much stronger the FIRAS bounds on the variance $A$ of curvature perturbations are in the limit of non-Gaussianity with chi-squared statistics.
    }
    \label{fig:non_pert_mu_dist}
\end{figure}
The constraints on $A$ in this case,
$A<\mu_{\rm FIRAS}/W^{\rm (NG2)}_\mu(k_*)$ and
$A<y_{\rm FIRAS}/W^{\rm (NG2)}_y(k_*)$, are shown in \cref{fig:pert_mu_dist}. Formally, these constraints are equivalent to those obtained in the previous section in the limit in which the non-Gaussian contribution dominates over the Gaussian one, that is, when the primordial spectrum has a single peak with a see-tooth shape and a maximum at $k=2k_*$. Once more, the constraints are the same as in the Gaussian case in the region where $W_{\mu,y}(k_*)$ are nearly independent of $k_*$, and become much stronger at large $k$ due to the $k<k_*$ tail of the non-Gaussian ${\cal P}_{\cal R}(k)$, or a bit stronger at small $k$ due to the peak of ${\cal P}_{\cal R}(k)$ being shifted from $k_*$ to $2k_*$. 

To show by how much the FIRAS bounds on $A$ differ in the Gaussian and chi-squared cases, we show in \cref{fig:non_pert_mu_dist} the ratio $W^{\rm (NG2)}_s(k_*)/W_s(k_*)$. We see for instance that for $k_*=4 \times 10^4\,{\rm Mpc}^{-1}$ (resp. $k_*=6 \times 10^4\,{\rm Mpc}^{-1}$), the bounds from $\mu$ distortions are 20 (resp. 300) times stronger in the non-Gaussian case.

\subsection{Higher-order non-perturbative non-Gaussianity \label{sec:NPNG_HO}}

In \cref{eq:Q_dot_full}, we stopped the expansion of $\dot{Q}$ at order three in the photon temperature anisotropy field $\Theta(\vec{x},\hat{n},\eta)$. Higher-order corrections would be of the type $\langle \Theta^n \rangle_{\hat{n},\vec{x}}$, or featuring multiple angular averages within the spatial average: for instance, at order four, one would get contributions from $\langle \Theta^4 \rangle_{\hat{n},\vec{x}}$ but also from $\left \langle \langle \Theta^2 \rangle_{\hat{n}} \langle \Theta^2 \rangle_{\hat{n}} \right\rangle_{\vec{x}}$.

Writing only powers of $\Theta$ is a simplification. When deriving \cref{eq:Q_dot_full}, we have seen that the final result can involve the full anisotropy field $\Theta$, but also other combinations of temperature and polarisation multipoles or of the baryon dipole, like $\Theta^{\rm B}$, $\Theta^{\rm C1}$, $\Theta^{\rm C2}$, or the comoving monopole $\tilde{\Theta}_{(0)}$.

We can actually make a more precise statement on the structure of each term of order $n$. In the sum of all contributions of order $n$ to $\dot{Q}$, it must be possible to factorize at least one power of $\Theta^{\rm B}$ and exactly one power of $(\Theta^{\rm C1}+\Theta^{\rm C2})$. The first statement comes from the structure of $\rho_{\rm ex}$ in \cref{eq:rho_ex_1}. In the baryon rest frame, $\rho_{\rm ex}$ must exactly vanish when the anisotropy field reduces to a simple monopole, $\Theta=\Theta_{(0)}$. Physically, this follows from the fact that in the absence of anisotropies there is no mixing of blackbodies with different temperatures. Mathematically, this is guaranteed by the fact that in the Taylor expansion of $\dot{Q}$, one can factorise at least one power of $(\Theta-\Theta_{(0)})$ in the contribution from each order. It is easy to check that this is the case at order $n=2,3$ in \cref{eq:rho_ex_2}, and we also checked it explicitly at the next orders. After going from the baryon frame to an arbitrary frame, $(\Theta-\Theta_{(0)})$ becomes $\Theta^{\rm B}$. The second statement comes from the fact that each contribution to $\dot{Q}$ features one power of $\dot\Theta$, which gives a factor $(\Theta^{\rm C1}+\Theta^{\rm C2})$ after using the Boltzmann equation. 

Moreover, we know that $\Theta^{\rm C2}$ remains small compared to $\Theta^{\rm C1}$, such the leading contribution of order $n$ is given by products like $\Theta^{n-2} \Theta^{\rm B} \Theta^{\rm C1}$. The first two fields are yet unspecified, but we can further use the fact that in an expansion in multipoles, $\Theta$ is dominated by the monopole while $\Theta^{\rm B}$ and $\Theta^{\rm C1}$ are dominated by the quadrupole. Therefore, at order $n$, the largest term always reads
\begin{equation}
\left\langle \tilde{\Theta}_{(0)}^{n-2} \Theta_{(2)}^{\rm B} \Theta_{(2)}^{\rm C1} \right\rangle_{\hat{n},\vec{x}}~.
\end{equation}
After expanding in Fourier space and Legendre multipoles, this correlation function can be expressed as a multiple integral over $n$ wavevectors involving at least 
\begin{equation}
\tilde{\Theta}_0(k_1) ... \tilde{\Theta}_0(k_{n-2}) \, \Theta_{2}(k_{n-1}) \, \Theta_{2}(k_{n})
\,\, \langle {\cal R} (\vec{k_1}) ... {\cal R} (\vec{k_n}) \rangle~. 
\end{equation}
In the previous sections, we have studied in detail the cases $n=2$ and $n=3$. We found that for $n=3$ the three-point correlation function is strongly suppressed with respect to the two-point correlation function for a reason that is independent on assumptions concerning the statistics of the primordial curvature field. The reason is that $\tilde{\Theta}_{(0)}$ is a quickly oscillatory function centred in zero, whose integral is thus very small. The conclusion would be the same with any odd $n$. Instead, for even $n$'s, the integral over ${\Theta}_0^{n-2}$ is not suppressed - like in the case $n=2$ where $\sin^2(k r_s)$ can be replaced by one half.

We can compare schematically the contribution of each even term of order $n=2m$ by writing it as
\begin{equation}
\Theta_{2}^2 
\,\, \langle {\cal R}^{2} \rangle~,
\qquad  
\tilde{\Theta}_0^{2} \, \Theta_{2}^2 
\,\, \langle {\cal R}^{4} \rangle~,
\qquad  
\tilde{\Theta}_0^{4} \, \Theta_{2}^2 
\,\, \langle {\cal R}^{6} \rangle~,
\qquad
\tilde{\Theta}_0^{2m-2} \, \Theta_{2}^2 
\,\, \langle {\cal R}^{2m} \rangle~.
\end{equation}
Since $\tilde{\Theta}_0^2$ is of order one, it does not contribute to the suppression of higher-order terms. $\Theta_{2}^2$ is a few orders of magnitude smaller, but this is irrelevant in the comparison between the different terms. The variance $\langle {\cal R}^2 \rangle$ must remain substantially smaller than one to avoid an overproduction of black holes. A term of order $n=2m$ could still be significant with respect to the usual term of order $n=2$ if $\langle {\cal R}^{2m} \rangle$ was at least as large as $\langle {\cal R}^{2} \rangle$. This condition is hardly compatible with the requirement that $\langle {\cal R}^{2} \rangle \ll 1 $. Usually, even strongly non-Gaussian pdfs have $\langle {\cal R}^{2m} \rangle \sim \langle {\cal R}^{2} \rangle^m$, such that higher powers are suppressed. One would need to invoke some very specific probability distributions to enhance high-order contributions compared to the usual $n=2$ contribution.

%\julien{Devil's advocate: if ${\cal R}={\cal R}_{\rm G}^p$, $\langle {\cal R}^{2} \rangle=(2p-1)!! \langle {\cal R}_{\rm G}^{2} \rangle^p$, $\langle {\cal R}^{2m} \rangle=(2mp-1)!! \langle {\cal R}_{\rm G}^{2} \rangle^{mp}$, can't the second term win even if $\langle {\cal R}_{\rm G}^{2} \rangle^p\ll1$, just because $[(2mp-1)!!/(2p-1)!!]$ can become so large? For instance, take $p=2$ (chi-squared non-Gaussianity). $\langle {\cal R}^{2} \rangle=3 \langle {\cal R}_{\rm G}^{2} \rangle^2$ (in fact should be a 2), $\langle {\cal R}^{2m} \rangle=(4m-1)!! \langle {\cal R}_{\rm G}^{2} \rangle^{2m}$. If $\langle {\cal R}_{\rm G}^{2} \rangle^2=\alpha$ with $\alpha\ll1$, $(4m-1)!!\alpha^m$ can still become very large...}

\subsection{Comparison with previous results}

Since we present the first accurate attempt to compute spectral distortions in the presence of primordial non-Gaussianity, we do not have previous results to compare with. To the best of our knowledge, the only exception is the appendix of reference \cite{Juan:2022mir}, which presents a simplified discussion of these potential effects. The authors of \cite{Juan:2022mir} acknowledge that their calculation is not meant to be rigorous. 

Reference \cite{Juan:2022mir} introduces a field $\theta$ accounting for the angle-averaged photon temperature fluctuation, $\theta=\delta \rho_\gamma / \bar{\rho}_\gamma$, and not for its anisotropies. In our notation, this would be the monopole $\Theta_{(0)}$. However, they express the $\mu$ distortions as an expansion over the same $\theta$, while the actual distortions depend mainly on the anisotropic part of the temperature distribution, $\Theta(\vec{x},\hat{n},\eta)-\Theta_{(0)}(\vec{x},\eta)$, and in particular on its quadrupole moment. Then, they expand $\mu$-distortions in $n$-point correlation functions $\langle \theta^n \rangle$. Next, they argue that in models leading to PBH formation, $\theta$ can be quite large (nearly of order one), such that several $n$-momenta could contribute to the total $\mu$. Finally, they assume a parametric from for the non-Gaussian statistics of $\theta$, and present approximate results for $\mu$-distortions. Their bounds differ from standard prediction mainly due to the role of high-order momenta $\langle \theta^n \rangle$.

In reality, we have seen that $\mu$ distortions are always sourced by at least two powers of the quadrupole $\Theta_{(2)}$, which remains very small compared to $\delta \rho_\gamma / \bar{\rho}_\gamma$ in the range of time and scales where $\mu$-distortions are generated. We distinguished between the role of the monopole and of the quadrupole in next-to-leading-order momenta. We were led to the unexpected conclusion that high-order momenta tend to remain negligible, even in the limit of pure chi-squared non-Gaussianity. We expect this conclusion to hold unless one invokes very special probability distribution functions for the primordial curvature field, with high-order momenta strongly enhanced compared to the variance. With this in mind, we concentrated on the impact of non-Gaussianity on the distortions seeded, like in the Gaussian limit, by the two-point correlation function of the photon quadrupole. At this level, there is still a significant effect, because local non-Gaussianity induces mode-mode coupling and broadens the shape of the peak in the primordial spectrum, such that $\mu$-distortion constraints apply to a wider range of peak scales. 

\section{Conclusions\label{sec:conclusions}}

Spectral Distortions of CMB photons have the potential of uncovering some of the mysteries of the early universe. At very early times, $\mu$-type distortions are sourced by the dissipation of acoustic waves over a wide range of scales, including much smaller wavelengths than those constrained by CMB temperature and polarisation anisotropies. This mechanism could provide some important information about initial conditions from inflation. In the range of scales constrained by CMB anisotropies, primordial fluctuations are known to be very small, $\langle {\cal R}^2 \rangle \sim {\cal O}(10^{-9})$, and highly Gaussian. As already shown by several works \cite{Byrnes:2018txb, Mishra:2019pzq, Zaballa:2006kh}, at smaller scales, the primordial spectrum may feature some peaks that could be well over seven orders of magnitude higher, causing the formation of PBHs and questioning the naturalness of Gaussianity on such scales.

In this paper, we comprehensively analyse various schemes in the current literature for computing the $\mu$-distortions, and extend them to the case of primordial perturbations featuring non-Gaussian statistics. For the first time, we demonstrate the impact of non-Gaussian seeds from inflation on $\mu$-distortions via a robust formalism.

Sticking first to the assumption of Gaussian perturbations, we confirm previous state-of-the-art calculations of  $\mu$-distortions, based on a semi-analytic approximation for the acoustic dissipation source function that is called (A2) in this work, see \cref{eq:Sac_A2}. Using the \CLASS{} code, we compute a more precise version of this source function, see \cref{eq:s4}, but we then prove that using (A2) is sufficient to derive accurate $y$- and $\mu$-distortion bounds from FIRAS. Instead, the most commonly used analytical fits to the source function, called here (F1) and (F2), see \cref{eq:Wmu-long,eq:Wmu-short}, are very crude. We provide a more accurate analytical fit called (F3), see \cref{eq:W_F3} and \cref{tab:W_NG2}, which can be safely used to derive FIRAS bounds in a very straightforward way for any given power spectrum, while achieving 10\% accuracy. 

%{\it We then discuss the inconsistencies of these schemes at some scales and introduce a more accurate scheme, taking into account the non-negligence of the baryon-photon fraction at the approach of recombination that is ignored by the rest approximations (see \cref{fig:S_sd_comparision}). We find that this causes a tiny suppression in the source function at $z< {\rm few} \times 10^3$ by a few per cent, but remains unimportant at higher redshifts. \textcolor{red}{Hence, not significant for mu distortions but may be important for y distortions?}}

%Moreover, the existing works begin with Gaussian initial conditions that source the heating rate of the distortions. This assumption might be justified at the PLANCK scales, however, as already shown by several works \cite{Byrnes:2018txb, Mishra:2019pzq, Zaballa:2006kh}, at much smaller scales, the primordial power spectrum may be well over seven orders of magnitude higher, causing the formation of PBHs and questioning the plausibility of Gaussianity at all scales. In this light, for the first time, we demonstrate the impact of non-Gaussian seeds from inflation on the $\mu$-distortions via a robust formalism. 

To analyse the case of non-Gaussian primordial fluctuations, we push for the first time the calculation of the heating rate that sources $y$- and $\mu$-distortions to next-to-leading order, that is, to the level of the three-point correlation function of temperature anisotropies, $\langle \Theta^3 \rangle$, using the second-order Boltzmann equation. At this order, the heating rate depends locally on the value of the temperature monopole $\Theta_0$. This may come as a surprise since we focus here on distortions generated by temperature anisotropies. Our result is nevertheless consistent because $\Theta_0$ modulates the absolute amplitude of photon anisotropies that an observer would measure at a given point at order two in perturbations, see \cref{eq:role_monopole}. We also check that our next-to-leading-order heating rate is gauge invariant.

We then express the heating rate and compute the resulting spectral distortions in the particular case of local non-Gaussianity in the statistics of primordial curvature fluctuations ${\cal R}$, parameterised by $\tilde{f}_{\rm NL}$. For the sake of simplicity and concision, we further assume that the Gaussian field ${\cal R}_{\rm G}$ that generates such fluctuations features itself a Dirac-delta function peaked at some arbitrary scale $k_*$. Note that with our formalism, it is trivial to obtain more general results in the case of broader peaks. Due to mode-mode-coupling, a sharp peak in the spectrum of ${\cal R}_{\rm G}$ leads to a broader one in that of ${\cal R}$, with a maximum amplitude at a slightly different scale. It also leads to a non-zero bispectrum for ${\cal R}$.

We then propagate this primordial non-Gaussianity to the level of the heating rate and spectral distortions. According to a rough estimate, depending on the precise value of $|\tilde{f}_{\rm NL}|$, the correction to the distortions caused by non-Gaussianity could be dominated either by the modified contribution of the two-point correlation function $\langle \Theta^2 \rangle$, or by that of the new contribution of the three-point correlation function $\langle \Theta^3 \rangle$. However, a detailed calculation of the latter contribution in \cref{sec:PNGC} shows that it is more suppressed than expected, due to integrations over highly oscillatory functions centred around zero. As a result, we show that for any value of $\tilde{f}_{\rm NL}$ such that non-Gaussian corrections are not totally negligible, the contribution from the two-point correlation function always dominates. Like in the Gaussian case, we provide an analytical fit to the window function that allows to easily compute non-Gaussian corrections to $y$- and $\mu$-distortions for any delta-peaked spectrum in the underlying Gaussian field ${\cal R}_{\rm G}$ and any parameter $\tilde{f}_{\rm NL}$, see \cref{eq:W_F3} and \cref{tab:W_NG2}.

We then compute the bounds on the primordial curvature spectrum amplitude $A=\langle {\cal R}^2 \rangle$ that can be inferred from FIRAS for an arbitrary value of $\tilde{f}_{\rm NL}$. The difference with respect to the  Gaussian case is caused by the broadening of the primordial spectrum peak due to mode-mode coupling. This effect does not change the bounds on $A$ in the range of scale where $y$- or $\mu$-distortions are maximal. However, it becomes relevant at the edges of this range, for which the bounds become stronger. Intuitively, when primordial fluctuations are generated on a scale that would in principle slightly evade the bounds from spectral distortions in the Gaussian limit, some power can propagate (through mode-mode coupling) to neighbouring scales sourcing higher distortions, and the bounds straighten. This is particularly true for $\mu$-distortions in the range $10^4 {\rm Mpc}^{-1} \leq k_* \leq 10^6 {\rm Mpc}^{-1}$.

Our final results can be summarised as follows. In the regime of perturbative non-Gaussianity, $\tilde{f}_{\rm NL} \sqrt{A} \leq 1$, corrections to the standard $\mu$-distortion bounds start to be relevant when $\tilde{f}_{\rm NL} \sqrt{A}$ approaches the order of 0.1, see \cref{fig:pert_mu_dist}. They grow very quickly for larger values. However, when $\tilde{f}_{\rm NL} \sqrt{A} \gg 1$, the effect saturates: this corresponds to the limit of non-perturbative non-Gaussianity with a pure $\chi^2$-statistics for ${\cal R}$. We also derive the FIRAS bound on $A$ as a function of $k_*$ in this limit, as shown in the same \cref{fig:pert_mu_dist}. For instance, in the $\chi^2$ limit, the bound on $A$ is stronger than in the Gaussian case by one order of magnitude at $k_*\sim 3.6 \times 10^{-4}{\rm Mpc}^{-1}$, two orders of magnitude at $k_*\sim 5.4 \times 10^{-4}{\rm Mpc}^{-1}$, or three orders of magnitude at $k_*\sim 6.8 \times 10^{-4}{\rm Mpc}^{-1}$. These numbers can be reproduced by simply taking the ratio of our fitting function in \cref{tab:W_F3,tab:W_NG2}, or directly by looking at \cref{fig:non_pert_mu_dist}.

%We show that, in the non-perturbative regime, for a local type of $\chi^2$ characterized non-Gaussianity, the $\mu$-distortions become tighter by roughly a factor of two (see Fig. \cref{fig:non_pert_mu_dist}). With the help of an intensive calculation in \cref{pert_NG_calc}, we find that $\mu$ does not change considerably in the next to leading order contribution in the perturbative limit, though the magnitude of smallness is scale-dependent. This feature becomes clear from \cref{fig:NGtoGaussian_ratio}, where we show the fractional magnitude change in the distortion amplitude for non-Gaussian cases with the scale for power spectra of different widths. Albeit, we see that in the maximum possible perturbative limit, the total $\mu$-distortion is stronger by roughly a factor of three to four as shown in \cref{fig:max_pert}. 

In a companion paper \cite{paper2}, we compute the density of PBHs generated by a peak in the primordial spectrum obeying local non-Gaussian statistics. We then use the results of the present paper to derive FIRAS bounds on the density of PBHs generated by such a mechanism.

We recall that, while our derivation of the heating rate at order three in photon anisotropies is general, our subsequent calculation of the effect of non-Gaussianity on spectral distortions is model-dependent. Specifically, our results only cover the case of local non-Gaussianity. Even in this case, they do not include the special limit in which the $\tilde{f}_{\rm NL}$ contribution is suppressed compared to the next-order $\tilde{g}_{\rm NL}$ contribution in the expansion of \cref{eq:Rg}. Nonetheless, our analysis provides a robust basis to start discussing in a quantitative way how $\mu$-distortion constraints on the abundance of PBHs are affected by non-Gaussianity.

%By neglecting the next-order contribution to the Taylor expansion of the curvature parameter \cref{eq:Rg}, we intrinsically assume that the trispectrum quantifier $g_{\rm NL} $ cannot be too larger than $\tilde{f}_{\rm NL} $. This is true for a majority of the inflationary models, however in some models, for example, a curvaton sourced non-Gaussianity \cite{Enqvist:2005pg, Enqvist:2008gk, Huang:2008zj}, one can generate an amplified magnitude of $g_{\rm NL} $, in which case, this method would under-calculate the $\mu$-distortion amplitude. On top of that, if one chooses to characterize local non-Gaussianity other than the usual $\chi^2$, it would in principle complicate our calculations, hence, probably changing the magnitude by an undetermined estimate. These interesting issues will be addressed in the forthcoming works.

%\printbibliography

\subsection*{Acknowledgements}

We thank Jens Chluba, Christian Fidler and David Seery for illuminating comments.  We thank the Royal Society for funding an International Exchanges Scheme which allowed the three authors to travel between Brighton and Aachen. DS is supported by the German Academic Exchange Service (DAAD) grant. CB is supported by STFC grants ST/X001040/1 and ST/X000796/1.

\appendix

\section{Averaging the local contribution to spectral distortion\label{app:averaging}}

To go from the local contribution to spectral distortions to the average contribution, one performs the average of \cref{eq:x_dot_energy}:
\begin{equation}
\dot s \propto \frac{d}{d\eta} \left\langle \frac{ \rho_{\rm ex} |_{\rm b} }{\rho_\gamma |_{\rm b}}\right\rangle_{\rho(\vec{x})}~.
\end{equation}
In principle, one could use either a simple average $\frac{1}{V} \int_V d^3x (...)~,$ an average weighted by the number density $\frac{1}{\bar{n}_\gamma} \int_V d^3x \, n_\gamma(\vec{x})(...)$, or an average weighted by the energy density $\frac{1}{\bar{\rho}_\gamma} \int_V d^3x \, \rho_\gamma(\vec{x})(...)$. Since a distortion created in a given point contributes more to the average in locations with more photons, the average must be weighted by the photon density. Weighting by the energy density is the most sensible prescription, given that this is equivalent to averaging only over the {\it extensive} quantity $\rho_{\rm ex}(\vec{x})$, see \cref{eq:x_dot_energy2}. However, we also considered the possibility of averaging over number density. Here, we report the consequences of this other prescription in case it would be useful for future work.

These averaging processes lead to the same expression for the heating rate $\dot Q$ at leading order, but to different factors at next-to-leading order. With an average weighted by the local number density, one needs to integrate
\begin{align}
    \frac{n_\gamma(\eta,\vec{x})|_{\rm b}}{\bar{n}_\gamma(\eta)}
    \frac{\rho_{\rm ex}(\eta, \vec{x})|_{\rm b}}{\rho_\gamma (\eta, \vec{x})|_{\rm b}}
    &
    = 
    \left[ \langle (1+\Theta)^3 \rangle_{\hat{n}} \right]_{\rm b}
    \frac{\left[ \langle (1+\Theta)^4 \rangle_{\hat{n}} -  \langle (1+\Theta)^3 \rangle_{\hat{n}}^{4/3}\right]_{\rm b}}
    {\left[ \langle (1+\Theta)^4 \rangle_{\hat{n}} \right]_{\rm b}}
    \\
    &
    = \left[ \left\langle 2 \Theta^2 
    -2 \Theta_{(0)}^2
    +\frac{8}{3} \Theta^3
    -6 \Theta_{(0)} \Theta^2
    + \frac{10}{3} \Theta_{(0)}^3
    \right\rangle_{\hat{n}} 
    + {\cal O}(\Theta^4) \right]_{\rm b}
\end{align}
over volume. After some algebra, this gives the same heating rate as in \cref{eq:Q_dot_full}, but with a factor one (instead of two) in front of the term $(\Theta_{(0)} + \frac{\dot{a}}{a} v_{\rm b}) \Theta^{\rm B} \Theta^{{\rm C}1}$. Then, the window function describing non-Gaussian corrections of order $\Theta^3$ is $\frac{1}{2}W^{\rm (NG3)}$ instead of $W^{\rm (NG3)}$. 

If instead one performs a simple average, not weighted by number nor energy density, one needs to integrate
\begin{align}
    \frac{\rho_{\rm ex}(\eta, \vec{x})|_{\rm b}}{\rho_\gamma (\eta, \vec{x})|_{\rm b}}
    &
    = 
    \frac{a_{\rm R} \bar{T}^4 \left[ \langle (1+\Theta)^4 \rangle_{\hat{n}} -  \langle (1+\Theta)^3 \rangle_{\hat{n}}^{4/3}\right]_{\rm b}}
    {a_{\rm R} \bar{T}^4 \left[ \langle (1+\Theta)^4 \rangle_{\hat{n}} \right]_{\rm b}}
    \\
    &
    = \left[ \left\langle 2 \Theta^2 
    -2 \Theta_{(0)}^2
    +\frac{8}{3} \Theta^3
    -12 \Theta_{(0)} \Theta^2
    + \frac{28}{3} \Theta_{(0)}^3
    \right\rangle_{\hat{n}} 
    + {\cal O}(\Theta^4) \right]_{\rm b}
\end{align}
over volume. The heating rate then has a minus sign in front of the term $2(\Theta_{(0)} + \frac{\dot{a}}{a} v_{\rm b}) \Theta^{\rm B} \Theta^{{\rm C}1}$, and the window function providing non-Gaussian corrections of order three is $-W^{\rm (NG3)}$ instead of $W^{\rm (NG3)}$.

As explained before, we expect that weighting the average by the energy density is the physically correct choice, but we stress that other choices would not change our results, since we show in \cref{sec:NG_SD_Q} that non-Gaussian corrections of order three are always negligible in the context of this work.

\section{Gauge invariance of the source of spectral distortions \label{app:gauge}}

Here we discuss the gauge invariance of our starting point for the calculation of spectral distortions, \cref{eq:x_dot_branching}.  

\subsection{Gauge transformations \label{app:gauge_intro}}

The temperature anisotropy field $\Theta(\vec{x},\hat{n},\eta)$ can be expanded in spherical harmonics. Here we denote as $\Theta_{(\ell)}$ the sum of all contributions at a given multipole order $\ell$, $\Theta_{(\ell)}=\sum_m a_{lm} Y_l^m$, such that $\Theta=\sum_\ell \Theta_{(\ell)}$. We write the subscript $\ell$ between parentheses to distinguish $\Theta_{(\ell)}$ from the Legendre multipole of a function of a single angle. The monopole $\Theta_{(0)}=\langle \Theta \rangle_{\hat{n}}$ accounts for a local density inhomogeneity in the photon field. The dipole $\Theta_{(1)}=\hat{n}\cdot\vec{v}_\gamma$ describes the local bulk velocity of photons. Since we are only interested in an inhomogeneous universe with scalar perturbations, the photon velocity is a pure gradient field, $\vec{v}_\gamma = \vec{\nabla} v_\gamma$, with a velocity potential $v_\gamma$ normalised by convention to $\langle v_\gamma \rangle_{\vec{x}}=0$. The same applies to the bulk velocity of baryons, $\vec{v}_{\rm b} = \vec{\nabla} v_{\rm b}$, with $\langle v_{\rm b} \rangle_{\vec{x}}=0$.

An infinitesimal coordinate transformation $x^\mu \rightarrow x^\mu+\epsilon^\mu(\vec{x})$ induces a gauge transformation that may change the monopole and the dipole while leaving higher-order multipoles invariant. A purely spatial gauge transformation, $x^i \rightarrow x^i+\epsilon^i(\vec{x})$, acts on metric perturbations but is irrelevant for the temperature anisotropy field. A purely temporal one, $x^0 \rightarrow x^0+\epsilon^0(\vec{x})$, acts on the monopole and dipole as
\begin{align}
\Theta_{(0)} &\longmapsto \Theta_{(0)} + \frac{\dot{a}}{a} \, \epsilon_0~,
\nonumber \\
\Theta_{(1)} &\longmapsto \Theta_{(1)} - \hat{n} \cdot \vec{\nabla} \epsilon_0~,
\end{align}
with the dot denoting a derivative with respect to $x^0$. Then, using $\bar{\rho}_\gamma=a_R \bar{T}^4$ with $\dot{\bar{T}} / \bar{T} = - \dot a / a$, one finds that the photon density perturbation and bulk velocity transform as
\begin{align}
\delta \rho_\gamma &\longmapsto \delta \rho_\gamma  - \epsilon_0 \, \dot{\bar{\rho}}_\gamma~,
\nonumber \\
\vec{v}_\gamma  &\longmapsto \vec{v}_\gamma -  \vec{\nabla} \epsilon_0~,
\end{align}
or simply $v_\gamma  \longmapsto v_\gamma - \epsilon_0$.
Similarly, the baryon density and velocity perturbations transform as
\begin{align}
\delta \rho_{\rm b} &\longmapsto \delta \rho_{\rm b}  - \epsilon_0 \, \dot{\bar{\rho}}_{\rm b}~,
\nonumber \\
\vec{v}_{\rm b} &\longmapsto \vec{v}_{\rm b} -  \vec{\nabla} \epsilon_0~,
\end{align}
or simply $v_{\rm b}  \longmapsto v_{\rm b} - \epsilon_0$.
Note that, out of these perturbations, we can build at least two gauge-invariant combinations:
\begin{equation}
\vec{v}_{\gamma} - \vec{v}_{\rm b}
\qquad {\rm and} \qquad
\Theta_{(0)} + \frac{\dot{a}}{a} \, v_{\rm b}~.
\end{equation}
Up to a factor 4, the second quantity is the comoving density perturbation of photons (see e.g. \cite{Kodama:1984ziu}).

\subsection{Comoving baryon gauge \label{app:gauge_baryon}}

Starting from a general gauge where $\vec{v}_{\rm b} \neq \vec{o}$ or $v_{\rm b} \neq 0$, it is possible to go to one of the comoving baryon gauges, that is, any gauge in which baryon velocities vanish everywhere, using a temporal gauge transformation with $\epsilon_0=v_{\rm b}$. The photon velocity potential, monopole and dipole then transform as 
\begin{align}
v_\gamma & \longmapsto v_\gamma - v_{\rm b} ~,\nonumber \\
\Theta_{(0)} & \longmapsto \Theta_{(0)} + \frac{\dot{a}}{a} \, v_b~, \nonumber \\
\Theta_{(1)} & \longmapsto \Theta_{(1)} - \hat{n} \cdot \vec{\nabla} v_{\rm b} = \hat{n} \cdot (\vec{v}_\gamma - \vec{v}_{\rm b})~.
\end{align}
The prescription $\epsilon_0=v_{\rm b}$ fixes uniquely the temporal part of the gauge transformation, but the spatial part could still be arbitrary. Thus, there is an infinity of baryon comoving gauges. However, the fact that $\rho_{\rm ex}$ in \cref{eq:x_dot_branching} must be evaluated in a baryon comoving gauge is sufficient to guarantee that $\dot{s}$ or $\dot Q$ are gauge-invariant. Indeed, these quantities can be expressed entirely as a function of the temperature field $\Theta$ (and its time derivative). Thus, $\rho_{\rm ex}$ is only affected by temporal gauge transformations, and the above prescription fixes it entirely.

\subsection{Role of the monopole \label{app:gauge_monopole}}

The fact that the monopole remains in the expression of the heating rate (see e.g. eq.~\cref{eq:Q_dot_full}) may sound surprising at first sight. Since a pure monopole does not contribute to a superposition of blackbodies, we may expect that it cannot source spectral distortions. However, when the field $\Theta$ features simultaneously a monopole and some anisotropies, the monopole becomes relevant. This comes from the fact that, for the purpose of studying perturbations in the whole universe, we define temperature anisotropies relative to the global spatial-averaged (and immeasurable) quantity $\bar{T}$, rather than to the local angular-averaged (and measurable) quantity $\langle T \rangle_{\hat{n}}=\bar{T}(1+\Theta_{(0)})$. Since in each point, we can write 
\begin{equation}
T = \bar{T} \left(1 + \Theta_{(0)} + \sum_{l\geq1} \Theta_{(l)} \right)
= \bar{T} (1 + \Theta_{(0)}) \left( 1 + \sum_{l\geq1} \frac{\Theta_{(l)}}{1+\Theta_{(0)}} \right)~,
\label{eq:role_monopole}
\end{equation}
we see that the anisotropies that are effectively measurable by a local observer read $\Theta_{(l)}/(1+\Theta_{(0)})$ for any $\ell \geq 1$. Thus, the local value of $\Theta_{(0)}$ modulates the local anisotropies that enter into the calculation of the excess energy density $\rho_{\rm ex}$. This monopole correction does not contribute to the leading-order expression of $\dot Q$, but it does appear in higher-order contributions. For instance, at next-to-leading order, there is a contribution from the term $\langle \Theta_{(0)} (\Theta-\Theta_{(0)})^2 \rangle_{\hat{n}, \vec{x}}$ in the expression of 
%{\color{purple}
$\langle \rho_{\rm ex}|_{\rm b} \rangle_{\vec{x}}$.
%}
%{\color{teal} $\langle \rho_{\rm ex}/ \rho_\gamma |_{\rm b}\rangle_{\vec{x}}$}.

The presence of such monopole terms does not compromise the gauge invariance of the heating rate, since the final expression only involves the gauge-invariant quantity $(\Theta_{(0)} + \frac{\dot{a}}{a} v_{\rm b} )$ that describes the local monopole measured in the baryon rest frame.

\section{Expansion of two-point correlation function in multipoles\label{app:two_point}}

In order to compute $\dot{Q}$ with \cref{eq:QBC1}, we need to evaluate the two-point correlation function $\langle \Theta^{\rm B} \Theta^{\rm C1}\rangle_{\hat{n},\vec{x}}$. We expand $\Theta^X$ with $X= {\rm B}, {\rm C}1$ in Fourier space:
\begin{align}
    \Theta^X(\eta, \vec{x}, \hat{n}) = \int \frac{d^3\vec{k}}{(2\pi)^{3/2}} e^{-i\vec{k}\cdot\vec{x}}
    \Theta^X(\vec{k},\hat{n},t)~.
\end{align}
As long as the CMB temperature anisotropy $\Theta$ obeys to a linear equation of evolution for scales entering the Hubble radius and smaller, we can write 
\begin{equation}
    \Theta^X(\vec{k},\hat{n},t) = \Theta^X(k,\mu,t) {\cal R}(\vec{k})~,
\end{equation}
where the transfer function $\Theta^X(k,\mu,t)$ only depends on the wavenumber, the cosine $\mu$ of the angle between $\vec{k}$ and $\hat{n}$, and time $t$. ${\cal R}(\vec{k})$ stands for the Fourier modes of the primordial curvature perturbation. This result remains true even when the primordial curvature ${\cal R}(\vec{k})$ obeys non-Gaussian statistics. Then, we can expand this transfer function in Legendre multipoles,
\begin{equation}
\Theta^X(k,\mu,t) = \sum_\ell (-i)^\ell (2\ell+1) \Theta^X_\ell(k) P_\ell(\mu)~,
\end{equation}
such that
\begin{align}
    \Theta^X(\eta, \vec{x}, \hat{n}) = \int \frac{d^3\vec{k}}{(2\pi)^{3/2}} e^{-i\vec{k}\cdot\vec{x}}
\sum_\ell (-i)^\ell (2\ell+1) \Theta^X_\ell(k) P_\ell(\mu) {\cal R}(\vec{k})~.
\label{eq:app:that_X_transfo}
\end{align}
We could also take the complex conjugate of the right-hand side of \cref{eq:app:that_X_transfo}, since the left-hand side is a real function.
Then, the function of $\Theta^{\rm B} \Theta^{{\rm C}1}$ of $(\eta, \vec{x}, \hat{n})$ can be expanded as
\begin{align}
    \Theta^{\rm B} \Theta^{{\rm C}1} =
     \int \frac{d^3\vec{k} d^3\vec{k}'}{(2\pi)^{3}} e^{-i(\vec{k}-\vec{k}')\cdot\vec{x}}
\sum_{\ell,\ell'} (-i)^{(\ell-\ell')} 
& (2\ell+1)(2\ell'+1) \Theta^{\rm B}_\ell(k) \Theta^{{\rm C}1}_{\ell'}(k') P_\ell(\mu) P_{\ell'}(\mu')
\nonumber \\
& \times {\cal R}(\vec{k}) {\cal R}^*(\vec{k}')~.
\end{align}
We want to evaluate the average of the left-hand side over $\hat{n}$, $\vec{x}$. As usual in cosmology, we make the assumption of ergodicity and  replace this average with one over all the realisations of the stochastic theory, which amounts to evaluating the left-hand side with 
$\langle \Theta^{\rm B} \Theta^{{\rm C}1} \rangle_{\vec{x}, \hat{n}}$ and the right-hand side with
$
\langle {\cal R}(\vec{k}) {\cal R}^*(\vec{k}')
\rangle_{\rm realisations}
$. In what follows, we remove the subscript denoting  an average over realisations and we replace the curvature covariance by
\begin{equation}
\langle {\cal R}(\vec{k}) {\cal R}^*(\vec{k}') \rangle = 
P_{\cal R} (k) \, \delta^{(3)}(\vec{k}-\vec{k'})~,
\label{eq:app:def_PR}
\end{equation}
which only assumes statistical homogeneity and isotropy in the universe and remains true even for non-Gaussian primordial perturbations.
%$P_{\cal R} (k)$ is the primordial curvature power spectrum, while ${\cal P}_{\cal R} (k)\equiv\frac{k^3}{2\pi^2} P_{\cal R} (k)$ denotes the primordial {\it dimensionless} power spectrum.
We can then perform one integral and obtain
\begin{align}\label{eq:app:quad_sq_G}
    \langle \Theta^{\rm B} \Theta^{{\rm C}1} \rangle =
     \int \frac{d^3\vec{k}}{(2\pi)^{3}}
\sum_{\ell,\ell'} (-i)^{(\ell-\ell')} (2\ell+1)(2\ell'+1) \Theta^{\rm B}_\ell(k) \Theta^{{\rm C}1}_{\ell'}(k) P_\ell(\mu) P_{\ell'}(\mu) P_{\cal R} (k)~.
\end{align}
We expand the integral on the remaining wave-vector as $d^3\vec{k} = k^2 dk \, d\hat{k}$, and perform the integration over the angular part using
\begin{align}
    \int d\hat{k} P_\ell(\mu) P_{\ell'}(\mu) &= 
    \int_0^\pi d\theta \, \sin\theta \int_0^{2\pi} d\phi
    P_\ell(\cos \theta) P_{\ell'} (\cos \theta) \\
    & = 2\pi \int_{-1}^{1} dx P_\ell(x) P_{\ell'} (x)
    = 2\pi \frac{2}{2\ell+1} \delta_{\ell \ell'}~.
\end{align}
We obtain the result shown in \cref{eq:theta2}:
\begin{align}
    \langle \Theta^{\rm B} \Theta^{{\rm C}1}  \rangle_{\vec{x}, \hat{n}} &=
     \int \frac{dk \, k^2}{(2\pi)^{3}}
\sum_{\ell} 4\pi  (2\ell+1) 
\Theta^{\rm B}_\ell(k) \Theta^{{\rm C}1}_\ell(k)
  P_{\cal R} (k)~.
\label{eq:app:theta2}
\end{align}

\section{Bispectrum of multipoles with local non-Gaussianity\label{app:pert_NG_calc}}

Here, our goal is to expand and compute the three-point correlator 
\begin{equation}
B_{\ell_1\ell_2\ell_3}\equiv \langle \Theta_{(\ell 1)}(\vec{x}, \hat{n},\eta)\, \Theta_{(\ell 2)}(\vec{x}, \hat{n},\eta)\,\Theta_{(\ell 3)}(\vec{x}, \hat{n},\eta) \rangle_{\vec{x}, \hat{n}}~,
\end{equation}
which appears in \cref{eq:Q_cubic}. Following the same steps as in \cref{sec:Q_order_two}, we expand this product as
\begin{eqnarray}
B_{\ell_1\ell_2\ell_3}
&=& \left\langle \int \frac{d^3\vec{k}_1 d^3\vec{k}_2 d^3\vec{k}_3}{(2\pi)^{9/2}} e^{-i(\vec{k}_1+\vec{k}_2+\vec{k}_3)\cdot\vec{x}} (-i)^{\ell_1 + \ell_2 + \ell_3} (2\ell_1+1)(2\ell_2+1)(2\ell_3+1)\right. \nonumber \\
&&\left. \times \, \Theta_{\ell_1}(k_1) \Theta_{\ell_2}(k_2) \Theta_{\ell_3}(k_3) P_{\ell_1}(\hat{k}_1\cdot \hat{n}) P_{\ell_2}(\hat{k}_2\cdot \hat{n}) P_{\ell_3}(\hat{k}_3\cdot \hat{n}) 
     {\cal R}(\vec{k}_1) {\cal R}(\vec{k}_2) {\cal R}(\vec{k}_3)\right\rangle~.\nonumber \\
\end{eqnarray}
Assuming ergodicity, we replace the average over $(\vec{x}, \hat{n})$ by an average over stochastic realisations of the curvature perturbation at a fixed position $\vec{x}=\vec{{\rm o}}$ and direction $\hat{n}$. Next, we use the leading-order curvature bispectrum of \cref{eq:three-point-fn-dirac} to express this as
\begin{eqnarray}
B_{\ell_1\ell_2\ell_3}
&=&  \tilde{f}_{\rm NL}\int \frac{d^3\vec{k}_1 d^3\vec{k}_2 d^3\vec{k}_3}{(2\pi)^{9/2}}  (-i)^{\ell_1 + \ell_2 + \ell_3} (2\ell_1+1)(2\ell_2+1)(2\ell_3+1) \nonumber \\
&&\times \, \Theta_{\ell_1}(k_1) \Theta_{\ell_2}(k_2) \Theta_{\ell_3}(k_3) P_{\ell_1}(\hat{k}_1\cdot \hat{n}) P_{\ell_2}(\hat{k}_2\cdot \hat{n}) P_{\ell_3}(\hat{k}_3\cdot \hat{n}) \nonumber \\
     && \times \, \left[P_{\cal R}(k_1)P_{\cal R}(k_2) + P_{\cal R}(k_2)P_{\cal R}(k_3) + P_{\cal R}(k_3)P_{\cal R}(k_1)\right]  \delta^{(3)} (\vec{k}_1+\vec{k}_2+\vec{k}_3)~.
     \nonumber \\
&& ~~~\label{eq:substitute-equal-bispec} 
\end{eqnarray}
We replace the Dirac delta-function in terms of spherical harmonics (in their complex representation), Wigner $3j$-symbols and spherical Bessel functions of the first kind (see e.g. Eq.~(6.30) in \cite{Pettinari:2013ieg}),
\begin{eqnarray}
    \delta^{3}(\vec{k}_1+\vec{k}_2+\vec{k}_3)&=&
    8 \sum_{L_i,M_i} i^{L_1+L_2+L_3} \sqrt{\frac{\prod_i \left(2 L_i+1\right)}{4 \pi}}\left(\begin{array}{ccc}L_1 & L_2 & L_3 \\ 0 & 0 & 0\end{array}\right)\left(\begin{array}{ccc}L_1 & L_2 & L_3 \\ M_1 & M_2 & M_3\end{array}\right) \nonumber \\
    && \times \, Y_{L_{1}}^{M_{1}}(\hat{k}_1)Y_{L_{2}}^{M_{2}}(\hat{k}_2)Y_{L_{3}}^{M_{3}}(\hat{k}_3) \int dr r^2 j_{L_1}(rk_1)j_{L_2}(rk_2)j_{L_3}(rk_3)~.\nonumber \\
\end{eqnarray}
Inserting this in \cref{eq:substitute-equal-bispec}, we find that the three-point correlator reads
\begin{eqnarray}
B_{\ell_1\ell_2\ell_3}
&=&
24 \tilde{f}_{\rm NL}\int \frac{d^3\vec{k}_1 d^3\vec{k}_2 d^3\vec{k}_3}{(2\pi)^{9/2}}  (-i)^{\ell_1 + \ell_2 + \ell_3} 
(2\ell_1+1)(2\ell_2+1)(2\ell_3+1)
\nonumber \\
&&\times \, \Theta_{\ell_1}(k_1) \Theta_{\ell_2}(k_2) \Theta_{\ell_3}(k_3) P_{\ell_1}(\hat{k}_1\cdot \hat{n}) P_{\ell_2}(\hat{k}_2\cdot \hat{n}) P_{\ell_3}(\hat{k}_3\cdot \hat{n}) 
P_{\cal R}(k_1)P_{\cal R}(k_2) \nonumber \\
     &&  \times \,  \sum_{L_i,M_i} i^{L_1+L_2+L_3} \sqrt{\frac{\left(2 L_1+1\right)\left(2 L_2+1\right)\left(2 L_3+1\right)}{4 \pi}} 
     \left(\begin{array}{ccc}L_1 & L_2 & L_3 \\ 0 & 0 & 0\end{array}\right) \left(\begin{array}{ccc}L_1 & L_2 & L_3 \\ M_1 & M_2 & M_3\end{array}\right)
     \nonumber\\ &&
     \times \, Y_{L_{1}}^{M_{1}}(\hat{k}_1)Y_{L_{2}}^{M_{2}}(\hat{k}_2)Y_{L_{3}}^{M_{3}}(\hat{k}_3)  \int dr r^2 j_{L_1}(rk_1)j_{L_2}(rk_2)j_{L_3}(rk_3)~,\label{quadrupole-wignersymbols}
\end{eqnarray}
where we used the fact that the three permutations inside the bispectrum are equal. 
We then expand each Legendre multipole in spherical harmonics, such that for each $i=1,2,3$,
\begin{eqnarray}
    \int d\hat{k}_i \, P_{\ell_i}(\hat{k_i}\cdot \hat{n})\,
    Y_{L_i}^{M_i}(\hat{k}_i)
    &=& \int d\hat{k}_i \left( \sum_{m_i=-l_i}^{l_i} \frac{4\pi}{(2 \ell_i+1)} 
    Y_{\ell_i}^{m_i}(\hat{n})
    Y_{\ell_i}^{m_i*}(\hat{k_i})
    \right) Y_{L_i}^{M_i}(\hat{k}_i) \nonumber \\
    &=& \sum_{m_i=-l_i}^{l_i} \frac{4\pi}{(2 \ell_i+1)} 
    Y_{\ell_i}^{m_i}(\hat{n}) 
    \delta_{\ell_i,L_i} \delta_{m_i,M_i}~,
\end{eqnarray}
where we have used the orthogonality of spherical harmonics. We are now left with
\begin{eqnarray}
B_{\ell_1\ell_2\ell_3}
&=&
24 \tilde{f}_{\rm NL} (4\pi)^3 \int \frac{dk_1 dk_2 dk_3}{(2\pi)^{9/2}} k_1^2 k_2^2 k_3^2 \,
 \Theta_{\ell_1}(k_1) \Theta_{\ell_2}(k_2) \Theta_{\ell_3}(k_3)  
P_{\cal R}(k_1)P_{\cal R}(k_2) \nonumber \\
     &&  \times \,  \sum_{m_i}
      \sqrt{\frac{\left(2 \ell_1+1\right)\left(2 \ell_2+1\right)\left(2 \ell_3+1\right)}{4 \pi}} 
     \left(\begin{array}{ccc}\ell_1 & \ell_2 & \ell_3 \\ 0 & 0 & 0\end{array}\right) \left(\begin{array}{ccc}\ell_1 & \ell_2 & \ell_3 \\ m_1 & m_2 & m_3\end{array}\right)
     \nonumber\\ &&
     \times \, Y_{\ell_{1}}^{m_{1}}(\hat{n})Y_{\ell_{2}}^{m_{2}}(\hat{n})Y_{\ell_{3}}^{m_{3}}(\hat{n})  \int dr r^2 j_{\ell_1}(rk_1)j_{\ell_2}(rk_2)j_{\ell_3}(rk_3)~.
\end{eqnarray}
Without loss of generality, we can assume that the unit vector $\hat{n}$ points in the direction $\theta=0$, such that $Y_{\ell_{i}}^{m_{i}}(\hat{n})=\sqrt{(2\ell_i+1)/4\pi} \,\, \delta_{m_i,0}$. Then we obtain a compact result, and
\begin{eqnarray}
B_{\ell_1\ell_2\ell_3}
&=&
24 \tilde{f}_{\rm NL} (4\pi) \int \frac{dk_1 dk_2 dk_3}{(2\pi)^{9/2}} k_1^2 k_2^2 k_3^2 
 \Theta_{\ell_1}(k_1) \Theta_{\ell_2}(k_2) \Theta_{\ell_3}(k_3)  
P_{\cal R}(k_1)P_{\cal R}(k_2) \nonumber \\
     &&  \times \,  
      \left(2 \ell_1+1\right)\left(2 \ell_2+1\right)\left(2 \ell_3+1\right)
     \left(\begin{array}{ccc}\ell_1 & \ell_2 & \ell_3 \\ 0 & 0 & 0\end{array}\right)^2
      \int dr r^2 j_{\ell_1}(rk_1)j_{\ell_2}(rk_2)j_{\ell_3}(rk_3)~.\nonumber\\
\end{eqnarray}
For a numerical evaluation, this quadruple integral can be conveniently reordered as
\begin{eqnarray}
B_{\ell_1\ell_2\ell_3}
&=&
\frac{48 \tilde{f}_{\rm NL}}{(2\pi)^{7/2}}
\left[ \prod_i \left(2 \ell_i+1\right)\right]
     \left(\begin{array}{ccc}\ell_1 & \ell_2 & \ell_3 \\ 0 & 0 & 0\end{array}\right)^2
\int dr \, r^2
\left[\int dk_1 \, k_1^2 P_{\cal R}(k_1) \Theta_{\ell_1}(k_1) j_{\ell_1}(rk_1)\right]     \nonumber \\
&& \times\, 
\left[\int dk_2 \, k_2^2 P_{\cal R}(k_2) \Theta_{\ell_2}(k_2) j_{\ell_2}(rk_2)\right]
\left[\int dk_3 \, k_3^2 \Theta_{\ell_3}(k_3) j_{\ell_3}(rk_3)\right]~.\nonumber\\
\end{eqnarray}
The integrals over $k_1$ and $k_2$ are identical. In practice, one can first evaluate the one-dimensional integrals over $k_1$ and $k_3$ for each $r$, and then integrate over $r$. 
%{\color{blue} (For $B_{222}$ the prefactor before the integrals reads
%\begin{equation}
%\frac{48 \tilde{f}_{\rm NL}}{(2\pi)^{7/2}} 5^3 \frac{2}{35}
%= \frac{8.75 \tilde{f}_{\rm NL}}{7.2^{3/2} \pi^{7/2}}~.
%\end{equation}}

The integral over $r$ could also be performed analytically (see \cite{Mehrem:2010qk,Jackson}). However, the analytical result can be quite cumbersome, unless one of the indices $(\ell_1, \ell_2, \ell_3)$ vanishes and the other two are identical. Fortunately, this applies to our case, since for our purposes we are only interested in the correlator $B_{022}=B_{202}=B_{220}$, which reads 
\begin{eqnarray}
B_{220}
&=&
\frac{15 \sqrt{2} \tilde{f}_{\rm NL}}{ \pi^{7/2}}
\int dk_1 \, k_1^2 \Theta_{2}(k_1) P_{\cal R}(k_1) 
\int dk_2 \, k_2^2 \Theta_{2}(k_2) P_{\cal R}(k_2) 
\int dk_3 \, k_3^2 \Theta_{0}(k_3) \nonumber \\
%\int dk_1 dk_2 dk_3 \, k_1^2 \, k_2^2 \, k_3^2
%\Theta_{\ell_1}(k_1) \Theta_{\ell_2}(k_2) \Theta_{\ell_3}(k_3)
%P_{\cal R}(k_1)P_{\cal R}(k_2)
&& \times J_{220}(k_1,k_2,k_3)~, 
\label{eq:B_220_PNG}
\end{eqnarray}
with
\begin{equation}
J_{220}(k_1,k_2,k_3)
\equiv \int dr r^2 j_{2}(rk_1)j_{2}(rk_2)j_{0}(rk_3)~.
\label{eq:def_J}
\end{equation}
Defining $\Delta = \frac{k_1^2+k_2^2-k_3^2}{2k_1k_2}$, we have
\begin{equation}
J_{220}(k_1,k_2,k_3) = 
\left\{
\begin{tabular}{ll}
$\frac{\pi}{4 k_1 k_2 k_3} P_2(\Delta)$
& if $0 \leq \Delta \leq 1$, \\
0 & otherwise.
\end{tabular}
\right.
\label{eq:res_J}
\end{equation}

\section{Numerical evaluation of the bispectrum of temperature multipoles \label{app:PNG_Integrand_precision_check}}

\begin{figure}[H]
    \centering
    \includegraphics[width=15cm,angle=0]{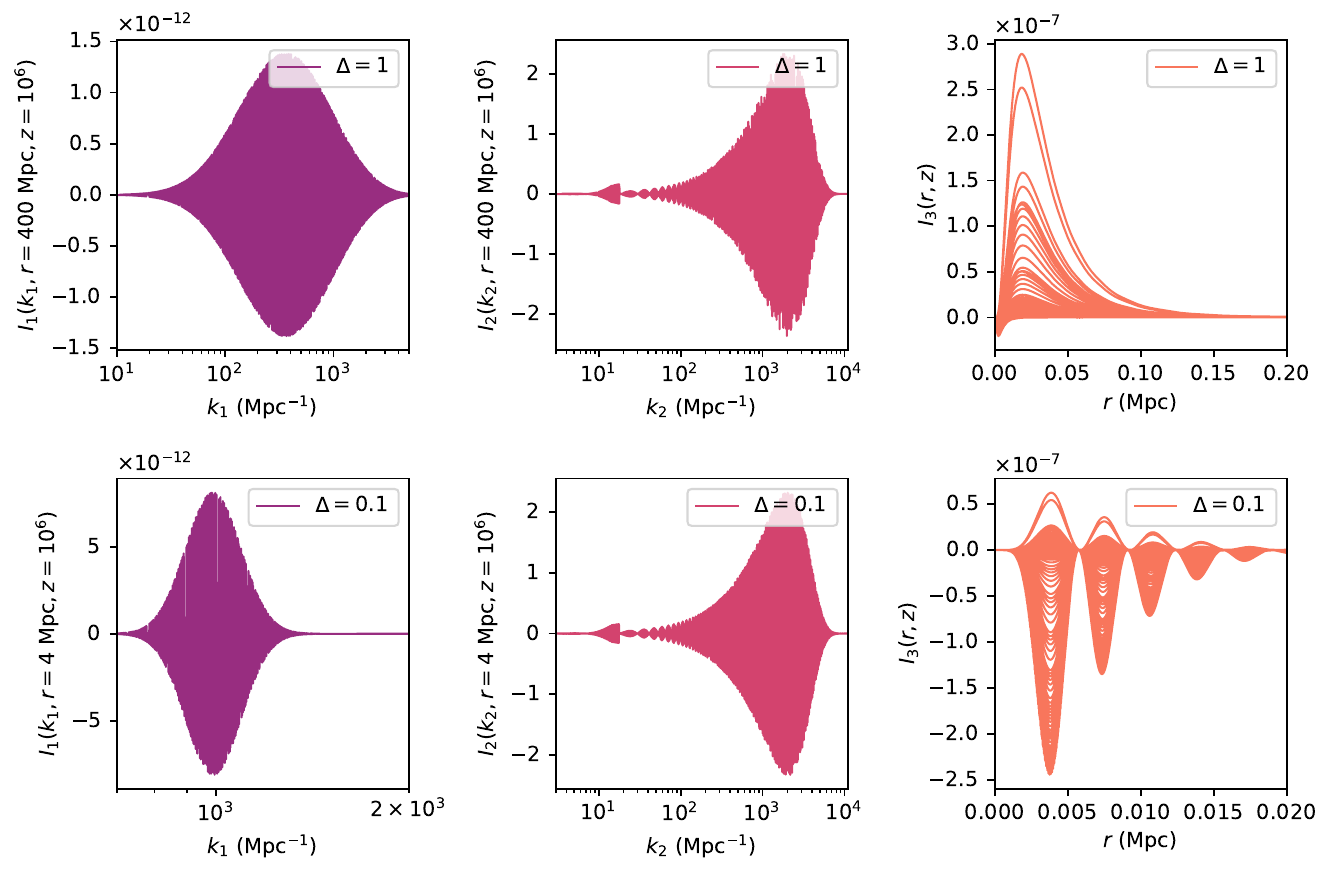}
    \caption{Integrands $I_1(k_1,r,z)$, $I_2(k_2,r,z)$ for selected values of $(r,z)$, and $I_3(r,z)$ for discrete values of $z$ in the range from $0$ to $5\times10^6$. Here we assume a log-normal spectrum ${\cal P}_{{\cal R}_{\rm G}}(k)$ with a peak at $k_*=10^3 \, {\rm Mpc}^{-1}$ and either $\Delta=1$ (top panels) or $\Delta=0.1$ (lower panels).}
    \label{fig:Integrand_kp_k1k2_k3_r}
\end{figure}

\begin{figure}[H]
    \centering
    \includegraphics[width=15cm,angle=0]{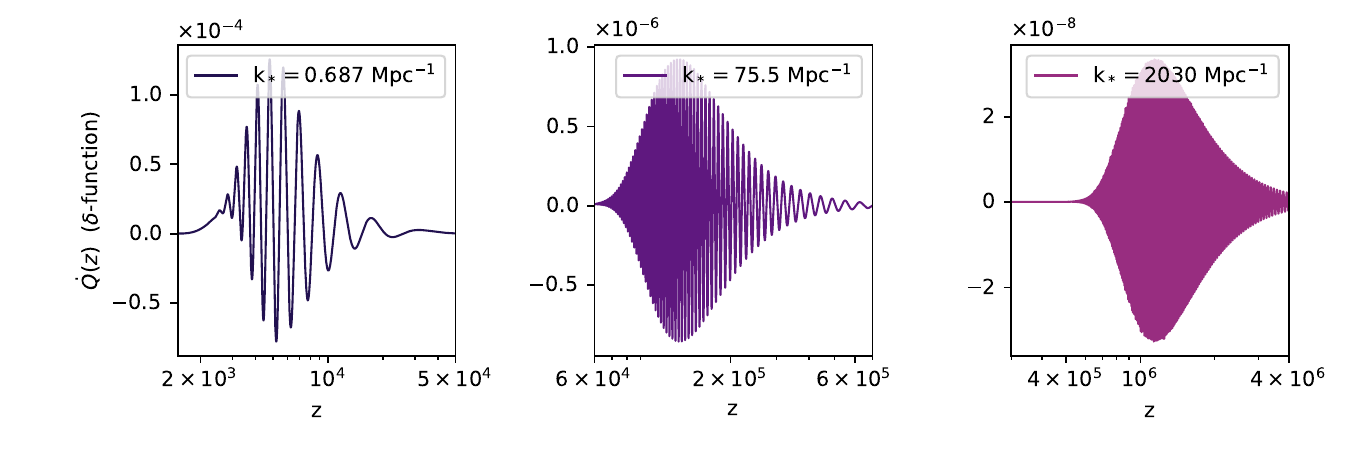}
    \caption{Heating rate $\dot{Q}(z)$ sourced by the bispectrum of temperature multipoles for three values of $k_*$, computed in the limit of a delta-function spectrum ${\cal P}_{{\cal R}_{\rm G}}(k)$ using \cref{eq:theta3_Dirac_analytic}.
    }
    \label{fig:Integrand_kp_z_diracDelta}
\end{figure}

Since the calculation of the window function $W^{\rm (NG3)}$ within \CLASS{} is numerically challenging, we provide here a few details on this calculation. The difficult task is to compute the term $\langle \tilde{\Theta}_{(0)} (\Theta_{(0)})^2 \rangle$, expressed as a multiple integrals over transfer functions, power spectra and spherical Bessel functions in \cref{eq:Integral-pert-NG}. When the power spectra are Dirac delta-functions, we can perform all integrals but one analytically and reduce the computation to \cref{eq:theta3_Dirac_analytic}. Our final result for $W^{\rm (NG3)}(k_*)$ was obtained using the latter formula. However, as a consistency check, we also integrated the full expression \cref{eq:Integral-pert-NG} for a log-normal power spectrum of width $\Delta$,
\begin{equation}
    {\cal P}_{{\cal R}_{\rm G}}(k)
    = \frac{1}{\sqrt{2\pi}\Delta}
    \exp
    \left[-\frac{1}{2} 
    \left(
    \frac{\ln (k/k_*)}{\Delta}
    \right)^2 
    \right]~,
\end{equation}
using either $\Delta=1$ or $\Delta=0.1$. We checked that when $\Delta$ decreases, the result from \cref{eq:Integral-pert-NG} converges towards that from  \cref{eq:theta3_Dirac_analytic}.
We introduce shorthand notations for the integrands in \cref{eq:Integral-pert-NG}: 
\begin{align}
    I_1(k_1,r,z) &=  P_{{\cal R}_{\rm G}}\!(k_1)  \,\,\, k_1^2 \,\,\, \Theta_{2}(k_1,z) \,\,\,j_{2}(rk_1)~, \label{eq:Ik1}\\
    I_2(k_2,r,z) &= k_2^2 \,\,\, \Theta_{0}(k_2,z) \,\,\, j_{0}(rk_2)~, 
    \label{eq:Ik3}\\
    I_3(r,z)  &=   \, \frac{15 \sqrt{2} \tilde{f}_{\rm NL}}{ \pi^{7/2}} \,\, r^2  \,\, \left(\int^{\infty}_0 \!\!\! dk_1 \, I_1(k_1,r,z) \right)^2 \,\,\, \left(\int^{\infty}_0 \!\!\! dk_2 \,  I_2(k_2,r,z) \right)~, \\
    \dot{Q}(z) &= -6 \bar{\rho}_\gamma \dot{\tau} \, \langle 
    \tilde{\Theta}_{(0)}
    (\tilde{\Theta}_{(2)})^2 \rangle
    = -6 \bar{\rho}_\gamma \dot{\tau} 
    \int_0^{\infty} \!\!\! dr \,\, I_3(r,z)~.
    \label{eq:Ir}
\end{align}
To illustrate the behaviour of these integrands, in
\cref{fig:Integrand_kp_k1k2_k3_r}, we show $I_1(k_1,r,z)$ and $I_2(k_2,r,z)$ as a function of $k_1$ or $k_2$ for selected values of $(r,z)$, and $I_3(r,z)$ as a function of $r$ for many values of $z$ ranging from $0$ to $5\times10^6$. These plots assume $k_*=10^3 \, {\rm Mpc}^{-1}$ and $\Delta=1$ (top panels) or $\Delta=0.1$ (lower panels).
Since the spherical Bessel functions are highly oscillatory and slowly decaying, it is important to choose some appropriate (logarithmic) step sizes and upper bounds in the numerical evaluation of the integrals.  Finally, we show in \cref{fig:Integrand_kp_z_diracDelta} the behaviour of $\dot{Q}(z)$ for three values of $k_*$, computed in the limit of a delta-function spectrum ${\cal P}_{{\cal R}_{\rm G}}(k)$ using \cref{eq:theta3_Dirac_analytic}. We see that even the heating rate sourced by the bispectrum of temperature multipoles is highly oscillatory.

\bibliographystyle{utphys}
\bibliography{references}

\end{document}